\documentclass[a4paper,12pt]{book}
\usepackage{lettrine} 
\usepackage{fancyhdr}
\usepackage{lipsum}
\usepackage{pdfpages}
\usepackage{siunitx}
\usepackage{mathtools}
\usepackage{tikz}

\newcommand{\initial}[1]{
	\lettrine[lines=3,lhang=0.33,nindent=0em]{
		\color{black}
     		{\textsc{#1}}}{}}

\usepackage{amssymb}
\usepackage{lastpage}
\usepackage{notoccite}
\usepackage{graphicx} 
\usepackage{url}

\usepackage[citestyle=numeric,style=numeric,backend=bibtex,sorting=none,maxbibnames=99]{biblatex}
\usepackage{amsmath,bm}
\usepackage{titlesec}
\usepackage{verbatim}
\usepackage{indentfirst} 

\usepackage[a4paper, margin=2.6cm]{geometry}
\usepackage{multicol}
\usepackage{multirow}
\usepackage{bm}
\usepackage{mathrsfs}
\usepackage{color}
\usepackage{hyperref}
\usepackage{lmodern}

\hypersetup{
    colorlinks,
    linkcolor=black,
    anchorcolor=black,
    citecolor=black,
    filecolor=black,
    menucolor=black,
    runcolor=black,
    urlcolor=black,}

\usepackage{tocloft}

\begin{document}


\begin{center}
\thispagestyle{empty}
\begin{Large}
\textbf{UNIVERSIDADE FEDERAL DO PARANÁ} \\
\end{Large}
\begin{large}
\vskip 0.2cm

\vskip 0.2cm

\end{large}

\vskip 7cm

\rule[0.5ex]{\linewidth}{2pt}\vspace*{-\baselineskip}\vspace*{3.2pt}
\rule[0.5ex]{\linewidth}{1pt}\\[\baselineskip] 
\begin{Huge} 
\vskip 1mm 
\textbf{Effects of the local dynamics in the synchronization of neural models}
\end{Huge} \vskip 3mm
\rule[0.5ex]{\linewidth}{1pt}\vspace*{-\baselineskip}\vspace{3.2pt}
\rule[0.5ex]{\linewidth}{2pt}\\

\vskip 2.5cm
\begin{large}
\textbf{Bruno R. R. Boaretto}
\end{large}
\vskip 5cm

\vskip 1.5cm
\large{Curitiba}\\
$24^\mathrm{th}$ February, 2022
\clearpage
\end{center}


\begin{center}
\thispagestyle{empty}
\begin{Large}
\textbf{UNIVERSIDADE FEDERAL DO PARANÁ} \\
\end{Large}
\begin{large}
\vskip 0.2cm

\vskip 0.2cm

\end{large}

\vskip 7cm

\rule[0.5ex]{\linewidth}{2pt}\vspace*{-\baselineskip}\vspace*{3.2pt}
\rule[0.5ex]{\linewidth}{1pt}\\[\baselineskip] 
\begin{Huge} 
\vskip 1mm 
\textbf{Effects of the local dynamics in the synchronization of neural models
}
\end{Huge} \vskip 3mm
\rule[0.5ex]{\linewidth}{1pt}\vspace*{-\baselineskip}\vspace{3.2pt}
\rule[0.5ex]{\linewidth}{2pt}\\

\vskip 2.5cm
\begin{large}
\textbf{Bruno R. R. Boaretto}
\end{large}
\vskip 2cm

\begin{flushright}
\begin{minipage}{8cm}
Thesis presented as a requirement for the degree of PhD in Science in the Graduate program in Physics, Exact Sciences Sector, of the Universidade Federal do Paraná.\\
Advisor: Prof. Dr. Sergio Roberto Lopes
\end{minipage}
\end{flushright}

\vskip 1.5cm
\large{Curitiba}\\
$24^\mathrm{th}$ February, 2022
\newpage
\end{center}




\thispagestyle{empty}
\begin{center}
\begin{flushright}
\begin{minipage}[b]{11cm}
\vskip 23cm 
\large{\it Aos meus pais, Claudia Reichert e Jacir Boaretto.}
\end{minipage}
\end{flushright}
\end{center}
\newpage


\begin{huge}
\thispagestyle{empty}
\begin{center}
{\textbf{AGRADECIMENTOS}}
\end{center}
\end{huge} \vskip 15mm

Agradeço aos meus pais, Claudia Reichert e Jacir Boaretto, por proverem todo o suporte e apoio necessários, permitindo que esta tese se concretizasse. Agredeço também ao meu irmão, Marco, pelo convívio nos últimos anos.

Agradeço à Mariana, minha companheira, que acompanhou-me durante toda a essa jornada, demonstrando  compreensão, paciência e amor em todos os momentos.

Agradeço ao meu orientador, Prof. Dr. Sergio Roberto Lopes, pelo permanente incentivo na execução desta tese e dos demais projetos produzidos desde a iniciação científica. Homenageando-o, agradeço aos demais colaboradores que com muita humildade e interesse acreditaram em mim e nas minhas capacidades.

Agradeço aos meus amigos e colegas de grupo, Roberto e Kalel, pela amizade e cumplicidade nas discussões sobre a pesquisa e a todos os colegas do programa pela amizade e apoio nos estágios desse processo.

Agradeço aos professores que leram revisaram e corrigiram atentamente esta tese em suas diferentes versões: Profa. Dra. Sabrina Borges Lino Araujo (UFPR), Prof. Dr. Ricardo Luiz Viana (UFPR), Prof. Dr. Marcus Werner Beims (UFPR),  Prof. Dr. Thiago de Lima Prado (UFPR), Prof. Dr. Giovani Vasconcelos (UFPR), Prof. Dr. Iberê Luiz Caldas (USP), Prof. Dr. Antonio Marcos Batista (UEPG), Prof. Dr. Elbert Einstein Nehrer Macau (UNIFESP).

Finalmente, agradeço à Universidade Federal do Paraná, ao Departamento de Física e ao Programá de Pós-Graduaçao em Física por toda a estrutura cedida. Registro meus agradecimentos também a CAPES pelo auxílio financeiro.

\newpage

\newpage
\begin{center}
\thispagestyle{empty}
\vfill
\begin{minipage}{15cm}
\begin{flushright}
\vskip 20cm
\hfill \textit{``O que aprendi, acima de tudo, é a seguir em frente, pois a grande ideia é a de que, como o acaso efetivamente participa de nosso destino, um dos importantes fatores que levam ao sucesso está sob o nosso controle: o número de vezes que tentamos rebater a bola, o número de vezes que nos arriscamos, o número de oportunidades que aproveitamos."}\\
\hfill \textit{Leonard Mlodinow - O Andar do Bêbado, 2009.}
\end{flushright}

\end{minipage}

\end{center}
\clearpage


\begin{huge}
\thispagestyle{empty}
\begin{center}
{\textbf{RESUMO}}
\end{center}
\end{huge} \vskip 15mm

O comportamento cooperativo de neurônios e áreas neuronais associadas ao comportamento de sincronização se apresenta como mecanismo fundamental para o funcionamento cerebral. Além disso, níveis anormais de sincronização têm sido relacionados a estados patológicos. Ao longo desta tese, abordam-se diferentes fenômenos de sincronização que surgem por meio da dinâmica coletiva de modelos de neurônios acoplados em uma rede. Primeiramente, mostra-se uma forte correlação entre a dinâmica individual do neurônio com o comportamento global da sincronização da rede, em que a periodicidade observada no neurônio isolado é refletida em uma sincronização de fase ao considerar um acoplamento fraco. Em segundo lugar, estuda-se o papel da biestabilidade na sincronização de uma rede de neurônios idênticos, acoplados através de um esquema de campo médio. Mostra-se que a simples existência de dois estados estáveis distintos pode levar a rede a diferentes estados de sincronização, dependendo da inicialização do sistema. Por fim, é investigado o mecanismo de sincronização explosiva de uma rede neural complexa composta por neurônios não-idênticos. A presença deste regime é acompanhada por um loop de histerese na dinâmica da rede, à medida que o parâmetro de acoplamento é adiabaticamente aumentado e reduzído. Demonstra-se que as transições de sincronização abruptas estão associadas a rotas para o caos e que os mecanismos dinâmicos para a região de biestabilidade são dados em termos de uma bifurcação de sela-nó e uma crise de fronteira. Portanto, os resultados desta tese mostram uma riqueza de comportamentos de sincronização associados a pequenas mudanças na dinâmica neuronal, trazendo novos insights para o estudo teórico das redes neurais.

\vskip 5mm

\textbf{Palavras-chave:} Redes neurais, Dinâmica local, Sincronização de Fase, Transição de Sincronização, Sincronização Explosiva.

\clearpage
\begin{huge}
\thispagestyle{empty}
\begin{center}
{\textbf{ABSTRACT}}
\end{center}
\end{huge} \vskip 15mm

%
The cooperative behavior of neurons and neuronal areas associated with the synchronization behavior proves to be a fundamental neural mechanism. In addition, abnormal levels of synchronization have been related to unhealthy neural behaviors. Throughout this thesis, it is explored different synchronization phenomena which emerge through the collective dynamics of models of neurons coupled in a network. Firstly, it is shown a strong correlation between the individual dynamics of the neuron with the global behavior of the synchronization, in which the periodicity seen in the isolated neuron is reflected in a phase synchronization in the weak coupling region. Secondly, it is studied the role of bistability in the synchronization of a network of identical neurons coupled through a mean-field scheme. It is shown that the simple existence of two distinct stable states can lead the network to different states of synchronization, depending on the initialization of the system. Lastly, it is investigated the mechanism for explosive synchronization of a complex neural network composed of non-identical neurons. The presence of this regime is accompanied by a hysteresis loop on the network dynamics as the coupling parameter is adiabatically increased and decreased.  It is shown that the abrupt synchronization transitions are associated with routes to chaos. The dynamical mechanisms for the bistability region, are given in terms of a saddle-node bifurcation and a boundary crisis. Therefore, the results of this thesis show a richness of synchronization behaviors associated with small changes of the neuronal dynamics bringing new insights to the theoretical study of neural networks. 

\vskip 5mm


\newpage


\vskip 1cm

\thispagestyle{empty}


\cleardoublepage
\begingroup
\makeatletter
\let\ps@plain\ps@empty
\makeatother

\pagestyle{empty}
\listoffigures
\cleardoublepage
\endgroup


\cleardoublepage
\begingroup
\makeatletter
\let\ps@plain\ps@empty
\makeatother

\pagestyle{empty}
\listoftables
\cleardoublepage
\endgroup


\cleardoublepage
\begingroup
\makeatletter
\let\ps@plain\ps@empty
\makeatother

\pagestyle{empty}
\tableofcontents
\cleardoublepage
\endgroup


\chapter{Introduction}\label{chap:intro}

\initial{T}{he} synchronization phenomenon is studied for centuries since the scientist Christiaan Huygens reported his observations that two weakly coupled pendulum clocks become synchronized in-phase \cite{huygens1980horologium} (English translation \cite{huygens1986pendulum}). After that, this phenomenon was detected in a wide range of biological systems \cite{pecora1990synchronization,goldbeter1997biochemical,pikovsky2003synchronization}, as in the rhythmic flashing of fireflies \cite{buck1976synchronous}, in the crickets which synchronize their chirps by responding to the preceding chirp of their neighbors \cite{walker1969acoustic}, groups of women whose menstrual periods become mutually synchronized  \cite{mcclintock1971menstrual}, in the synchronous of rabbit sino-atrial pace-maker cells \cite{jalife1984mutual}, and also in the action potentials of the nervous system \cite{fell2011role,glass2001synchronization}. 

A healthy human brain is composed of $\sim 10^{11}$ neuronal cells interconnected by $\sim 10^{15}$ synapses creating groups of connected neurons divided into brain regions, each with specific functions \cite{kandel2013principles}. In particular, the complex behaviors seen in the brain are directly related to the emergence of spatial-temporal activation patterns that comes spontaneously as a result of the cooperative interaction among neurons. The role of neuroscience is to understand how does the behaviors produced by the brain like perception, movement, language, thought, memory, etc; can be explained in terms of the activity patterns of neurons since all the behavioral disorders that characterize psychiatric illness are disturbances on the brain functioning \cite{kandel2013principles}. In this context, the cooperative behavior of neurons and neuronal areas associated with the synchronization behavior proves to be a fundamental neural mechanism \cite{fell2011role}. It supports memory process \cite{fell2011role,klimesch1996memory}, information process \cite{roelfsema1997visuomotor}, and is relevant for many cognitive processes \cite{rodriguez1999perception,cavanagh2009prelude}. In addition, abnormal levels of synchronization have been related to unhealthy neural behaviors \cite{kandel2013principles}. While a high degree of synchronization is detected in epileptic seizures \cite{mormann2000mean}, where the increase in the synchronization of some groups of neurons generates seizure episodes, and in Parkinson's disease, where there is an excessive synchronization in basal ganglia \cite{hammond2007pathological,popovych2014control}, reduced levels of synchronization among cortical areas can be associated with brain disorders such as autism \cite{dinstein2011disrupted}, and Alzheimer's disease \cite{greicius2004default}.

From the theoretical point of view, the use of complex networks proves to be useful for the study of an ensemble of coupled sites, in which the global behavior of the system results from the interaction among the sites achieving an ample possibility of phenomena even in the case of simple interactions \cite{boccara2010modeling,green1993emergent}. Particularly, for neuronal systems, each site of the network is composed of a neuron and the edges of the network represent its synaptic connections. This approach can be used to improve our knowledge and give theoretical insights into the understating of brain functioning. In studies of neuronal systems, there are always two critical choices: what model describes the firing dynamics of each neuron and how the neurons are connected \cite{izhikevich2004model}. While several works are focused on understanding the role of the connection architecture, also called topology of connection, in the synchronization of dynamical systems \cite{strogatz2001exploring,arenas2008synchronization,rodrigues2016kuramoto,townsend2020dense,budzinski2018nonstationary,budzinski2019investigation,budzinski2019phase,budzinski2019synchronous}, this thesis, it is studied the sensibility of the synchronization features to the dynamics of the isolated neurons.

In the last decades, after the success of the Hodgkin-Huxley model, being the first quantitative description of the regenerative currents generating the action potential \cite{ermentrout2010mathematical,hodgkin1939action,hodgkin1952quantitative}, dozens of models were created to reproduce more complicated firing patterns exhibited by neurons \cite{izhikevich2004model,izhikevich2007dynamical}. From the simplest models described by iterated maps \cite{chialvo1995generic,rulkov1995generalized}, to more sophisticated models of several non-linear differential equations \cite{hodgkin1952quantitative,fitzhugh1961impulses,hindmarsh1984model,braun1998computer,izhikevich2003simple}. Often, by changing parameters of the models it is possible to change the dynamical behavior of the neuron, from regular activity to a chaotic one, where the activation patterns of the neurons occur in a non-periodic way \cite{ermentrout2010mathematical}. In addition, some dynamical models also can exhibit multistable states, where a neuron initialized with different initial conditions can present different stable states, with different firing patterns, frequencies, regularity, and chaoticity \cite{feudel2008complex,foss1996multistability,sainz2004influence,ma2007multistability}.

Throughout this thesis, it is studied how the individual dynamics of the neurons affect the synchronization of the neuronal network. For neurons that individually exhibit chaotic behavior, synchronization occurs similarly to the transition of chaotic oscillators known in the literature \cite{xu2018synchronization,kuramoto1975self}. However, the knowledge about the synchronization of non-chaotic neurons that lose these features when coupled, acquiring characteristics of chaotic neurons, is still being studied \cite{xu2018synchronization}. It is shown that, in some cases, these neurons tend to synchronize in phase to weak coupling regimes (a coupling parameter close to zero) due to the influence of the regularity of the individual dynamics. This synchronized state is lost with the increment of the coupling strength, due to the influence of the collective behavior of the dynamics \cite{prado2014synchronization,budzinski2017detection,boaretto2017suppression}, characterizing a {\it non-monotonic} evolution of the synchronization phenomenon as a function of coupling. 

On the other hand, for neurons that individually present periodic dynamics, i.e., a well-defined frequency, the phase synchronization of a network with identical neurons can be achieved for any non-zero coupling strength, with more weakly-coupled networks needing more time to reach the phase-synchronized state. In this sense, the existence of identical bistable neurons in the network can delay the achievement of a complete-phase-synchronized state, demanding transitions to a unique and identical state for all neurons in the network. These transitions are induced by the coupling of the network and occur according to the stability of each state, in which the less stable state transitions to the more stable state.

Moreover, the synchronization of periodic neurons also can be disturbed with the existence of heterogeneity in the network, in which non-identical-periodic neurons are coupled. Neurons simulated with different parameters may depict different activation patterns. This dissimilitude in the neuronal dynamics can influence both frequency and amplitude of each neuron generating a chaotic non-synchronized regime instead of a complete-periodic synchronization. It is show that the increase of the coupling strength can transits the system abruptly to the phase-synchronized state, characterizing an {\it explosive synchronization} \cite{gomez2011explosive,leyva2012explosive,ji2013cluster,zou2014basin,avalos2018emergent}.

Given this context, this thesis aims to understand the effect of individual dynamics in the synchronization of neural networks. This objective can be divided into three independent topics, described as:
\begin{itemize}
    \item Understand the non-monotonic evolution of the synchronization as a function of the coupling in neuronal networks where the chaoticity arises from the synaptic currents.
    \item Investigate the role of bistability in the synchronization of a network of identical bursting neurons coupled through a generic electrical mean-field scheme.
    \item  Study the mechanism for explosive synchronization of a complex neural network composed of non-identical spiking neurons and coupled through a small-world network. 
\end{itemize}

This thesis is divided into two parts, the first part, from Chapters 2 to 5, is devoted to introducing all the theoretical frameworks used in this work, while the results and conclusions are presented in the second part from Chapters 6 to 9.

Chapter \ref{chap:sistemas} presents a general review of the main concepts related to dynamical systems and nonlinear dynamics, like fixed points, stability of fixed points, dynamical features of chaotic systems, Lyapunov exponent, and routes to chaos. Chapter \ref{chap:modelos} is focused on the presentation of properties of physiological neurons. Moreover, it is introduced the formulation ideas of dynamical systems which display a qualitative behavior of real neurons. The models can be described by differential equations, like the Hodgkin-Huxley model, and also by iterated maps like the Chialvo model. Also, it is introduced the synaptic structures that permit the communication between presynaptic and postsynaptic neurons. After this, Chapter \ref{chap:redes} shows some concepts of the graphs theory used to construct the complex networks which rule the connections between the neurons. Finally, Chapter \ref{chap:ferramentas} explores some properties of coupled oscillators, and also it is presented a powerful tool used to quantify phase synchronization, the Kuramoto order parameter.

In Chapter \ref{chap:role}, it is studied the effect called non-monotonic synchronization, where using a Hodgkin-Huxley-like (HH$\ell$) neuron, proposed by Braun {\it et al.} \cite{braun1997low}, which in a periodic-bursting regime, it is possible to achieve phase synchronization for small values of coupling. It shows a clear relation between the individual dynamics of the isolated neuron with the type of phase synchronization transition of the network. After that, it is shown that the same correlation occurs for neurons under external perturbations and with other similar neuronal models.

By investigating the parameter space of the HH$\ell$ model, it is found a parameter region where the neuron exhibits bistability behavior. One of the states named state I is always periodic, while the other state, named state II, goes through a sequence of period-doubling bifurcations from periodic to chaotic behavior. In this scenario, Chapter \ref{chap:bistability} investigates the stability of each state, and how the existence of bistability can change the phase synchronization of a network. To isolate this behavior, the neurons are coupled through a generic mean-field scheme (all-to-all connection) where the only difference between the neurons are the initial conditions. After all, the simple existence of two distinct stable states in the network can produce distinct synchronization states.

Chapter \ref{chap:mechanism} analyzes the mechanism for an abrupt transition to phase synchronization, called explosive synchronization, which is found in a complex network composed of non-identical spiking neurons, simulated with the neuronal model proposed by Chialvo \cite{chialvo1995generic}. It is shown that this regime is accompanied by a hysteresis loop on the network dynamics as the coupling parameter is adiabatically increased and decreased, characterizing a bistability regime. The abrupt synchronization transitions are associated with routes to chaos, and the dynamical mechanisms for the bistability region are given in terms of a saddle-node bifurcation and a boundary crisis.

At last, the conclusions based on the results and open questions for further investigation are presented in Chapter \ref{chap:cons}.

\chapter{Dynamical systems and chaos
}\label{chap:sistemas}

\begin{figure}[t]
    \centering
    \includegraphics[width=\columnwidth]{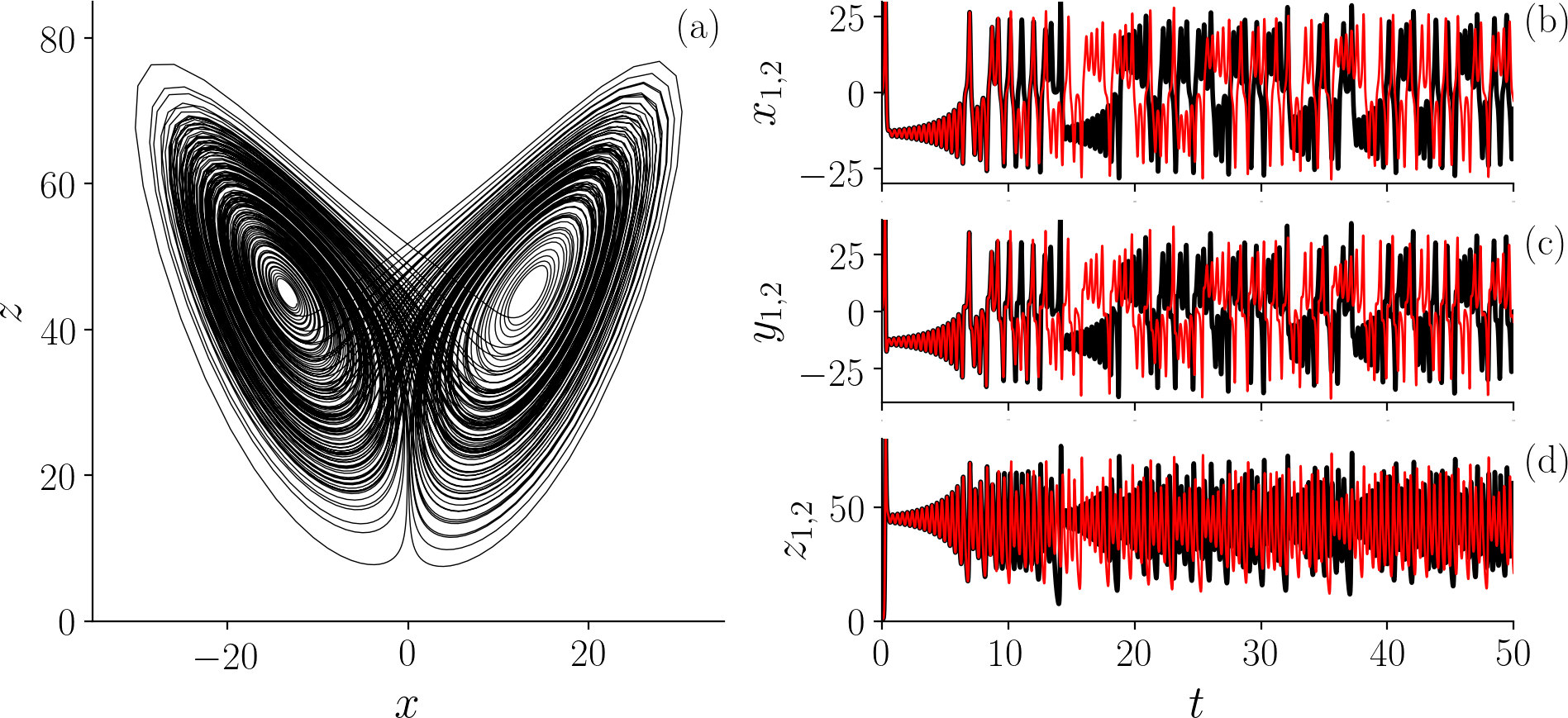}
    \caption[Projection of the Lorenz attractor in the phase space and the sensibility to initial conditions.]{\textbf{Projection of the Lorenz attractor in the phase space and the sensibility to initial conditions.} (a) Projection of the Lorenz attractor $x \times z$. (b -- d) Dynamical variables of the Lorenz system using $x_1(0)=x_2(0)=0,\,y_1(0)=y_2(0)=1\,,z_1(0)=1.05,\,z_2(0)=1.051$ and $\sigma=16$, $r=45.92$ and $b=4$, with an integration timestep of $0.01$. After some integration steps due to the chaotic behavior, even for close initial conditions the orbits tend to diverge from each other.}
    \label{fig:lorenz_a}
\end{figure}

\initial{T}{his} chapter presents some concepts about dynamical systems, which are fundamental for the understanding of the time evolution of real systems, both regular and chaotic systems. In this sense, Edward Lorenz has studied a simplified mathematical model for atmospheric convection \cite{lorenz1963deterministic,wolf1985determining,ott2002chaos}. The model is a system of three ordinary differential equations where $x$, $y$, and $z$ denote variables proportional to convective intensity, horizontal, and vertical temperature differences
\begin{eqnarray}\label{eq:lorenz}
\dot{x}&=& \sigma (y - x),\\
\dot{y}&=& x(r-z) - y,\\
\dot{z}&=& xy -bz,
\end{eqnarray}
where the $(\cdot)$ symbol represents the derivative of a function of time and $\sigma$, $r$, and $b$ are parameters, called the Prandtl number, Rayleigh number, and a geometric factor, respectively. This system becomes very popular due to the ``butterfly effect'' which emphasizes the sensitive dependence on initial conditions in which a small change in one state of a deterministic nonlinear system can result in large differences in a later state. 
Figure \ref{fig:lorenz_a} (a) presents the projection of the Lorenz attractor in the phase space, with the shape that may also be seen to resemble a butterfly, and panels (b -- d) show the time evolution of two slightly different initial conditions. The sensibility to initial conditions which evolve in such a way that their trajectories diverge exponentially is one of the signatures of a chaotic motion. This chapter shows how a dynamical system can achieve chaotic dynamics and how to quantify it.

\section{Dynamical systems}

A dynamical system is a concept in which a mathematical function describes the behavior of particles under the action of a set of laws. In this sense, for a given configuration, called initial condition, a dynamical system is deterministic and presents only one possible solution. There are two ways to describe the evolution of a dynamical system: with differential equations, which describe the evolution of the system considering time as a continuous variable, such systems are called flows; and through difference equations that consider time as a discrete variable, such systems are called maps.

Considering a continuous system with $D$ dimensions characterized by the variables $(x_1,\,x_2,$ $\,\cdots,\,x_D)$, the temporal evolution of these variables are described by the set of $D$ equations in which,
\begin{eqnarray*}
    \dot{x_1} &=& f_1(x_1,x_2,\cdots,x_D), \\ 
    \dot{x_2} &=& f_2(x_1,x_2,\cdots,x_D), \\ 
    &\vdots& \\
    \dot{x_D} &=& f_D(x_1,x_2,\cdots,x_D), 
    \end{eqnarray*}
%
or rewriting 
\begin{equation}\label{eq:cont}
    \dot{\mathbf x} = \mathbf f(\mathbf x).
\end{equation}
%
In this sense, if the elements of $\mathbf f$ presents products, power, and functions of $\mathbf x$, such as $x_i x_j$, $x_i^3$, or even $\cos(x_i x_j)$ the system is characterized nonlinear, or linear, otherwise \cite{strogatz2001nonlinear}. The system is periodic if exists a time $\mathcal T$ where:
\begin{equation}
    \mathbf x(t+\mathcal T) = \mathbf x(t).
\end{equation}

\begin{figure}[t]
    \centering
\includegraphics[page=1]{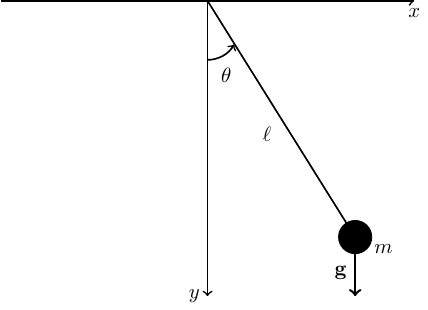}
\caption[Simple pendulum.]{\textbf{Simple pendulum}. A particle with mass $m$ attached to an inextensible wire with length $\ell$ that oscillates with an angle $\theta (t)$ around a fixed point due to the gravitational acceleration $\mathbf g$.}\label{fig:simple_pendulum}  
\label{fig:pendulo}
\end{figure}

The simple pendulum is an example of a system where the time is a continuous variable, represented in Fig. \ref{fig:pendulo}, where a particle of mass $m$ is attached to a wire of length $\ell$ which under the action of the gravitational acceleration oscillates with an angle $\theta(t)$ around a fixed point. Even though it is two-dimensional system, when considering the constraint (wire), the movement of the particle is restricted to an arc of angle $\theta$ of radius $\ell$. Consequently 
\begin{equation}
    \ddot{\theta}= -\frac{g}{\ell} \sin \theta,
\end{equation}
where $g$ is the absolute value of the gravitational acceleration, and $\ell$ is the length of the wire. The therm $\sin \theta$ characterizes the nonlinearity of the system.

The nonlinearity in the equations makes it difficult to find an analytic solution to the problem, that is, writing $\theta$ as a function of $t$. To make the things simple, the small oscillation approximation is used, where $\sin\theta \approx \theta$, 
\begin{equation}\label{eq:pendulo}
    \ddot{\theta}\approx -\frac{g}{\ell} \mathrm \theta.
\end{equation}
Considering the angular frequency $\omega = \sqrt{g/\ell}$, and as initial conditions $\theta(0) = \theta_0$, and $\dot \theta(0) = 0$, the analytical solution is described by
\begin{equation}\label{eq:linear_pendulum}
    \theta (t) = \theta_0 \cos(\omega t),
\end{equation}
and the period of oscillation is  
\begin{equation}
    \mathcal T_\mathrm{linear} = 2\pi \sqrt{\frac{\ell}{g}},
\end{equation}
such equation (Eq.(\ref{eq:pendulo})), now linear, is equivalent to the harmonic oscillator equation, where the period of oscillation is independent of the amplitude of the system.

The fact that a system does not present an analytical solution is not necessarily a limiting factor. Using any integration method it is possible to find a numerical solution (point-by-point) for the system. The nonlinear pendulum is one of the special cases in which can be solved analytically, the solution is given in terms of elliptic integrals \cite{marion2013classical}, where the period of oscillation is described by
\begin{equation}
    \mathcal T_\mathrm{nonlinear} = 4\sqrt{\frac{\ell}{g}}\mathcal F\left (\sin \frac{\theta_0}{2},\frac{\pi}{2}\right ),
\end{equation}
where $\mathcal F$ represents the elliptic function of Legendre of the first kind \cite{marion2013classical} defined as  
\begin{equation}
    \mathcal F(k,\varphi) = \int_{0}^\varphi \frac{1}{\sqrt{1-k^2\sin^2\theta}}d\theta.
\end{equation}

\begin{figure}[t]
    \centering
\includegraphics[page=2]{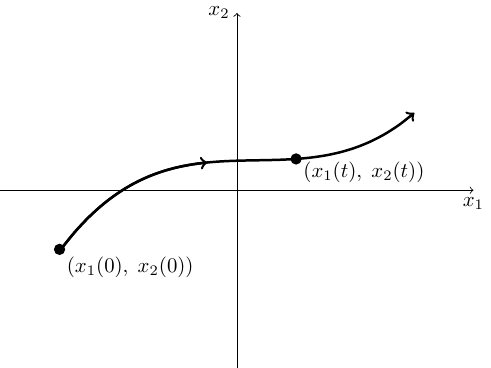}
\caption[Generic phase space ($x_1,x_2$).]{\textbf {Generic phase space ($\mathbf{x_1,x_2}$)}. For a set of initial conditions ($x_1(0),\;x_2(0)$) the solution of the system ($x_1(t),\;x_2(t)$) corresponds to a curve in the phase-space called trajectory.}\label{fig:phasespace}
\end{figure}

Considering $x_1 = \theta$ and $x_2 = \dot \theta$ the Eq. (\ref{eq:pendulo}) that has a second-order derivative, can be rewritten in terms of two first-order derivatives.
\begin{eqnarray}
    \dot{x_1} &=& x_2,\\
    \dot{x_2} &=& -\frac{g}{\ell}\sin x_1,
\end{eqnarray}
that is, a two-variable system, where $x_1$ is the position of the particle, and $x_2$ is the velocity. Assuming that the solution of this system for a given initial condition is known, this solution will be a set of functions $x_1(t)$ (position) and $x_2(t)$ (velocity). Considering an abstract space with coordinates $(x_1,\;x_2)$, then the solution $(x_1(t),\;x_2(t))$ corresponds to a point moving along a curve in that space. Figure \ref{fig:phasespace} represents this abstract space, known as \textit{phase space}, in which the curve represents the evolution of the system solution, called \textit{trajectory}.

The phase space is completely composed of trajectories and every point in phase space can be considered an initial condition. When analyzing the system equations, several restrictions or transitions can be found in its dynamics in phase space. For example, there are solutions known as equilibrium solutions where for $\mathbf x(t) = \mathbf x^*$ the temporal evolution of the system is
\begin{equation}
    \frac{d\mathbf x}{dt} \bigg |_{\mathbf x = \mathbf x ^*} = \mathbf f(\mathbf x ^*)=0
\end{equation}
and the point $\mathbf x ^*$ is denominated \textit{equilibrium point}. When the trajectory reaches such a point it is confined to it. In this sense, the equilibrium points of a system can be \textit{stable} or \textit{unstable}. When disturbing the system in the vicinity of the point, if the trajectory is attracted (repelled) the equilibrium point is considered stable (unstable). For the case of the pendulum, the equilibrium points are $\mathbf x^* = (x_1^*=j\pi,x_2^*=0)$ where $j \in \mathbb Z$. However, physically there is no difference between $0$ and $2\pi$, so only two equilibrium points are considered $\mathbf x^{(*,1)} = (0,0)$ and $\mathbf x^{ (*,2)} = (\pi,0)$, that is, the lower and upper extremes of the pendulum, respectively. As this is an idealized system (without friction, or damping), intuitively, it is possible to notice that there is a difference between the two equilibrium points. When disturbing the system at the equilibrium point $\mathbf x^{(*,2)}$, that is, the pendulum at the top end, the system is automatically repelled. On the other hand, when perturbing the point $\mathbf x^{(*,1)}$, the trajectory will be kept very close to it. Note that for this case without damping, the system will never stabilize at the equilibrium point again, because even if it returns to the point $x_1 = 0$, the velocity will be $x_2 \neq 0$.

In general, the stability of the equilibrium points can be determined by analyzing the vicinity of the points. Considering a equilibrium point $\mathbf x^*$ of a $D$-dimensional system, and a perturbation in the vicinity of the point $\delta \mathbf x = \mathbf x(t) - \mathbf x^*$, the temporal evolution of the disturbance can be described in terms of
\begin{equation*}
    \dot{\delta \mathbf x} = \frac{d}{dt}(\mathbf x(t) - \mathbf x^*),
\end{equation*}
and since $\mathbf x^*$ is a constant $\dot{\delta \mathbf x} = \dot{\mathbf x}$, and $\dot{\delta \mathbf x} = \mathbf f(\mathbf x)=\mathbf f(\mathbf x^* + \delta \mathbf x)$. The Taylor series expansion is 
\begin{equation*}
    \mathbf f (\mathbf x^* + \delta \mathbf x) = \mathbf f (\mathbf x^*) + \frac{\partial \mathbf{f}(\mathbf x)}{\partial \mathbf x}\bigg |_{\mathbf x^*}\delta \mathbf x + \mathcal O(\delta \mathbf x^2),
\end{equation*}
where $\mathbf f (\mathbf x^*) = 0$ due to the equilibrium definition and $\mathcal O(\delta \mathbf x^2)$ representing the higher-order terms of $\delta \mathbf x$. Disregarding the terms of order greater than or equal to 2, as long as the disturbance is sufficiently small,
\begin{equation}
    \dot{\delta \mathbf x} \approx \frac{\partial \mathbf f(\mathbf x)}{\partial \mathbf x}\bigg |_{\mathbf x^*}\delta \mathbf x,
\end{equation}
the partial derivative of the equation is known as \textit{Jacobian matrix} and is defined by
\begin{gather}
\mathcal J(\mathbf x^*) \equiv \frac{\partial \mathbf f(\mathbf x)}{\partial \mathbf x}\bigg |_{\mathbf x^*} 
 =
  \begin{bmatrix}
   \frac{\partial f_1}{\partial x_1} & \frac{\partial f_1}{\partial x_2} & 
   \cdots & \frac{\partial f_1}{\partial x_D} \\ 
      \frac{\partial f_2}{\partial x_1} & \frac{\partial f_2}{\partial x_2} & 
   \cdots & \frac{\partial f_2}{\partial x_D} \\
   \vdots  & \vdots  & \ddots  & \vdots \\
   \frac{\partial f_D}{\partial x_1} & \frac{\partial f_D}{\partial x_2} & 
   \cdots & \frac{\partial f_D}{\partial x_D} \\
   \end{bmatrix}_{x_1^*,x_2^*,\cdots,x_D^*},
\end{gather}
so
\begin{equation}
    \dot{\delta \mathbf x} = \mathcal J(\mathbf x^*) \delta \mathbf x,
\end{equation}
where $\mathcal J(\mathbf x^*)$ is the Jacobian matrix evaluated at the equilibrium points $\mathbf x^*$. According to the theory of differential equations, the solution can be described in terms of the complex eigenvalues of the Jacobian matrix. If the real part of all eigenvalues of $\mathcal J$ is negative, the point is stable. However, if at least the real part of one of the eigenvalues is positive, the point is unstable \cite{ott2002chaos}.

For discrete systems, instead of differential equations, it is used the iterative maps
\begin{equation}\label{eq:disc}
    \mathbf x_{t+1} = \mathbf f(\mathbf x_t), 
\end{equation}
in which the index $t$ represents the $t$-th iteration of the map. That is, the state of the variable $\mathbf x_{t+1}$ is described in terms of $\mathbf x_{t}$, where $t \in \mathbb N$. Compared to a continuous system
\begin{equation*}
    \frac{d \mathbf x}{dt} =   \lim_{\Delta t \rightarrow 0} \frac{\mathbf x(t + \Delta t) - \mathbf x(t)}{\Delta t} \approx  \frac{\mathbf x(t + \Delta t) - \mathbf x(t)}{\Delta t},
\end{equation*}
that is, ignoring the  limit of $\Delta t \rightarrow 0$, Eq. (\ref{eq:cont})
\begin{equation*}
    \frac{d \mathbf x}{dt} = \mathbf f (\mathbf x) \approx \frac{\mathbf x(t + \Delta t) - \mathbf x(t)}{\Delta t},  
\end{equation*}
isolating the term $\mathbf x(t + \Delta t)$
\begin{equation*}
    \mathbf x(t + \Delta t) \approx \mathbf x(t) + \mathbf f(\mathbf x)\Delta t,
\end{equation*}
that is, on the left, there is a state $\mathbf x(t + \Delta t)$, and on the right-side terms that depend only on $\mathbf x (t)$, something analogous to Eq. (\ref{eq:disc}). Therefore, maps can be understood as discrete approximations of continuous systems. In addition, the fact that an integrator is not required to perform the numerical calculations the maps are known to exhibit high performance in the computation of the system.

\begin{figure}[t]
    \centering
    \includegraphics[width=\columnwidth]{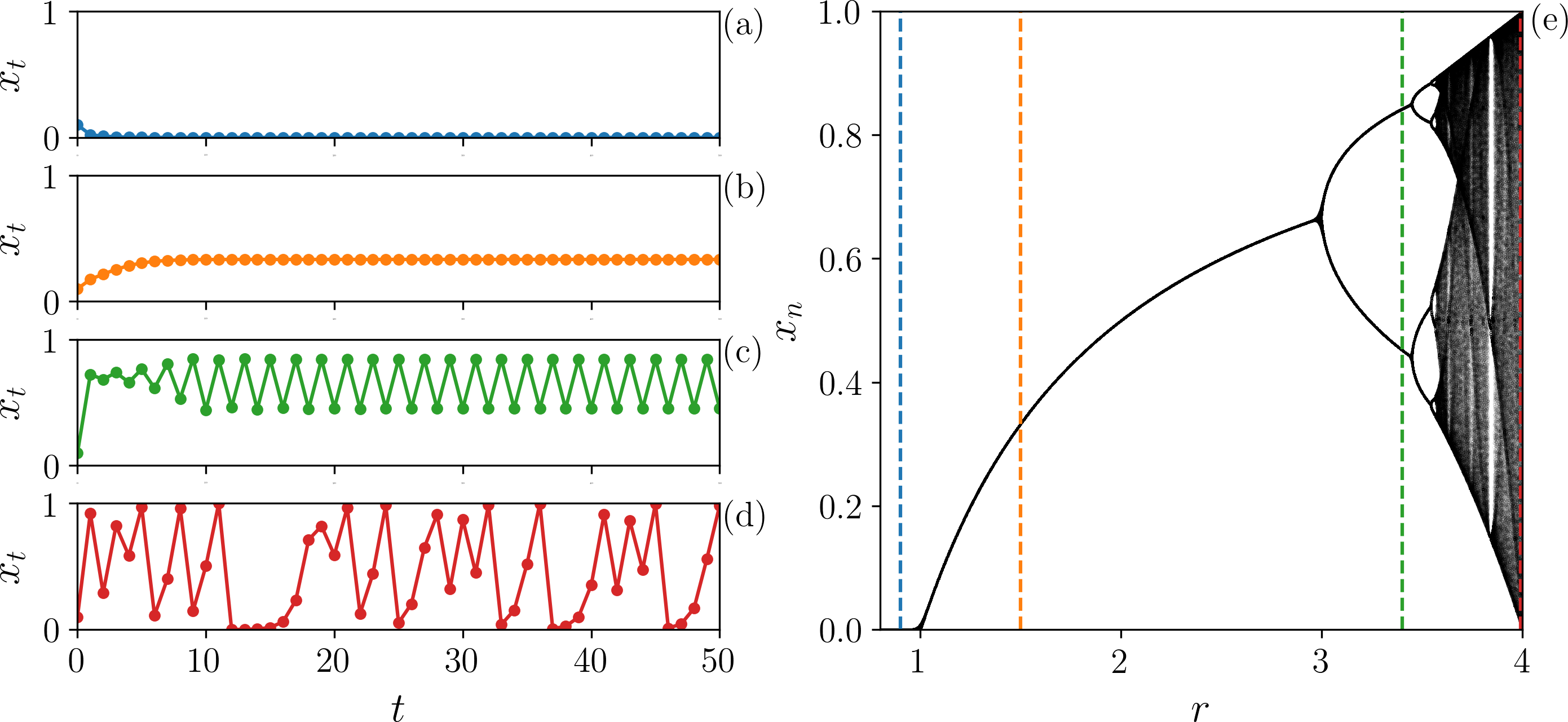}
    \caption[Iterations of the logistic map for different values of $r$.]{\textbf{Iterations of the logistic map for different values of $\mathbf r$.} (a) For $r=0.9$ after a transient time the system reaches $x_t=0$. (b) For $r=1.5$ the system reaches a fixed point. (c) For $r=3.5$ the system presents a $2$-periodic dynamics. (d) For $r=4$, there is no specific period. (e) Bifurcation diagram of the logistic map, after discarding the first $100$ iterations to avoid the transient effects, is plotted for all iterations $x_t$ as a function of $r$.} 
    \label{fig:log_1}
\end{figure}

The classic example of a map is the logistic map \cite{may2004simple}, which, despite being a discrete and one-dimensional model, by varying the single parameter of the map, different dynamical regimes can be reached, having applications in several areas such as physics \cite{ott2002chaos,strogatz2001nonlinear}, biology \cite{kendall1998spatial}, economy \cite{miskiewicz2004logistic}, electronics \cite{suneel2006electronic}, etc. The map is described by the following equation
\begin{equation}\label{eq:log}
    x_{t+1} = rx_t(1-x_t),
\end{equation}
where $r$ is a positive parameter, and $x_t$ is the ratio of existing population to the maximum possible population which varies from $0\leq x_t\leq 1$ if $0\leq r \leq 4$. In this sense, the logistic equation describes population growth, where the growth rate is controlled by the parameter $r$. In Fig. \ref{fig:log_1} it is studied the iterations for the logistic map considering different values of $r$. In panel (a) $r=0.9$, the map starts from the initial condition and evolves to a $x_t = 0$, that is, the extinction of the population. In panel (b) $r=1.5$, the map evolves to an equilibrium point $x_t=1/3$. In panel (c) $r=3.5$, the system evolves to an orbit of period $2$. In panel (d) $r=4$ the dynamics of the map presents a great variability of values of $x_t$, indicating an absence of period.

A more robust way to understand the role of a control parameter is through a \textit{bifurcation diagram}, which is a graphical representation of the qualitative behavior of a dynamical variable ($x_t$) as a function of a control parameter ($r$). Figure \ref{fig:log_1} (e) presents the bifurcation diagram as a function of $r$, in which for each value of $r$, the system evolves by $500$ iterations. Discarding the first $100$ iterations, more than the necessary time for the system to reach the equilibrium point, known as transient time, from this point all map iterations are plotted as a function of $r$. For $r<1$ the system evolves to $x_t=0$. For $1<r<3$ the system reaches an equilibrium point. For $3<r<1+\sqrt 6$ the orbit that was periodic is doubled, generating an orbit of period 2, where $x_{t+2} = f(f(x_t))$. For $1+\sqrt 6<r\lesssim 3.54409 $ each orbit splits into two, generating orbits of period $4$. As $r>3.54409$ is increased orbits of period $8$ appear, and in sequence $16,\,32,\,\dots$ This cascade of period doublings occurs repeatedly until $r \approx 3.56995$ where an infinite number of values of $x_t$ appear. This aperiodicity is one of the characteristics of chaotic behavior. For $r = 1 + \sqrt 8$ the system loses chaoticity, and presents a $3$-period dynamics, these orbits become $6$-periodic orbit, until for $3.57 \lesssim r\leq 4$ the system presents, again, chaotic behavior \cite{ott2002chaos}.
%

For discrete systems, a fixed point is defined as the point that maps itself, from Eq. (\ref{eq:disc})
\begin{equation} \label{eq:ptofixo}
 \mathbf x^* = \mathbf f(\mathbf x^*).    
\end{equation}
Considering a point in the vicinity of the fixed point $\delta \mathbf x_t = \mathbf x_t - \mathbf x^*$, and expanding $\mathbf f(\mathbf x)$ in Taylor series
\begin{equation*}
 \mathbf f(\mathbf x_t) = \mathbf f(\mathbf x^*+\delta \mathbf x_t) = \mathbf f(\mathbf x^*) + \frac{\partial \mathbf f(\mathbf x_t)}{\partial \mathbf x_t}\bigg |_\mathbf{x^*}\delta \mathbf x_t + \mathcal O(\delta \mathbf x_t^2), 
\end{equation*}
disregarding the second-order (and higher) terms, and isolating the derivative
\begin{equation*}
\mathbf f(\mathbf x^*+\delta \mathbf x_t) - \mathbf f(\mathbf x^*) \approx \frac{\partial \mathbf f(\mathbf x_t)}{\partial \mathbf x_t}\bigg |_\mathbf{x^*}\delta \mathbf x_t,
\end{equation*}
and
\begin{equation}\label{eq:subs}
    \mathbf x_{t+1} = \delta \mathbf x_{t+1} + \mathbf x^*,
\end{equation} 
isolating the term $\delta \mathbf x_{t+1}$ and substituting the Eqs. (\ref{eq:disc}) and (\ref{eq:ptofixo}) at Eq. (\ref{eq:subs}),
\begin{equation}
     \delta \mathbf x_{t+1} =  \frac{\partial \mathbf f(\mathbf x_t)}{\partial \mathbf x_t}\bigg |_\mathbf{x^*}\delta \mathbf x_t,
\end{equation}
and taking into account that the term of the partial derivatives is the Jacobian matrix of the map, evaluated at the fixed point $\mathbf x^*$
\begin{equation}
     \delta \mathbf x_{t+1} =  \mathcal{J}(\mathbf x^*)\delta \mathbf x_t,
\end{equation}
and, analogously to a flow, the stability of the fixed points can be studied from the eigenvalues of the Jacobian matrix, in which, if the modules of the eigenvalues are $< 1$, it implies that the trajectory converges to the fixed point for $t\rightarrow \infty$ (stable). If at least one of the eigenvalues is $>1$, $\delta \mathbf x_t$ diverges. In this sense, considering the case of the logistic map Eq. (\ref{eq:log}), the fixed points satisfy the following equation,
\begin{equation*}
    x^* = r x^*(1-x^*),
\end{equation*}
so, the logistic map presents two fixed points $x^{(*,1)} = 0$ and $x^{(*,2)} = 1 - 1/r$. Due to the restriction on the map domain ($0\leq x_t\leq 1$), $x^{(*,2)}$ is fixed point for $r\geq 1$.

Differentiating the Eq. (\ref{eq:log})
\begin{equation}
    f'(x_t) = r(1-2x_t),
\end{equation}
and, the first fixed point,
\begin{equation}
    f'(x^{(*,1)}=0) = r,
\end{equation}
and $x^{(*,1)}$ is stable for $r<1$, and unstable $r>1$. The second fixed point,
\begin{equation}
    f'(x^{(*,2)}=1-1/r) = 2-r,
\end{equation}
that is, $x^{(*,2)}$ is stable for $1<r<3$, and unstable for $r>3$. It is noteworthy that for the study of orbits with period $>1$, it is necessary to analyze the stability of the fixed points of subsequent iterations, e.g., $x_{t+2} = f(f(x_t))$ for the case of period orbits $2$, where a fixed point is the point that maps itself every two iterations $x^* = f(f(x^*))$.

\section{Chaos in dynamical systems}

As shown at the beginning of this chapter with the Lorenz system, one of the main characteristics of a chaotic motion is the sensitivity to initial conditions, being two conditions arbitrarily close, after a sufficiently long time the trajectories tend to diverge exponentially from each other \cite{ott2002chaos,strogatz2001nonlinear,lichtenberg2013regular}. In nonlinear dynamics, systems that exhibit such behavior are often studied. 

\begin{figure}[t]
    \centering
\includegraphics[page=3]{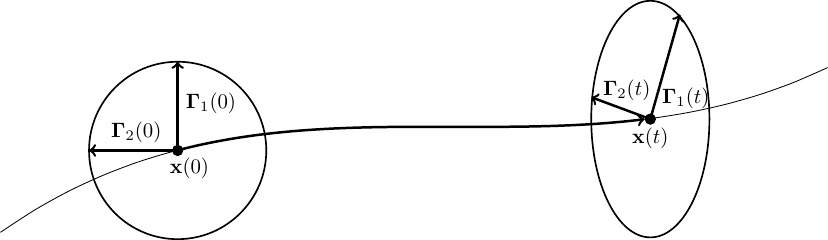}
\caption[Temporal evolution of a bidimensional sphere of initial conditions.]{\textbf{Temporal evolution of a bidimensional sphere of initial conditions.} For a set of initial conditions in a sphere of the radios $\mathbf\Gamma_0$, the expansion and the retraction of the directions $\mathbf{\Gamma}_1(t)$, $\mathbf{\Gamma}_2(t)$ indicates the chaotic divergence in the initial conditions.   
}
\label{fig_lyap}
\end{figure}

Considering a $D$-dimensional system $\dot{\mathbf x} = \mathbf f (\mathbf x)$. Given an arbitrary initial condition $\mathbf x(0)$, exists a hypersphere of radius $\Gamma_0$ of infinitesimal volume centered on $\mathbf x(0)$ that involves infinitely close initial conditions to $\mathbf x(0)$. The temporal evolution of this system comes from both $\mathbf x(t)$ and the rate of expansion (or retraction) of the axis of the hypersphere, transforming it into an ellipsoid whose directions are given by the set of $D$ vectors $\mathbf {\Gamma}_1(t),\mathbf{\Gamma}_2(t),\cdots,\mathbf{\Gamma}_D(t)$, as illustrated in the two-dimensional example in Fig. \ref{fig_lyap}.

To distinguish regular and chaotic dynamics for a multidimensional system, the Lyapunov spectrum is computed using Benettin’s algorithm \cite{wolf1985determining,benettin1980lyapunov} and including a Gram–Schmidt re-orthonormalization procedure \cite{wolf1985determining}. The method consists of the computation in parallel of the evolution of the system $\dot{\mathbf x} = \mathbf f(\mathbf x)$ and the set of vectors $\mathbf \Gamma(t)$. It is defined a matrix $\mathcal M$ in which the elements $\mathcal M_{ij}$ describe the $j$-th component of the $i$-th vector $\mathbf{\Gamma}_i(t)$,
\begin{gather}
\mathcal M(t)
 =
  \begin{bmatrix}
   \mathcal M_{1,1}(t) & \mathcal M_{1,2}(t) & 
   \cdots & \mathcal M_{1,D}(t) \\ 
   \mathcal M_{2,1}(t) & \mathcal M_{2,2}(t) & 
   \cdots & \mathcal M_{2,D}(t) \\ 
   \vdots  & \vdots  & \ddots  & \vdots \\
   {\mathcal M_{D,1}(t)} & {\mathcal M_{D,2}(t)} & 
   \cdots & {\mathcal M_{D,D}(t)} \\ 
   \end{bmatrix},
\end{gather}
%
where the temporal evolution of the matrix is based on the Jacobian matrix
\begin{equation}
    \dot{\mathcal M} = \mathcal J(\mathbf x) \mathcal M,
\end{equation}
and the initialization of $\mathcal M(0)$ is equal to a identity matrix $\mathbf I$.

In order to kept the vectors $\mathbf \Gamma(t)$ in an orthogonal direction, in each integration step it is used the Gram-Schmidt reorthonormalization procedure on the vector frame \cite{wolf1985determining}. After discard the transient effects, the Lyapunov spectrum are evaluated using the norm of the vectors
\begin{equation}
    \lambda_i = \frac{1}{t_\mathrm f-t_\mathrm i}\sum_{t=t_\mathrm i}^{t_\mathrm f} \ln |\mathbf \Gamma_i(t)|, \;\;\;\;\;\;\;\;\;\;\;\;\; i = 1,2,\cdots,D,
\end{equation}
where $t_\mathrm i$ and $t_\mathrm f$ are the initial and final times of computation, respectively, to obtain the stationary solutions of the dynamical system. If at least one of the exponents $\lambda_i > 0$, it usually taken as an indication that the system is chaotic, since there is an exponential divergence at the $i$-th direction, even if $\lambda_j \leq 0 ,\, \forall \, j\neq i$. In this sense, it is useful to define the largest Lyapunov exponent ($\Lambda$) where
\begin{equation}
    \Lambda = \mathrm{max} \{\lambda_1,\lambda_2,\cdots,\lambda_D\},
\end{equation}
which means that if $\Lambda > 0$ the system depicts a chaotic behavior.

\begin{figure}[t]
    \centering
    \includegraphics[width=\columnwidth]{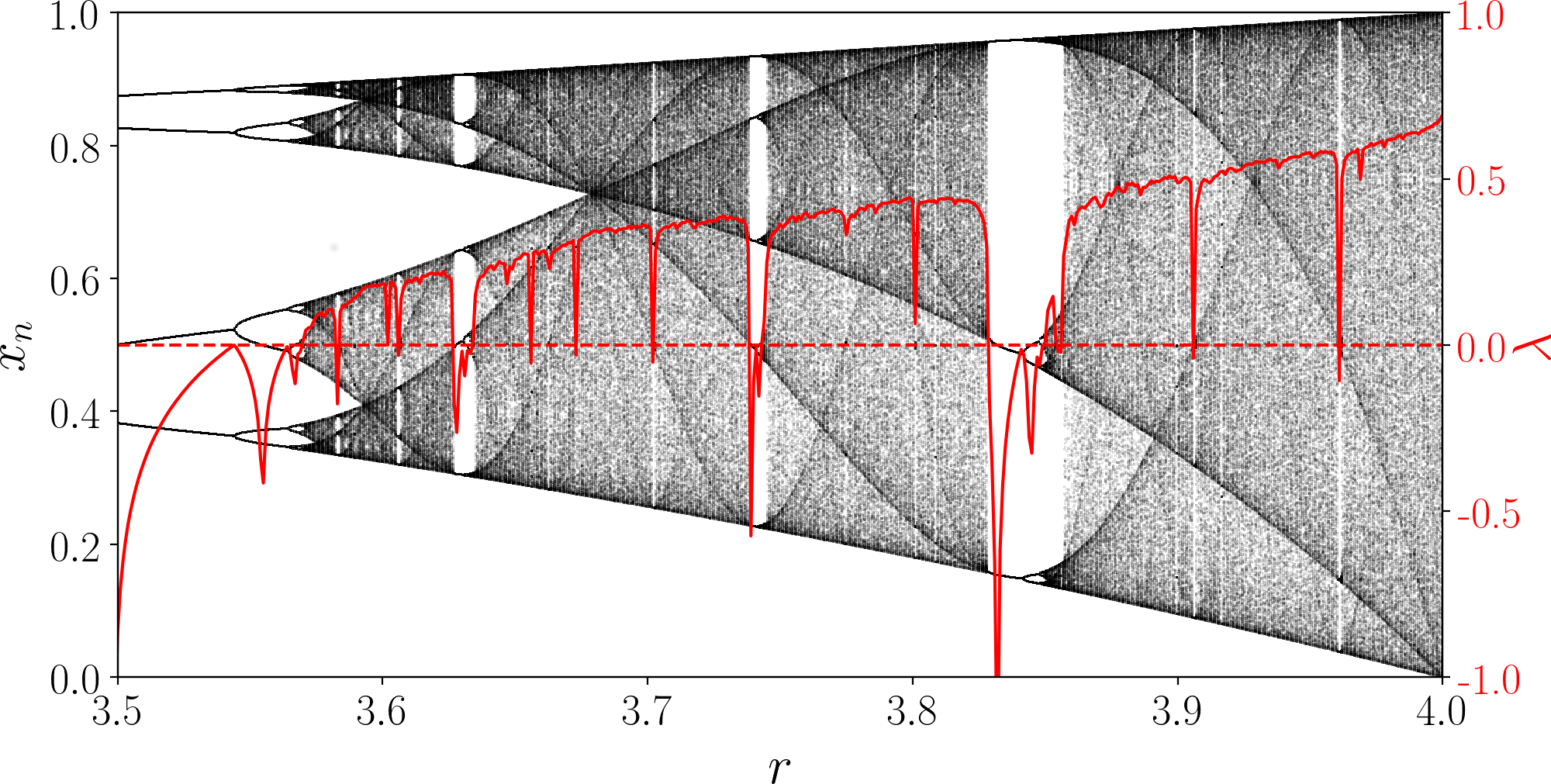}
    \caption[Lyapunov exponent of the logistic map.]{\textbf{Lyapunov exponent of the logistic map.} Bifurcation diagram of the logistic map (left scale) and the Lyapunov exponent $\lambda$ (right scale).}
    \label{fig:log_3}
\end{figure}

In the case of a unidimensional discrete system described by Eq. (\ref{eq:disc}), the ($n+1$)-th iteration of the map can be written in terms of the initial condition $x_0$
\begin{equation}
    x_{n+1} = f(x_n) = f^n(x_0).
\end{equation}
Consider a nearby point $x_0 + \delta_0$, where the initial separation $\delta_0$ is infinitesimal and $\delta_n$ the separation after $n$ iterations which can be described by
\begin{equation}\label{eq:lyap_aux}
    \delta_n = f^n(x_0 + \delta_0) - f^n(x_0),
\end{equation}
and assuming that the separation evolves exponentially 
\begin{equation}
    |\delta_n| \approx |\delta_0|e^{n\lambda},
\end{equation}
isolating $\lambda$
\begin{equation*}
    \lambda \approx \frac 1 n \ln \left |\frac{ \delta_n}{ \delta_0} \right|,
\end{equation*}
and by Eq. (\ref{eq:lyap_aux})
\begin{equation}\label{eq:lyap_aux_2}
    \lambda \approx \frac 1 n \ln \left |\frac{ f^n(x_0 + \delta_0) - f^n(x_0)}{ \delta_0} \right|,
\end{equation}
taking the limit $\delta_0 \rightarrow 0$ Eq. (\ref{eq:lyap_aux_2}) yields
\begin{equation}
    \lambda \approx \frac{1}{n}  \ln \left | \lim_{\delta_0\rightarrow 0} \frac{ f^n(x_0 + \delta_0) - f^n(x_0)}{ \delta_0} \right| = \frac 1 n \ln |(f^n(x_0))'| ,
\end{equation}
expanding the logarithm term using the chain rule
\begin{equation*}
\lambda \approx \frac 1 n \ln \left | \prod_{i=0}^{n-1} f'(x_i)\right | = \frac 1 n \sum_{i=0}^{n-1}\ln |f'(x_i)|,     
\end{equation*}
and, if the limit of $n \rightarrow \infty$ exists, this limit is defined to be the Lyapunov exponent of the map
\begin{equation}
    \lambda = \lim_{n\rightarrow \infty}\left \{\frac 1 n \sum_{i=0}^{n-1}\ln |f'(x_i)| \right \}. 
\end{equation}
In the same way that for the Lyapunov spectrum, a positive exponent $\lambda>0$ is a signature of chaos. In the case of the logistic map, explored in the last section,  
\begin{equation}
    \lambda = \lim_{n\rightarrow \infty} \left \{ \frac 1 n \sum_{i=0}^{n-1} \ln|r(1-2x_i)|\right \},
\end{equation}
the lyapunov exponent depends on the control parameter $r$, which is expected since as shown in Fig. \ref{fig:log_1}, different values of $r$ depict different dynamical behaviors. Figure \ref{fig:log_3} depicts the bifurcation diagram (left scale) and the lyapunov exponent $\lambda$ (red line, right scale) as a function of $r$. The chaotic dynamics takes place at $r\approx 3.57$ due to a known route to chaos called period doubling cascade \cite{ott2002chaos} which will be presented in more details in the next section.

\section{Routes to chaos}

As shown in the last section, the chaotic behavior can emerge in the dynamical system with the simplest change of a parameter, which causes a change in the stability of fixed points of the system. In this section, it is explored three of the most famous route to chaos: period doubling, intermittency, and crisis \cite{ott2002chaos}.

\subsection{The period doubling cascade}

\begin{figure}[t]
    \centering
    \includegraphics[width=.9\columnwidth]{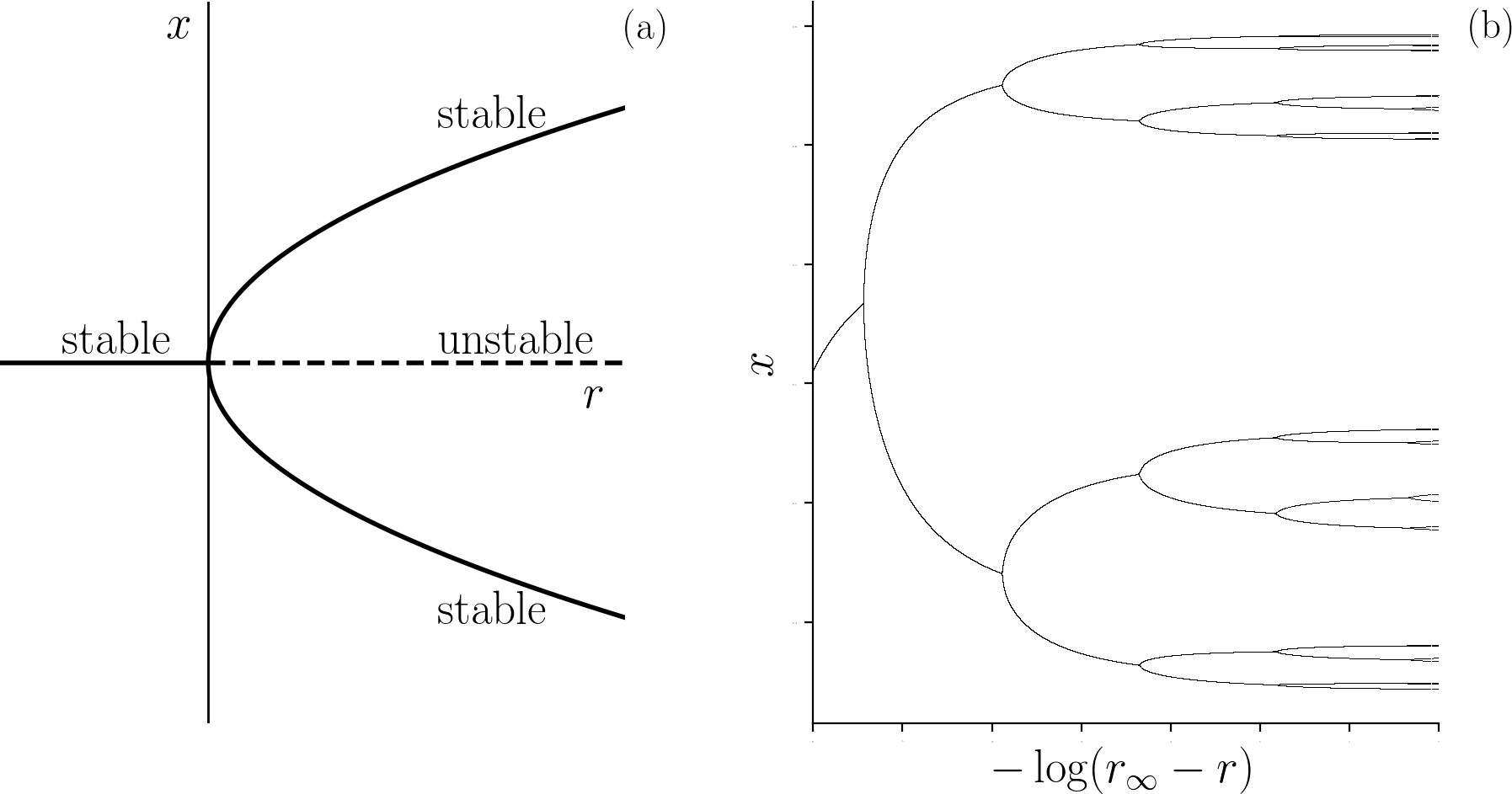}
    \caption[Period doubling route to chaos.]{\textbf{Period doubling route to chaos.} (a) A stable point loses stability when a control parameter reaches a critical value, creating a 2-periodic stable fixed point. (b) The evolution of successive doubling of periods. The length of $r$ of the range of stability for an orbit of period $2^m$ decreases as $m$ increases, collapsing an accumulation of an infinite number of period doublings when $r_\infty$ making the system converges to a chaotic attractor.
}
    \label{fig:doubling}
\end{figure}

In a doubling period bifurcation, a stable fixed point loses stability to an attracting 2-period orbit as the parameter reaches a critical value \cite{ott2002chaos}. An illustration of this route can be seen in panel (a) of Fig. \ref{fig:doubling}. In the logistic map, the fixed point $x^{(*,2)}=1-\frac 1 r$ is stable for $1<r<3$ since $|f'(x^{*,2})|=|2-r|$ is $<1$ at this range. For $r>3$ the fixed point loses stability since $|f'(x^{*,2})|>1$, and, simultaneously, $f(f(x))$ leads the creation of a 2-periodic stable fixed point (which is not a fixed points for $f(x)$). As $r$ is increased to $1+\sqrt(6)\approx 3.449$, the 2-period fixed point loses stability leading the creation to a 4-period fixed point (due to the stability of $f^4(x)$), etc.  This process continues indefinitely; at each bifurcation, the periodic orbit is replaced by a new attracting periodic orbit of twice the period, producing an infinite cascade of period doublings with ranges $r_{m-1}<r<r_m$, in which a $2^m$ orbit is stable. Panel (b) of Fig. \ref{fig:doubling} presents the evolution of successive doubling of periods on a logarithmic scale. Since the length of $r$ of the range of stability for an orbit of period $2^m$ decreases as $m$ increases, there is a saturation value $r_\infty$ where there is an accumulation point of an infinite number of period-doubling bifurcations
\begin{equation}
    r_\infty = \lim_{m\rightarrow \infty} r_m \approx 3.5699,
\end{equation}
with the property of 
\begin{equation}
    \lim_{m\rightarrow \infty} \frac{r_m - r_{m-1}}{r_{m+1}-r_m} \equiv \hat{\delta} \approx 4.6692,
\end{equation}
where $\hat \delta$ is called Feigenbaum constant \cite{ott2002chaos,feigenbaum1978quantitative}. It is remarkable that $\hat \delta$ is a universal constant for functions approaching chaos via period doubling.

\subsection{Intermittency route to chaos}

\begin{figure}[t]
    \centering
    \includegraphics[width=.9\columnwidth]{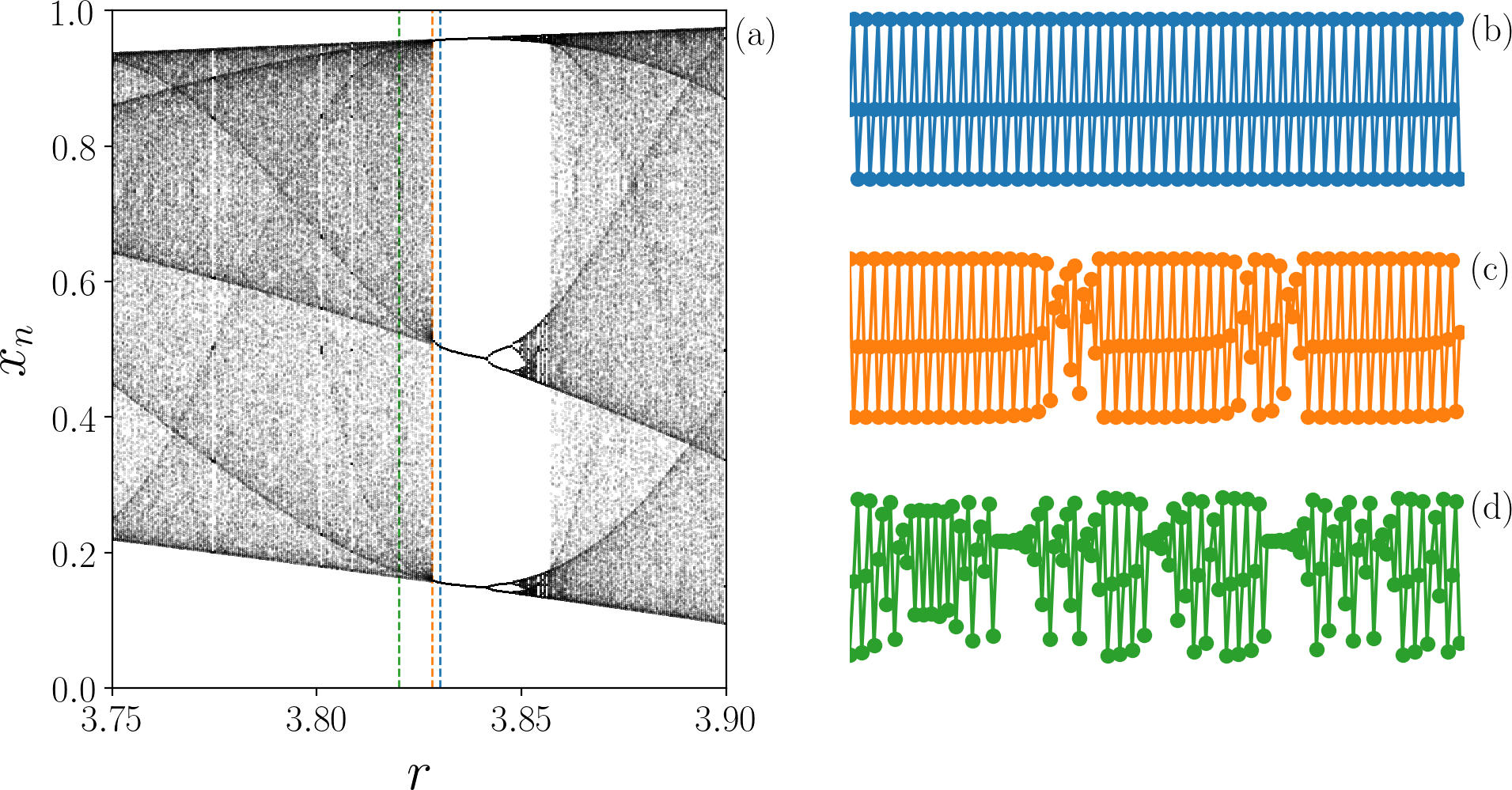}
    \caption[Intermittency route to chaos.]{\textbf{Intermittency route to chaos.} (a) Bifurcation diagram for the logistic map as a function of $r$. (b) 3-periodic orbit for $r=2.83$. (c) For $r=2.8282\approx r^*$  a remnant nearly-3-period orbit is seen (the 3-periodic orbit no longer exists), this orbit is interrupted by chaos that occurs intermittently. (d) For $r=2.82$ there is no apparent evidence of the 3-period orbit anymore.}
    \label{fig:log_intermittency}
\end{figure}

Considering a generic system, the intermittency route to chaos refers to how a periodic orbit is replaced by a chaotic attractor when the control parameter, namely $r$ reaches a critical value $r^*$ \cite{ott2002chaos,pomeau1980intermittent}. Supposing that the periodic orbit exists for $r>r^*$, for values $r\lesssim r^*$ the periodic orbit no longer exists and it is possible to see ``nearly-period'' orbits which are intermittently interrupted by a finite chaotic behavior. An example of this behavior can be seen in the logistic map where for $r^*= 1 + 2\sqrt 2 \approx 3.8284$ a 3-periodic orbit is created, and the chaotic behavior occurs for $r<r^*$. Figure \ref{fig:log_intermittency} (a) exhibits a magnification of the bifurcation diagram of the logistic map in the critical value $r^*$ when the 3-period orbit appears. Panels (b -- d) show an illustration of the intermittent behavior, where the system presents a 3-periodic orbit for $r=2.83$ (b). At panel (c) for $r=2.8282$, it is possible to see a remnant nearly-3-period (approximately regular), this orbit is interrupted by a chaotic behavior that occurs intermittently. And panel (d) or $r=2.82$ there is no apparent evidence of the 3-period orbit anymore.    

\begin{figure}[t]
    \centering
    \includegraphics[width=.9\columnwidth]{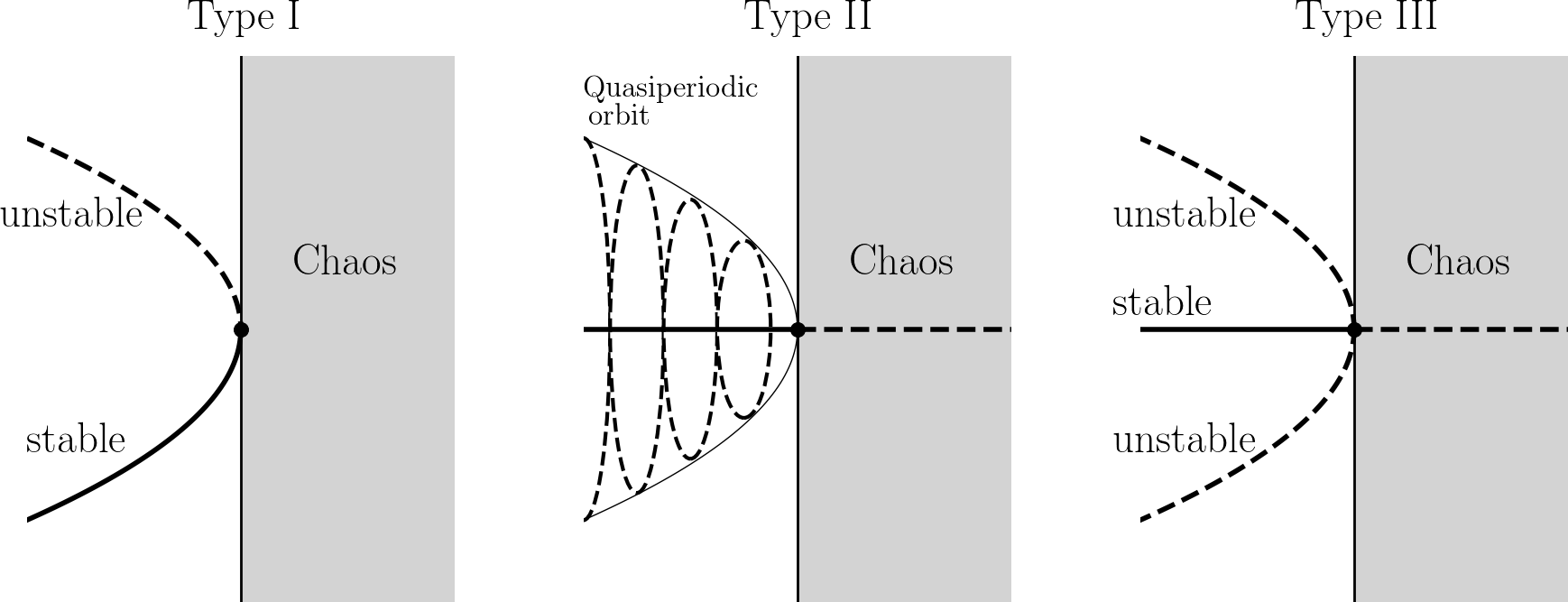}
    \caption[Three types of intermittency route to chaos.]{\textbf{Three types of intermittency route to chaos.} In this route to chaos, the chaotic state is reached in three different ways. Figure inspired by Fig. 8.6 of Ref \cite{ott2002chaos}.}
    \label{fig:int_1}
\end{figure}

Different from the doubling period cascade, where the stable fixed point loses stability and creates a 2-periodic stable fixed point, and the chaotic attractor is achieved due to the successive creation of an infinite number of period-doubling bifurcations. In this route, the stable fixed point either becomes unstable or is destroyed as the control parameter reaches a critical parameter. In this sense, three types of intermittency are distinguished corresponding to three types of bifurcations \cite{pomeau1980intermittent}. Figure \ref{fig:int_1} depicts an illustration of the three types of intermittency route to chaos. Type I: saddle-node bifurcation, where stable and unstable fixed points of a dynamical system collide and annihilate each other; Type II: Hopf bifurcation, where a stable fixed point of a dynamical system loses stability; Type III: inverse period-doubling bifurcation, where there is a stable fixed point with 2-periodic unstable fixed point, the stable fixed point collides with the unstable fixed points and lost its stability.

It is possible to define $\langle \tau_\mathrm{int} (r)\rangle$ as the mean characteristic time of occurrence of an intermittency behavior with 
\begin{equation}
    \lim_{r\rightarrow r^*} \langle \tau_\mathrm{int}(r)\rangle = \infty    
\end{equation}
where $r$ is studied in the chaotic regime. In this sense, for each type of intermittency case, this mean time decays with    
\begin{equation}
    \langle \tau_\mathrm{int} (r) \rangle\sim \begin{dcases}
    |r-r^*|^{-1/2}\;\; &\mathrm{for\;Type\;I},\\
    |r-r^*|^{-1} \;\; &\mathrm{for\;Type\;II\; and\;III}.
    \end{dcases}
\end{equation}

\subsection{Crisis}

\begin{figure}[t]
    \centering
    \includegraphics[width=.9\columnwidth]{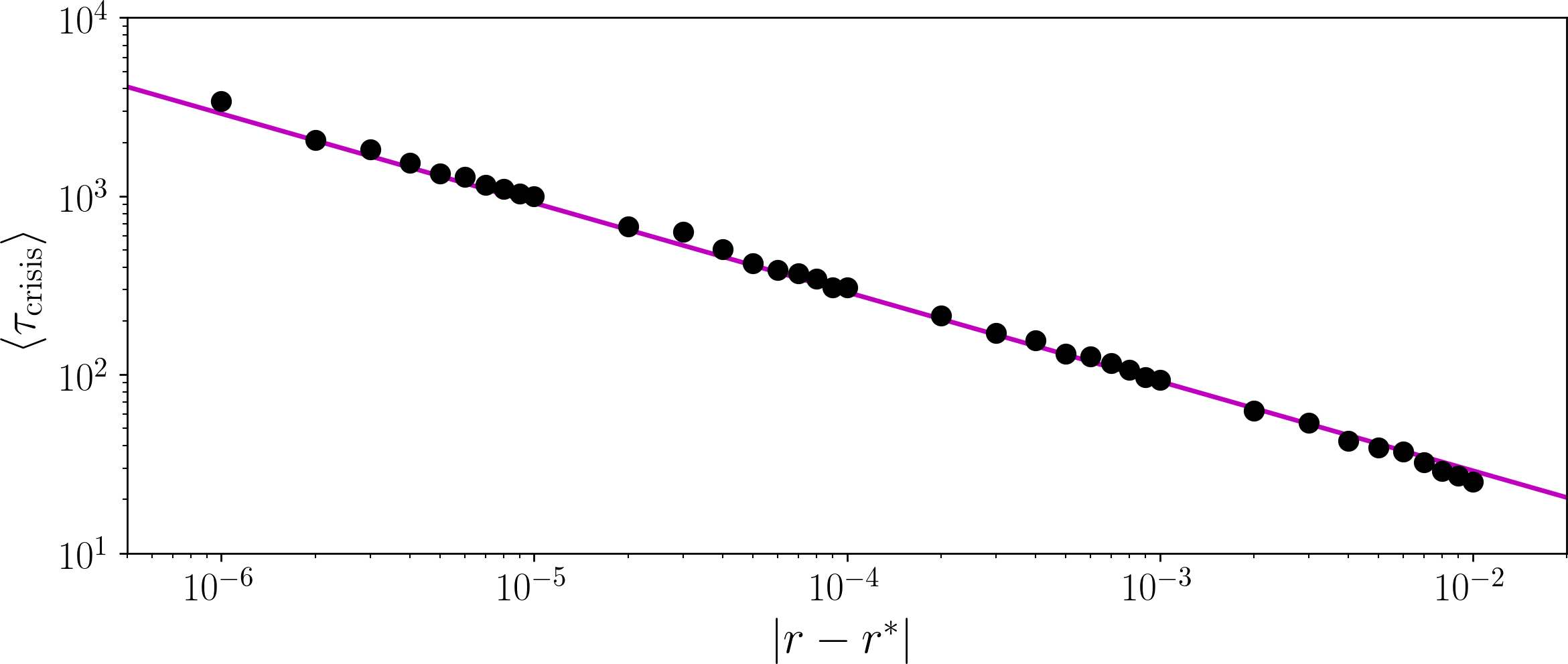}
    \caption[Crisis route in the logistic map.]{\textbf{Crisis route in the logistic map.} The mean transient life time of that the system spent near the chaotic transient $\langle \tau_\mathrm{crisis}\rangle$. The theoretical result $\propto |r-r^*|^{-1/2}$ is plotted as a line.}
    \vspace*{8in} 
    \label{fig:log_crisis}
\end{figure}

In the crisis route, the chaotic attractor is changed with the variations in the control parameter \cite{ott2002chaos}. In particular, in the boundary crisis, the chaotic attractor is annihilated with the collision of an unstable periodic orbit on its basin boundary \cite{ott2002chaos,grebogi1986critical,grebogi1987critical}. For example, for a dynamical system that is chaotic for $r<r^*$ and periodic, otherwise. For values $r$ sightly greater than the critical value the attractor no longer exists but is replaced by a chaotic saddle that does not attract trajectories but allows a chaotic transient. This means that an initialization of the system in some region of the phase space inside the chaotic saddle spend some time behaving as chaotic until leaving the attractor and never return. The mean transient lifetime that the system spent near the chaotic transient $\langle \tau_\mathrm{crisis}\rangle$ becomes longer as $r\rightarrow r^*$, the decay with the distance of the critical point following
\begin{equation}
        \langle \tau_\mathrm{crisis} (r) \rangle \propto |r-r^*|^{-\kappa}
\end{equation}
where $\kappa$ is called critical exponent of the crisis \cite{ott2002chaos}. A simple example of a crisis is given by the logistic map, where the chaotic attractor exists for $r=4$, but for $r>4$ the chaotic attractor is replaced by a chaotic transient and for a initial condition inside the basin of attraction, i.e. $x_0$ $\in [0,1]$, after $\langle \tau_\mathrm{crisis}\rangle$ the system diverges to $-\infty$. Figure \ref{fig:log_crisis} presents the numerical result of $\langle \tau_\mathrm{crisis} (r)\rangle$ for the logistic map, in this case as well as for other one dimensional maps the critical exponent of the crisis is $\kappa = 1/2$ (solid magenta line) \cite{ott2002chaos}. In the case of multidimensional system is expected $\kappa > 1/2$ \cite{ott2002chaos}.

\chapter{Neuronal models and synapses} \label{chap:modelos}

\initial{T}{his} chapter presents some properties of the neuron, the main cell of the nervous system \cite{kandel2013principles,ermentrout2010mathematical}. Then, it is presented the Hodgkin-Huxley (HH) model \cite{hodgkin1952quantitative}, where the equations to modeling the time-evolution of the membrane potential are created based on an electrical circuit. After that, it is presented one of the adaptations of the HH model proposed by Braun \textit{et al.} \cite{braun1998computer,braun1997low}, here called the Hodgkin-Huxley-like (HH$\ell$) model, where the adaptation consists of the addition of two ionic currents and some temperature-dependence parameters, that makes it possible the neuron to depict a different dynamical feature, called burst dynamics. Then, it is presented the Hindmarsh-Rose (HR) model \cite{hindmarsh1984model}, a model of neuron of three coupled first-order differential equations. Therefore, the model proposed by Dante R. Chialvo \cite{chialvo1995generic}, in which, with two discrete equations, the dynamics of the action potential seen in the neuron can be reproduced.  Lastly, it is briefly presented the equations that ruled the synaptic interactions among neurons, which enables the transfer of information between the presynaptic neuron to the postsynaptic neuron
 \cite{kandel2013principles}.

\section{The action potential}

\begin{figure}[t]
\centering
\includegraphics[width=.85\columnwidth]{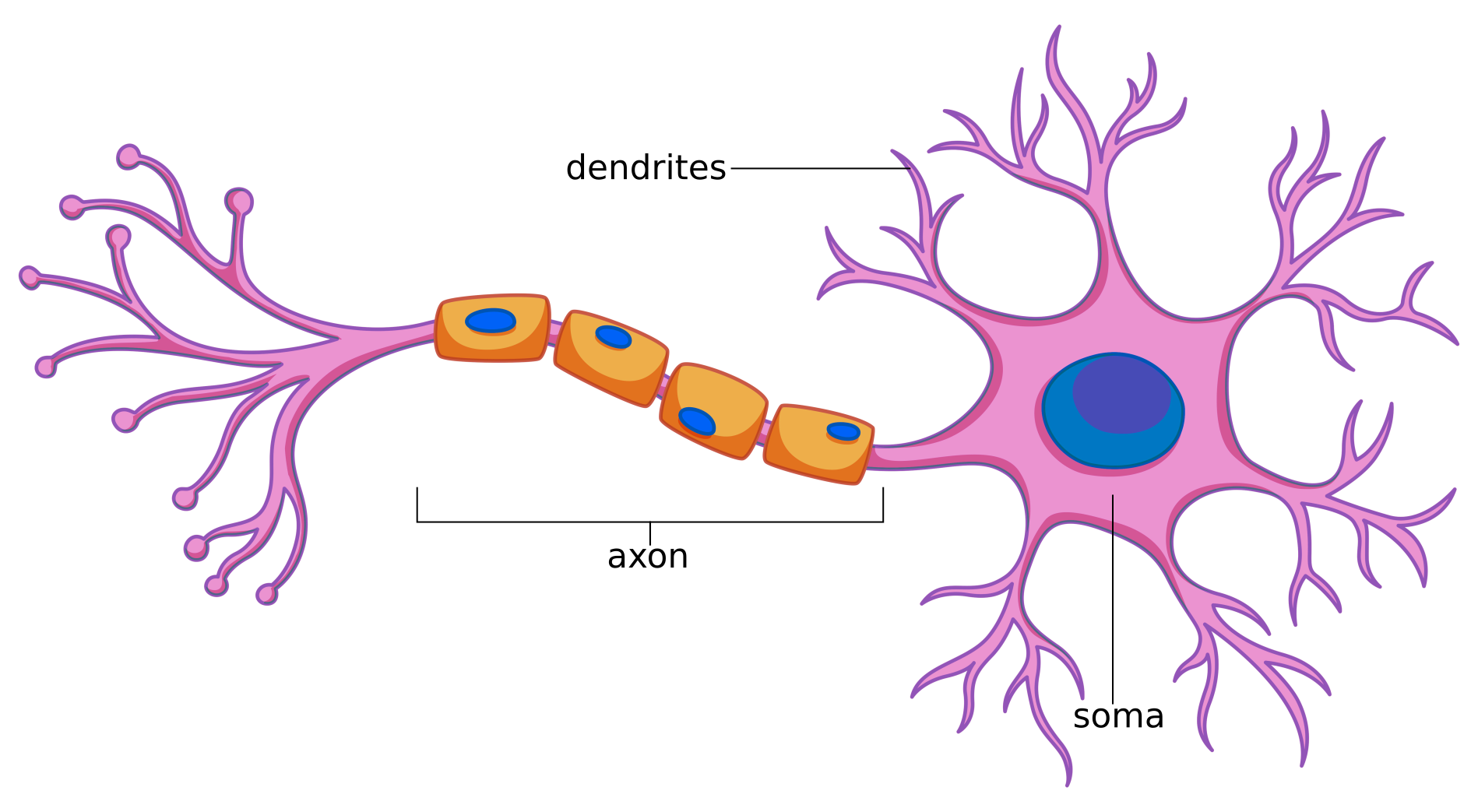}
 \caption[Basic anatomy of a neuron.]{\textbf{Basic anatomy of a neuron.} A typical neuron can be separated into three basic regions: the soma, dendrites, and axons \cite{kandel2013principles}. This figure is adapted from \cite{neuronfigure}.}
\label{fig:figneur}
\end{figure}

The neuron is one of the main cells of the nervous system, being responsible for the conduction of electrical impulses. In a general point of view, the neuron can be divided into three regions: soma, which contains the nuclei and cytoplasm; dendrites, whose main function is the reception of signals which comes from other neurons; axons, which carry the signals from the soma to other neurons \cite{kandel2013principles} (Fig.~\ref{fig:figneur}).

The main function of the neuron is the transmission of action potentials, which are electrical signals which propagate information at the nerve system. The action potential consists of a depolarization followed by repolarization, depicting a spike shape. These variations at the membrane potential occur due to the variations of the ion concentrations between the intracellular and extracellular media. In this sense, the membrane potential can be defined as
\begin{equation}
    V(t) = V_\mathrm{int}-V_\mathrm{ext},
\end{equation}
where $V_\mathrm{int}$ and $V_\mathrm{ext}$ are the potentials of the intracellular and extracellular media, respectively.

\begin{figure}[t]
\begin{center}
\includegraphics[width=.9\columnwidth]{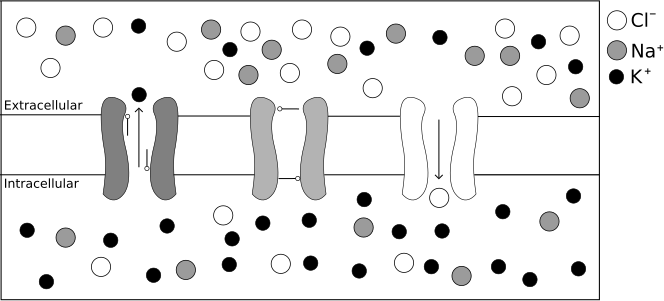}
\caption[Cell membrane representation]{\textbf{Cell membrane representation.} The cell membrane contains proteins, that function as channels that allow the ions to move through it. The ion flux changes the concentration of the intracellular (and extracellular) side, inducing variations in the membrane potential.}
\label{fig:figmemb}
\end{center} 
\end{figure}

The ions permeate the neural membrane by proteins which work as ion channels. There are gated channels and non-gated channels. While non-gated channels are always open, allowing the entry and exit of ions from the intracellular side, gated channels have potential-dependence gates, that is, the permissiveness of the channels depends directly on the potential of the membrane. The predominant ions found on either side of the cell membrane are potassium ions ($\mathrm K^+$), sodium ($\mathrm{Na}^+$), and chlorine ($\mathrm{Cl}^-$). One illustration of the ionic channels is presented in Fig.~\ref{fig:figmemb}, where one channel is open allowing the exit of a K$^+$ ion, the other is closed, and a non-gated channel allows the entrance of a Cl$^-$ ion.  

\begin{figure}[t]
\begin{center}
\fbox{\includegraphics[width=.5\columnwidth]{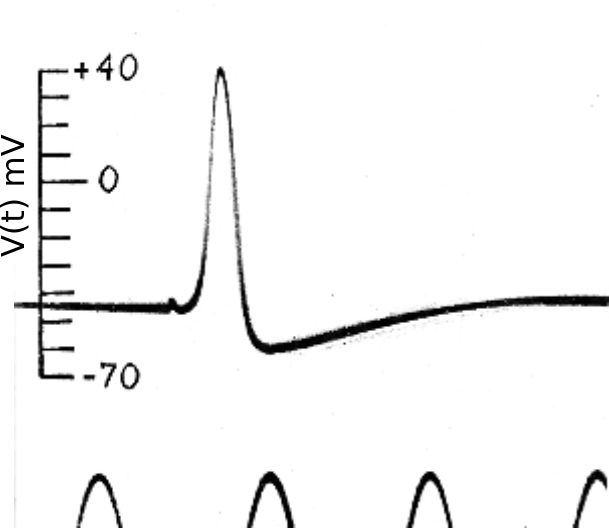}}
\caption[Action potential recorded between inside and outside axon]{\textbf{Action potential of the giant axon of squids.} Action potential between inside and outside of axon was recorded by Hodgkin and Huxley at $1939$ using a micro-electrode consisted of a glass tube filled with seawater. Figure taken from \cite{hodgkin1939action}. Adapted.}
\label{fig:spike_original}
\end{center} 
\end{figure}

At the equilibrium point also called the resting state, there are no ionic changes between the media. In this case, the intracellular side contains in majority  K$^+$ ions in comparison with Na$^+$ and Cl$^-$ that are more abundant in the extracellular side \cite{ermentrout2010mathematical}. Hence, the membrane potential is determined primarily by the K$^+$ resting potential,
\begin{equation*}
    V(t) =  V_\mathrm{int}-V_\mathrm{ext} = V_\mathrm{eq} \approx -\SI{70}{\milli\volt}.
\end{equation*}
When the cell is stimulated above a threshold, the Na$^+$ channels open allowing the entrance of ions inside the cell, this influx of Na$^+$ tends to depolarizes the cell, resulting in a positive variation in $V(t)$. This abrupt increase of $V$ inverts the polarity of the cell, closing the Na$^+$ channels and opening the K$^+$ channels, allowing the efflux of K$^+$ ions to repolarizes the cell, which takes the membrane potential to a level below the $V_\mathrm{eq}$ (hyperpolarization). Until the K$^+$ channels close up again, the membrane is in a refractory stage. During this time, the cell pumps the exchange excess Na$^+$ ions inside the cell with excess K$^+$ ions outside the cell, and a new cycle can be started if the stimulus is kept \cite{ermentrout2010mathematical}.  The Fig.  \ref{fig:spike_original} represents the first intracellular record of an action potential \cite{kandel2013principles} of the giant axon of squids \textit{Loligo forbesi}, published by Hodgkin and Huxley at 1939 \cite{hodgkin1939action} where temporal pulses was applied each $\SI{2}{\milli \second}$. 

\section{Neuron models}

\subsection{The Hodgkin-Huxley (HH) model} \label{sec_hh}

\textit{Alan Lloyd Hodgkin} and \textit{Andrew Fielding Huxley} were the firsts to describe mathematically a regenerative current that generates an action potential. They were awarded the Nobel Prize in Physiology or Medicine in 1963, together with \textit{Sir John Carew Eccles} for their discoveries about the ionic mechanisms involved in the excitation and inhibition of the membrane of nerve cells \cite{nobel}.

\begin{figure}[t]
\begin{center}
\includegraphics[width=.8\columnwidth]{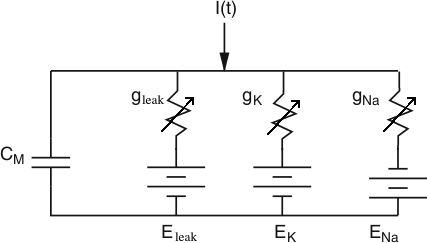}
\caption[Equivalent circuit model]{\textbf{Equivalent circuit model.} The circuit consists of three components: resistors, which represent the ion channels; sources, which represent the ion concentration gradient; and capacitors, which represent the charge that can be stored in the membrane \cite{ermentrout2010mathematical}.}
\label{fig:fig_circ}
\end{center} 
\end{figure}

To describe the action potential of the cell, Hodgkin and Huxley have used a circuit model, considering the contribution of two ionic currents (K$^+$ and Na$^+$) plus a passive current which takes the contribution of the non-gated channels and the less abundant ions in the media. The circuit, illustrated in Fig. \ref{fig:fig_circ}, is composed of three components: resistors, which represent the ion channels; sources, which represent the ion concentration gradient; and capacitors, which represent the charge that can be stored in the membrane \cite{ermentrout2010mathematical}.

The charge stored in the capacitor is proportional to the membrane potential, where the proportionality constant is the capacitance
\begin{equation}
\label{qcv}
Q=C_{\mathrm{M}}V,
\end{equation}
where $C_{\mathrm{M}}$ is the specific capacitance of the membrane, measured in
\SI{}{\micro\farad/\centi\meter^{2}}. $V$ is the membrane potential measured in \SI{}{\milli\volt}. Differentiating both sides in relation to time
\begin{equation}
    \label{eqaaa}
    \frac{dQ}{dt} = i_\mathrm{cap} = C_\mathrm{M}\frac{dV}{dt}
\end{equation}
where $i_{\mathrm{cap}}$ is the specific current referring to the capacitance divided by the area.

Applying Kirchhoff's law to the circuit at Fig.~\ref{fig:fig_circ} implies that
\begin{equation}
0=i_{\mathrm{cap}}+I_\mathrm{K}+I_\mathrm{Na}+I_\mathrm{leak} - I(t), 
\end{equation}
where $I_\mathrm{K}$ and $I_\mathrm{Na}$ are the ionic currents, $I_\mathrm{leak}$ is the passive current, and $I(t)$ is an external applied current. Since the currents are \textit{Ohmic} it can be rewritten as $I_\mu=g_\mu(V-E_\mu)$. Substituting $i_\mathrm{cap}$ with Eq.(\ref{eqaaa}) 
\begin{equation}\label{hh1}
    C_{\mathrm{M}}\frac{dV}{dt}=-g_{\mathrm{K}}(V-E_{\mathrm{K}})-g_{\mathrm{Na}}(V-E_{\mathrm{Na}})-g_{\mathrm{leak}}(V-E_{\mathrm{leak}})+I(t),
\end{equation}
in which $E_\mathrm{K}$, $E_\mathrm{Na}$ and $E_\mathrm{leak}$ are the resting potential of each channel, and $g_\mathrm{K}$, $g_\mathrm{Na}$ and $g_\mathrm{leak}$ are the conductances of each channel.

The conductances of the gated channels are voltage-dependent and are directly related to the permissively of the ion channels, which means that it rules the probability of opening/close the channels. Using voltage-clamp techniques Hodgkin and Huxley have fitted the equations for each conductance as
\begin{eqnarray}
g_{\mathrm{K}}&=&\overline{g}_{\mathrm{K}}n^{4}
\label{gk}, \\
g_{\mathrm{Na}}&=&\overline{g}_{\mathrm{Na}}m^{3}h, \label{gna}\\
g_{\mathrm{leak}}&=&\overline{g}_{\mathrm{leak}} \label{gvaz},
\end{eqnarray}
where $\overline{g}_{\mathrm{K}}$, $\overline{g}_{\mathrm{Na}}$ and $\overline{g}_{\mathrm{leak}}$ are the maximum values of conductance of each current, $n$, $m$ are activation functions and $h$ an inactivation function, these functions are related to the probabilities of opening and closing the channels. 

Considering $\alpha_\mu(V)$ as the probability of the channel being open, and $\beta_\mu(V)$ closed (with $\mu = m,\,n,\,h$) these functions satisfy the following equations
\begin{equation}
\frac{d\mu}{dt}=\alpha _{\mu}(V)(1-\mu)-\beta _{\mu}(V)\mu,  \;\;\;\;\;\;\;\;\;\;\;\;\;\;\;\;\;\;\;   \mu=n,{m},{h},
\label{mu}
\end{equation}
in which $\alpha_{\mu}$ and $\beta_{\mu}$ are described
\begin{eqnarray}
\alpha_{\mathrm{n}}(V)&=&\frac{0.01(V+55)}{(1-\exp[-(V+55)/10])},\\
\alpha_{\mathrm{m}}(V)&=&\frac{0.1(V+40)}{(1-\exp[-(V+40)/10])},\\
\alpha_{\mathrm{h}}(V)&=&0.07\exp[-(V+65)/20],\\
\beta_{\mathrm{n}}(V)&=&0.125\exp[-(V+65)/80],\\
\beta_{\mathrm{m}}(V)&=&4\exp[-(V+65)/18],\\
\beta_{\mathrm{h}}(V)&=&\frac{1}{(1+\exp[-(V+35)/10])},
\end{eqnarray}
taking into account that $V$ is measured in $\SI{}{\milli\volt}$, and, for simplicity, the dimensions required in the other values are omitted.

Substituting the Eqs. (\ref{gk}--\ref{gvaz}) at Eq. (\ref{hh1}), it is achieved the Hodgkin-Huxley equations \cite{ermentrout2010mathematical}
\begin{equation}\label{hfinal}
C_{\mathrm{M}}\frac{dV}{dt}=-\overline{g}_{\mathrm{K}}n^{4}(V-E_{\mathrm{K}})-\overline{g}_{\mathrm{Na}}m^{3}h(V-E_{\mathrm{Na}})-\overline{g}_{\mathrm{leak}}(V-E_{\mathrm{leak}})+I(t).
\end{equation}
\begin{figure}[t]
    \centering
    \includegraphics[width=.9\columnwidth]{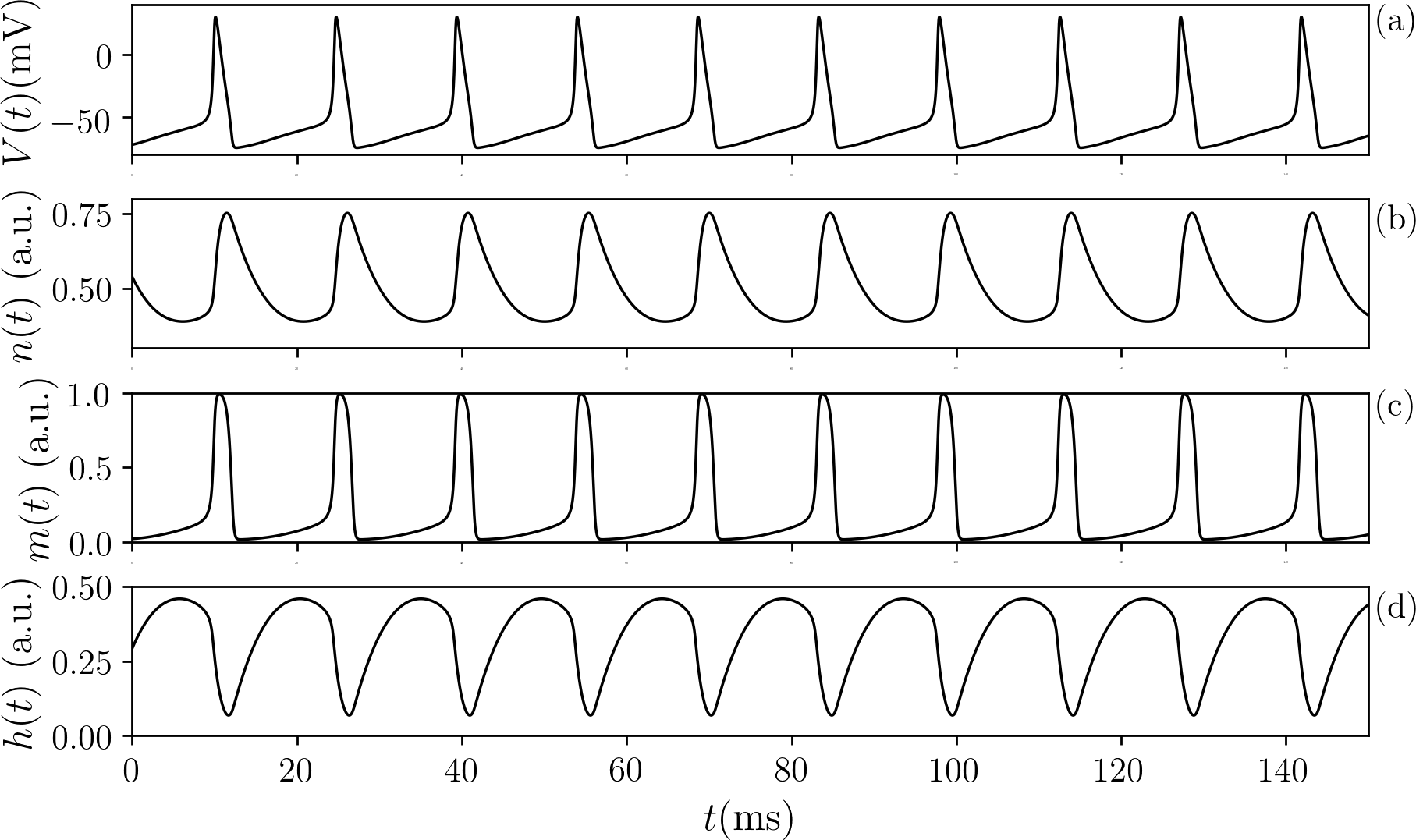}
    \caption[Time-evolution of the variables of the Hodgkin-Huxley model.]{\textbf{Time-evolution of the variables of the Hodgkin-Huxley model.} For $I=\SI{10}{\micro\ampere/\centi\meter^2}$ the membrane potential presents a sequence of action potentials.}
    \label{fig:hh_din}
\end{figure}

The Fig. \ref{fig:hh_din} demonstrates the dynamics of the HH model using the constants of Table \ref{tablehh} and $I=\SI{10}{\micro\ampere/\centi\meter^2}$. Each panel depicts the time evolution of a variable of the model. $V$ is presented in panel (a) and exhibits a sequence of action potentials (spikes). The other panels (b - d) depict the activation variables $n$ and $m$, and inactivation $h$, respectively. It is noted that when $V$ reaches a threshold, $n$ and $m$ increases quickly reaching their maximum values. In contrast, $h$ has the opposite effect, characterizing its inactivation effect, repolarizing the cell starting a new cycle. The neuron continues with this dynamics of periodic spikes while $I$ is kept active.

\begin{figure}[t]
\begin{center}
\includegraphics[width=\columnwidth]{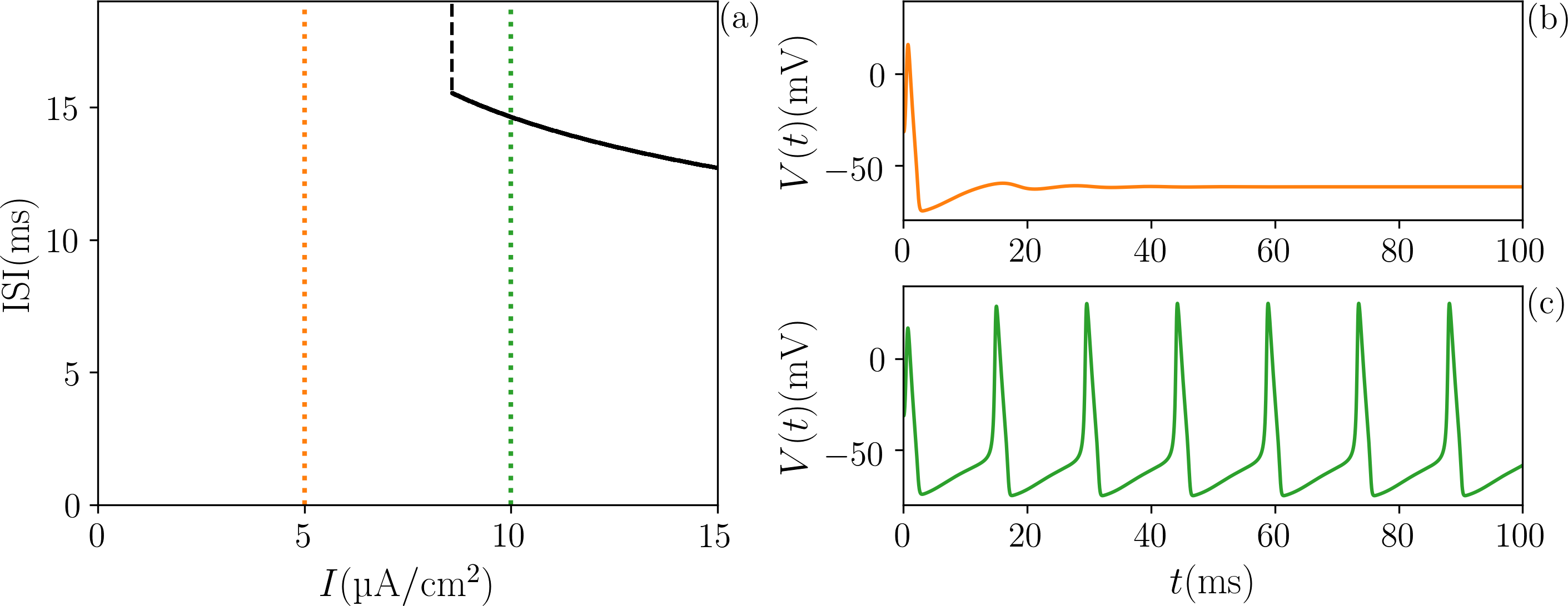}
\caption[Bifurcation diagram of the Hodgkin-Huxley model.]{\textbf{Bifurcation diagram of the Hodgkin-Huxley model.} (a) Inter-spike-interval (ISI) as a function of the applied current $I$, for $I\approx 8.39$ there is a Hopf bifurcation point where the ISI jumps from $\infty \rightarrow \SI{15.55}{\milli\second}$. (b) For $I=5$ after a transient time, the membrane potential reaches an equilibrium point. (c) $I=10$ the membrane potential depicts periodic spikes.}
\label{fig:fig_hh_2}
\end{center} 
\end{figure}

To understand the role of the applied current $I$, Figure \ref{fig:fig_hh_2} presents the bifurcation diagram of the inter-spike-interval (ISI) of the neuron. The occurrence of a spike is defined when $V$ reaches a threshold value of $-20$ $\SI{}{\milli\volt}$ (with positive derivative).
After that, it is evaluated the time between two consecutive spikes as a function of $I$ disregarding the first second to avoid the transient effect. It is noted that for $I<8.39$ the current is not strong enough to keep the neuron spiking. For $I\approx 8.39$ the model pass through a Hopf bifurcation \cite{ott2002chaos} and the ISI jumps from $\infty$ to  $\SI{15,55}{\milli\second}$. Panel (b) and (c) depict the time evolution of $V$ for $I=5$ and $I=10$, respectively. (b) The neuron depicts one spike, but after this, the equilibrium state is recovered since the current is not strong enough to keep the neuron spiking, as seen in panel (c). 

\begin{table}[htb!]
\centering
\caption{Constants used for the simulation of the Hodgkin-Huxley model \cite{ermentrout2010mathematical}}
\vspace*{0.0cm}
\label{tablehh}
\begin{tabular}{l l r}
    \hline \hline
     {Membrane capacitance
     ($\SI{}{\micro\farad/\centi\meter^{2}}$)} & $C_\mathrm{M}$ & $1$ \\ \hline
    \multirow{3}{*}{Maximum conductances
($\SI{}{\milli\siemens/\centi\meter^{2}}$)}  
&$\overline{g}_{\mathrm{Na}}$&$120$  \\ 
&$\overline{g}_{\mathrm{K}}$&$36$  \\ 
&$\overline{g}_{\mathrm{leak}}$&$0.3$\\
\hline
    \multirow{3}{*}{Resting potentials
 (\SI{}{\milli\volt})}  & $E_{\mathrm{Na}}$&$50$ \\  & $E_{\mathrm{K}}$&$-77$ \\   & $E_{\mathrm{leak}}$&$-54.4$    \\
    \hline
\end{tabular}
\label{tablehh}
\end{table}

\subsection{A Hodgkin-Huxley-like (HH$\ell$) model} \label{sec_hb}

The model proposed by Braun \textit{et al.} \cite{braun1998computer} consists in an adaptation of the HH model in order to reproduce similar patterns observed in thermally sensitive electroreceptors of the catfish (\textit{Ictaluris nebulosus}) \cite{braun1997low}. This dynamical behavior is called burst, is characterized by a silent phase of near-steady-state resting behavior alternating with an active phase of rapid spike oscillations. Bursting occurs in the activity of some thalamic cells, e.g., can implicate in the generation of sleep rhythms, whereas patients with parkinsonian tremors exhibit increased bursting activity in neurons within the basal ganglia \cite{ermentrout2010mathematical}. 

The adaptation consists of the addition of two sub-threshold currents with slower frequency, and some temperature-dependence parameters. The main equation of the model is defined as
\begin{equation} \label{eq_1}
    C_\mathrm{M}\frac{dV}{dt} = -I_\mathrm{d} - I_\mathrm{r} - I_\mathrm{sd} - I_\mathrm{sr} - I_\mathrm{l},
\end{equation}
where, again, $C_\mathrm{M}$ is the membrane capacitance; $I_\mathrm{d}$ and $I_\mathrm{r}$ represent the classical HH ionic currents related to Na$^+$ and K$^+$, respectively. $I_\mathrm{sd}$ and $I_\mathrm{sr}$ are sub-threshold currents of depolarization and repolarization, respectively, which can be related to the Ca$^{2+}$ ions \cite{shorten2000hodgkin}. $I_\mathrm{l}$ represents the passive current which takes the contribution of the non-gated channels. The currents are described by
\begin{eqnarray}
I_{\mathrm{d}}&=&\varrho g_{\mathrm{d}}a_{\mathrm{d}}(V-E_{\mathrm{d}}), \label{eq_hb_1}\\
I_{\mathrm{r}}&=&\varrho g_{\mathrm{r}}a_{\mathrm{r}}(V-E_{\mathrm{r}}) ,\\
I_{\mathrm{sd}}&=&\varrho g_{\mathrm{sd}}a_{\mathrm{sd}}(V-E_{\mathrm{sd}}), \\
I_{\mathrm{sr}}&=&\varrho g_{\mathrm{sr}}a_{\mathrm{sr}}(V-E_{\mathrm{sr}}), \\
I_{\mathrm{l}}&=&g_{\mathrm{l}}(V-E_{\mathrm{l}}),\label{eq_hb_2}
\end{eqnarray}
in which $g_{\mathrm{d}}$, $g_{\mathrm{r}}$, $g_{\mathrm{sd}}$, $g_{\mathrm{sr}}$, and $g_{\mathrm{l}}$ are the maximum values of the respective conductances, and $a_{\mathrm{d}}$, $a_{\mathrm{r}}$, $a_{\mathrm{sd}}$, and $a_{\mathrm{sr}}$ are the activation functions of the ion channels. $\varrho$ is a temperature-dependent parameter defined by
\begin{equation}
\varrho=1.3^{\frac{T-T_{0}}{\tau_{0}}},
\end{equation}
where $T$ is the temperature of the system, $T_{0}$ and $\tau_{0}$ are constants.

The temporal evolution of each activation term follows a differential equation
\begin{eqnarray}
\frac{da_{\mathrm{d}}}{dt}&=&\frac{\phi}{\tau_{\mathrm{d}}}(a_{\mathrm{d},\infty}-a_{\mathrm{d}}), \\
\frac{da_{\mathrm{r}}}{dt}&=&\frac{\phi}{\tau_{\mathrm{r}}}(a_{\mathrm{r},\infty}-a_{\mathrm{r}}),\\
\frac{da_{\mathrm{sd}}}{dt}&=&\frac{\phi}{\tau_{\mathrm{sd}}}(a_{\mathrm{sd},\infty}-a_{\mathrm{sd}},)\\
\frac{da_{\mathrm{sr}}}{dt}&=&\frac{\phi}{\tau_{\mathrm{sr}}}(-\eta I_{\mathrm{sd}}-\gamma a_{\mathrm{sr}}),  \label{a_sr_eq}
\end{eqnarray}
in which $\phi$ represents another temperature-dependent parameter given by $\phi = 3^{(T-T_0)/\tau_0}$; $\tau_{\mathrm{d}}$,  $\tau_{\mathrm{r}}$,  $\tau_{\mathrm{sd}}$,  $\tau_{\mathrm{sr}}$ are characteristic times which correspond to each activation function \cite{braun1998computer}. The term $\eta$ is a factor which relates the mixed Na/Ca current to the increment of the intracellular Ca$^{2 +}$, and $\gamma$ is the decrease rate of Ca$^{2+}$ \cite{shorten2000hodgkin}.

\begin{figure}[t]
    \centering
    \includegraphics[width=0.88\columnwidth]{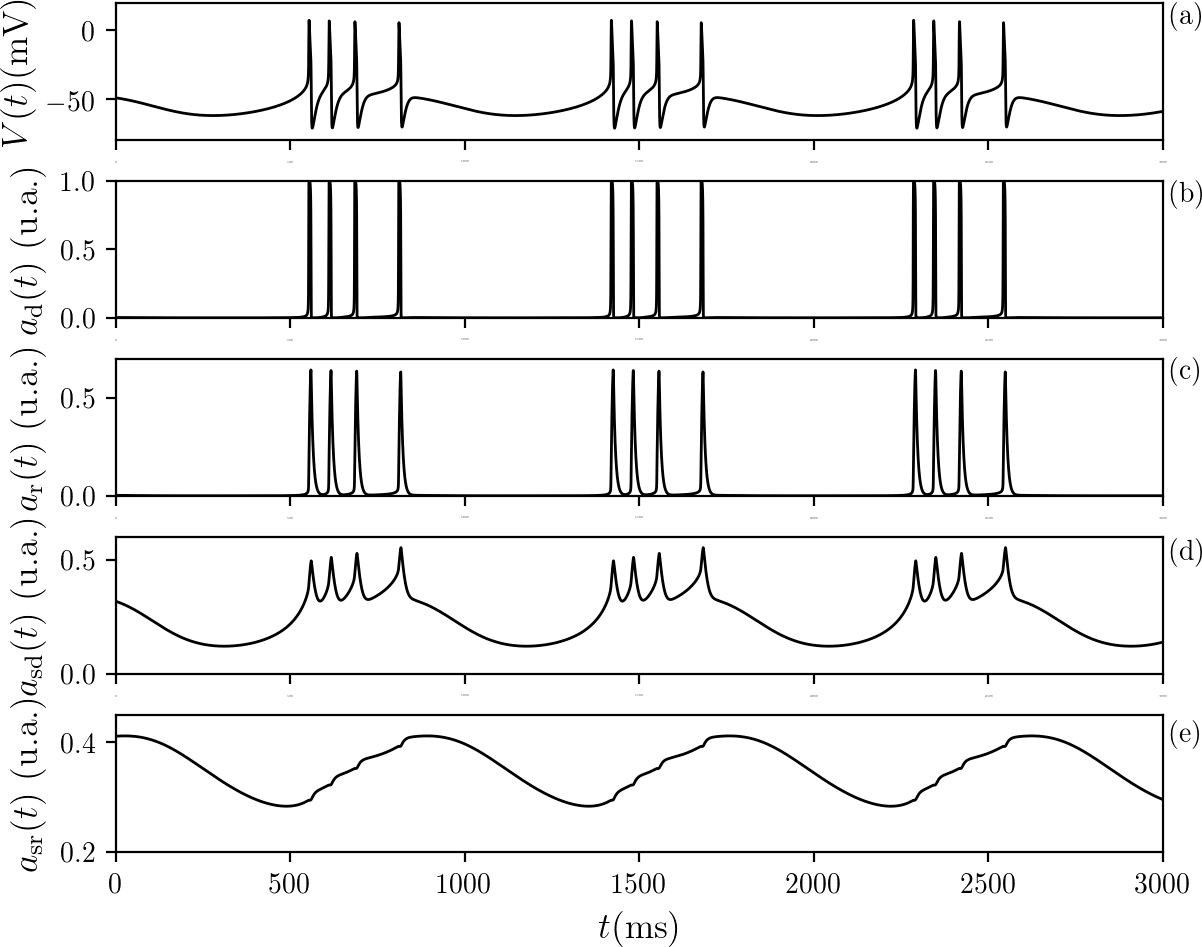}
    \caption[Time-evolution of the variables of the HH$\ell$ model.]{\textbf{Time-evolution of the variables of the HH$\ell$ model.} For the values of the Table \ref{tabeladeconstantes} the membrane potential $V(t)$ (a) depicts the burst dynamics: a sequence of spikes followed by a resting time. The activation variables $a_\mathrm{d}$ (b) and $a_\mathrm{r}$ (c) follow the $V$ fast dynamics, while $a_\mathrm{sd}$ (d) and $a_\mathrm{sr}$ (e) present a slower dynamics. In particular, $a_\mathrm{sr}$ oscillates with the bursting frequency slightly influenced by the spike occurrences.}
    \label{fig:evol_hb_din}
\end{figure}

And finally the functions
\begin{eqnarray}
a_{\mathrm{d},\infty}&=&\frac{1} {1+\exp[-s_{\mathrm{d}} (V-V_{0\mathrm{d}}) ] },\\
a_{\mathrm{r},\infty}&=&\frac{1} {1+\exp[-s_{\mathrm{r}} (V-V_{0\mathrm{r}}) ] },\\
a_{\mathrm{sd},\infty}&=&\frac{1}{1+\exp[-s_{\mathrm{sd}}(V-V_{0\mathrm{sd}})] },
\label{aaaa}
\end{eqnarray} 
where $s_{\mathrm{d}}$, $s_{\mathrm{r}}$, $s_{\mathrm{sd}}$, $V_{0\mathrm{d}}$, $V_{0\mathrm{r}}$ and $V_{0\mathrm{sd}}$ are parameters. The Fig. \ref{fig:evol_hb_din} presents the time-evolution of the $5$ variables of the model, using the parameters of the Table \ref{tabeladeconstantes}. In panel (a) the membrane potential $V$ depicts a bursting oscillation, a sequence of $4$ spikes followed by a resting time. At panels (b) and (c) the activation variables $a_\mathrm{d}$ and $a_\mathrm{r}$ present a similar dynamics. In panels (d) and (e) the activation variables $a_\mathrm{sd}$ and $a_\mathrm{sr}$ show a slower oscillation which follow the bursting oscillation. 

\begin{table}[htb!] 
\begin{center}
\caption{Values of the parameters of the HH$\ell$ model \cite{postnova2010computational}}
\begin{tabular}{l l r}
\hline\hline
{Membrane Capacitance ($\SI{}{\micro\farad/\centi\meter^{2}}$)} & $C_\mathrm{M}$ & $1$ \\ \hline
\multirow{5}{*}{Maximum conductances
($\SI{}{\milli\siemens/\centi\meter^{2}}$)}  &${g}_{\mathrm{d}}$&$1.5$   \\ 
&${g}_{\mathrm{r}}$&$2$    \\ &${g}_{\mathrm{sd}}$&$0.25$ \\ &${g}_{\mathrm{sr}}$&$0.4$  \\
&${g}_{\mathrm{l}}$&$0.1$  \\ \hline
\multirow{4}{*}{{Characteristic times ($\SI{}{\milli\second}$)}} & 
$\tau_{\mathrm{d}}$&$0.05$ \\ 
&$\tau_{\mathrm{r}}$&$2$ \\ 
&$\tau_{\mathrm{sd}}$&$10$ \\
&$\tau_{\mathrm{sr}}$&$20$  \\
\hline

\multirow{8}{*}{Reversal potentials $(\SI{}{\milli\volt})$} & $E_{\mathrm{d}}$&$50$   \\ 
&$E_{\mathrm{r}}$&$-90$  \\
&$E_{\mathrm{sd}}$&$50$  \\ 
&$E_{\mathrm{sr}}$&$-90$  \\
&$E_{\mathrm{l}}$&$-60$ \\
&$V_{0\mathrm{d}}$&$-25$ \\
&$V_{0\mathrm{r}}$&$-25$   \\ 
&$V_{0\mathrm{sd}}$&$-40$ \\ \hline

\multirow{3}{*}{Temperature parameters (\SI{}{\celsius})} & 
$T_{0} $&$ 25$ \\  &
$T $&$ 13$ 
\\
& $\tau_{0} $&$ 10$ \\ \hline

\multirow{5}{*}{Other parameters} & 
$s_{\mathrm{d}}$&$\SI{0.25}{\milli\volt^{-1}}$\\ &
$s_{\mathrm{r}}$&$\SI{0.25}{\milli\volt^{-1}}$ \\& 
$s_{\mathrm{sd}}$&$\SI{0.09}{\milli\volt^{-1}}$ \\
&
$\eta $&$ \SI{0.012}{\centi\meter^{2}/\micro\ampere}$ \\ & 
$\gamma$&$0.17$\\  
              \hline\hline
            \end{tabular}
            \label{tabeladeconstantes}
\end{center}
\end{table}

\begin{figure}[htb!]
\begin{center}
\includegraphics[width=.95\columnwidth]{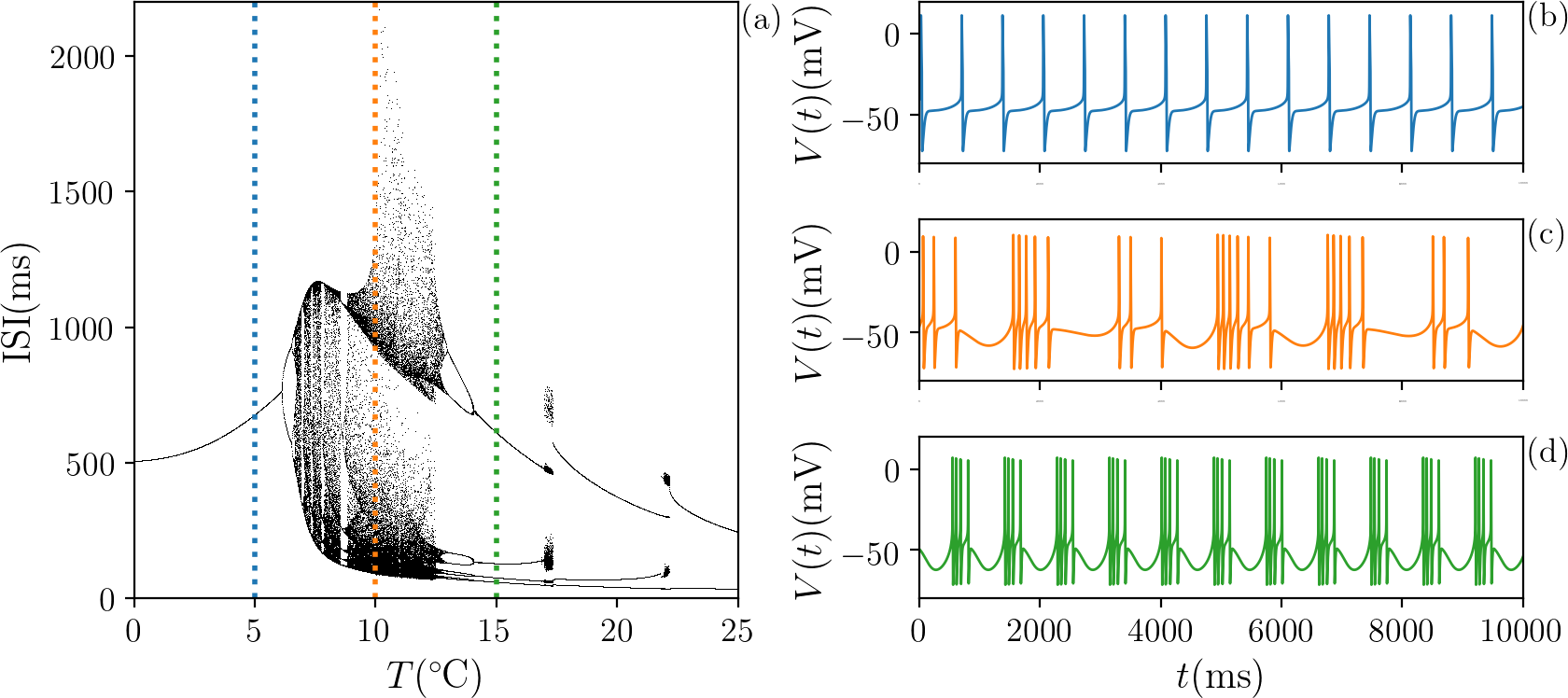}
\caption[Bifurcation diagram of the HH$\ell$ model.]{\textbf{Bifurcation diagram of the HH$\ell$ model.} (a) Inter-spike-interval (ISI) as a function of the temperature $T$ measured $\SI{}{\celsius}$. The membrane potential dynamics consists of (b) periodic spikes $T=5$, (c) chaotic bursts $T=10$, and (d) periodic bursts $T=15$.}
\label{fig_hb_1}
\end{center} 
\end{figure}

Considering different parameters it is possible to change the dynamics of the model. Figure~\ref{fig_hb_1} displays the bifurcation diagram of the ISI as a function of the temperature $T$ measured in $\SI{}{\celsius}$. The spike is computed in the same way as the HH model when $V$ crosses $-20$ $\SI{}{\milli\volt}$ (with positive derivative). The bifurcation diagram represents a classical route to chaos called doubling period \cite{ott2002chaos}. The first doubling occurs at $T \approx 6.765$ and the second $T \approx 7.195$, and for $T>7.31$ the system reach the chaoticity \cite{feudel2000homoclinic}. For $T \approx 10.66$ the ISI depicts extremely higher values, in Ref. \cite{feudel2000homoclinic}, the authors show evidences that this behavior is associated with a homoclinic bifurcation. For $T \gtrsim 13$ the system loses its chaoticity but presents two branches of ISI associate with the fast and slow temporal scales, associated with the spikes and bursts respectively. Panels (b -- d) depict the dynamics of $V$ for $T=5,\,10,\,15$, respectively. For $T=5$ the neuron depicts periodic spikes in a similar way that the HH model. For $T=10$ the neuron presents chaotic bursts, and $T=15$ periodic bursts. It should be noted that the transitions to chaos can also be achieved with the change of other parameters, as the maximum conductances for example.

\subsection{The Hindmarsh-Rose model}

\begin{figure}[htb!]
    \centering
    \includegraphics[width=.95\columnwidth]{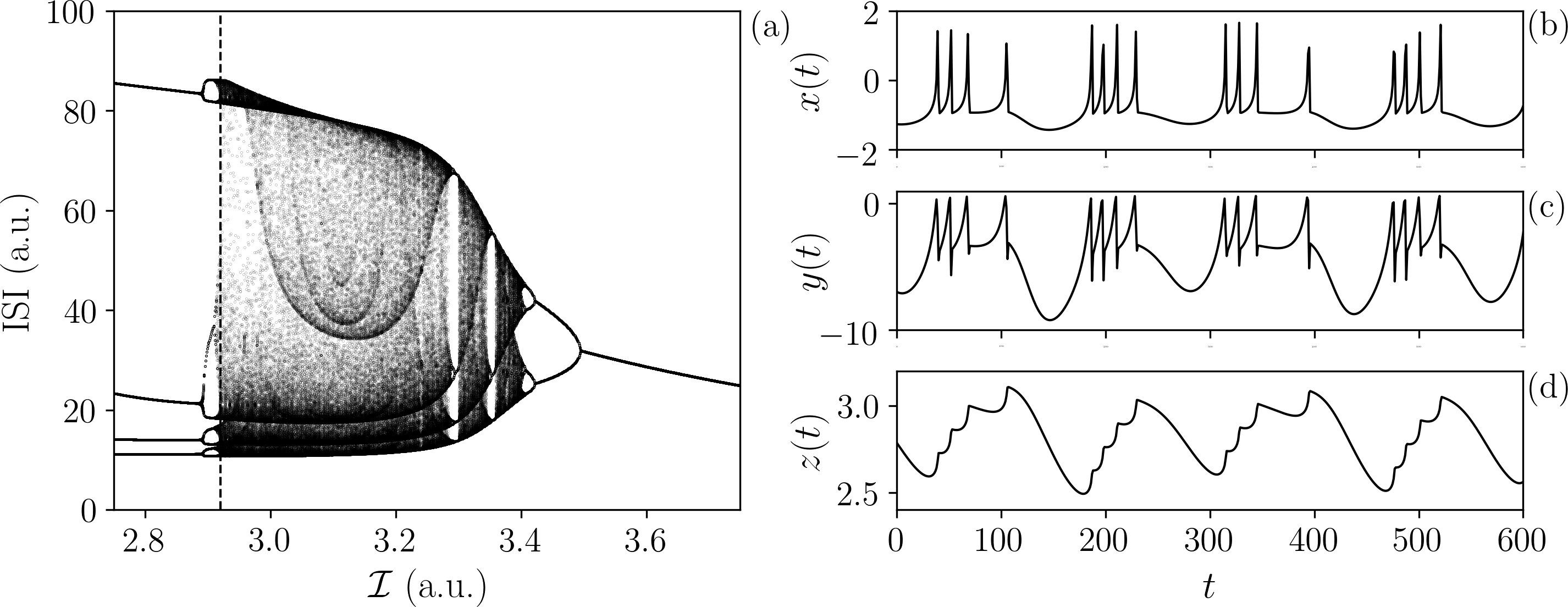}
    \caption[Bifurcation diagram of the HR model.]{\textbf{Bifurcation diagram of the HR model.} (a) Inter-spike-interval (ISI) as a function of the applied current $\mathcal I$. The time of each spike is calculated when $x$ crosses $0.0$ with a positive derivative. (b -- d) Dynamics of each variable of the model, $x$, $y$ and $z$ respectively. Using $\mathcal I=2.92$ and parameters of Table \ref{tablehr}.}
\label{fig:hr_2}
\end{figure}

In addition to the model described in the previous section, in the literature, it is possible to find several models that present the bursting dynamics \cite{ermentrout2010mathematical,rulkov1995generalized,hindmarsh1984model,izhikevich2003simple}. Most of the models are purely dynamical, which means that, besides the dynamics, there is no relation with the real neuron. In this sense, the Hindmarsh-Rose (HR) model \cite{hindmarsh1984model} is composed of three dimensionless variables ($x,y,z$) in which the $x$ variable depicts a similar-bursting dynamics observed in the membrane potential of real neurons \cite{ermentrout2010mathematical}. The model is described by three differential equations
\begin{eqnarray}
\frac{dx}{dt}&=&y-ax^{3}+bx^{2}-z+\mathcal{I},\label{hr_I}\\
\frac{dy}{dt}&=&c-dx^2-y, \label{hr_II}\\
\frac{dz}{dt}&=&r[s(x-x_\mathrm r)-z],\label{hr_III}
\end{eqnarray}
where $a$, $b$, $c$, $d$, $r$, $x_\mathrm r$, $s$, and $\mathcal{I}$ are parameters of the model. The $\mathcal I$ parameter acts similarly to an external current being applied to the neuron, often used as a control parameter \cite{hindmarsh1984model}. In this sense, Fig.~\ref{fig:hr_2} depicts a bifurcation diagram of the ISI as a function of $\mathcal I$, using the other parameters of Table \ref{tablehr}. The spike is evaluated when $x$ crosses $0$ (with positive derivative). This model depicts periodic bursts for $\mathcal I< 2.92$, chaotic bursts for $2.92< \mathcal I \lesssim 3.4$, and periodic spikes for $\mathcal I\gtrsim 3.4$. Panels (b -- d) present the dynamics of the three variables of the model for $\mathcal I=2.92$ to clarify the chaoticity of the bursting dynamics of this model. 

\begin{table}[t]
\centering
\caption{Constants used for the simulation of the Hindmarsh-Rose model \cite{hindmarsh1984model}.}
\vspace*{0.0cm}
\label{tablehr}
\begin{tabular}{c c c c c c c}
    \hline \hline
    $a = 1$ & $b=3$ & $c=1$ &
    $d=5$ & $x_\mathrm r=-8/5$ &
    $s=4$ & $r=0.006$ \\
    \hline
\end{tabular}
\end{table}

\subsection{The Chialvo model} \label{sec_chialvo}

\begin{figure}[t]
\begin{center}
\includegraphics[width=\columnwidth]{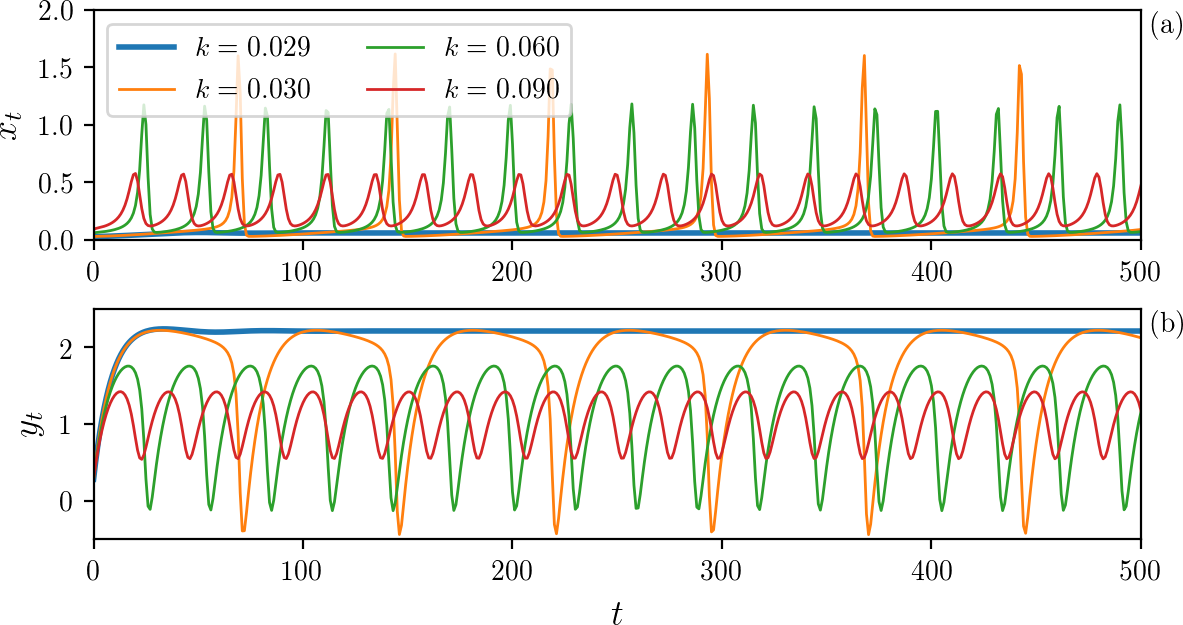}
 \caption[Time evolution of the variables of the Chialvo model.]{\textbf{Time evolution of the variables of the Chialvo model.} Evolution of the variables $x_t$ (a) and $y_t$ (b). For $a=0.89$, $b=0.6$, $c=0.28$, and different values of $k$. For $k<0.03$ the systems stays at a fixed point while $k>0.03$ the model depicts a periodic motion.}
\label{fig_chialvo}
\end{center} 
\end{figure}

The model proposed by Dante R. Chialvo \cite{chialvo1995generic} contains two  dimensionless map equations described by
\begin{eqnarray}
x_{t+1} &=& x_{t}^2\exp(y_{t}-x_{t})+k, \label{chialvo1}\\
y_{t+1} &=& a y_{t}-bx_{t}+c,\label{chialvo2}
\end{eqnarray}
where $x_t$ acts like the potential of a membrane, and $y_t$ like an recovery variable. The model depends of four parameters, $a$, $b$, $c$ e $k$. In particular $k$ acts like an external current. In Fig. \ref{fig_chialvo}, the time evolution of both variables $x_t$ and $y_t$ is shown using $a = 0.89$, $b = 0.6$, $c = 0.28$, for different values of $k$. For $k<0.03$ (blue line) the system is at an equilibrium state. For $k\geq 0.03$ the dynamics of the model consists of periodic spikes, but for higher values of $k$, smaller is the amplitude of the spikes and higher the frequency \cite{chialvo1995generic}.

\section{Synaptic current}

The synapse is the region responsible for carrying electrical signals initiated in the presynaptic neuron that propagate within a postsynaptic neuron \cite{kandel2013principles}. This connection is often represented by a current term called synaptic current. The synapses generally consist of three components: the axon terminals of the presynaptic neuron, a target on the postsynaptic neuron, and a zone of apposition. The structure of the apposition rules the type of the synapse, which can be electrical or chemical synapses \cite{kandel2013principles}. While electrical synapses provide almost instantaneous signal transmission, chemical synapses can amplify the signal.

The electrical synapses are characterized by a direct interaction among ions of the two cells, where the presynaptic terminal and the postsynaptic cell are in very close apposition at regions called gap junctions. The signal produced by the action potential of the presynaptic neuron reaches the postsynaptic neuron at a gap junction which connects both neurons \cite{kandel2013principles}. In this sense, the postsynaptic neuron receives a synaptic current described as
\begin{equation}
    I_\mathrm{syn} = g_\mathrm{syn}(V_\mathrm{pre}-V_\mathrm{post}),
\end{equation}
where $g_\mathrm{syn}$ is the conductance of the channels, $V_\mathrm{pre}$ and $V_\mathrm{post}$ are the membrane potentials of the presynaptic and postsynaptic neuron, respectively.  

In a chemical synapse, the connection does not occur directly. The action potential of the presynaptic neuron leaves neurotransmitters, the most common in cortical neurons are glutamate and $\gamma$-Aminobutyric acid (GABA) \cite{ermentrout2010mathematical}. These neurotransmitters diffuse through the synaptic cleft to reach the postsynaptic neuron, inducing the opening or the closing of channels. For these cases, the synaptic current is given by
\begin{equation}
        I_\mathrm{syn} = g_\mathrm{syn}r(t)(E_\mathrm{syn}-V_\mathrm{post}),
        \end{equation}
where $E_\mathrm{syn}$ is the reversal synaptic potential which characterizes if the synapse is excitatory or inhibitory, and $r(t)$ is a function that simulates the neurotransmitter kinetics \cite{destexhe1994efficient}.

A way of modeling $ r (t) $ is considered a sigmoidal function \cite{ermentrout2010mathematical}  
\begin{equation}
    r(t) = \frac{1}{1+\exp(-\lambda_\mathrm s(V_\mathrm{pre}(t)-\beta_\mathrm s))},
\end{equation}
where $ \lambda_\mathrm s$ and $\beta _\mathrm s $ are constants related to chemical synapses, adjusted for the respective neuronal model.

Another way to model $r(t)$ is with a kinetic function, which takes into account the fraction of open channels that allow the transmission of neurotransmitters,
\begin{equation} 
    \frac{dr}{dt} = \left(\frac{1}{\tau_\mathrm{r}}-\frac{1}{\tau_\mathrm{d}}\right)\frac{1-r}{1+\exp[-s_0(V_\mathrm{pre}(t)-V_{0})]}-\frac{r}{\tau_\mathrm{d}}, \label{eq_r_2}
\end{equation}
in which $\tau_\mathrm{r}$ and $\tau_\mathrm{d}$ are characteristic times, $s_0 = 1/\SI{}{\milli\volt}$ is a unitary constant, and $V_0$ is a reversal potential \cite{destexhe1994efficient}.

\chapter{Complex networks}\label{chap:redes}

\initial{T}{he} study of the global behavior of a group of connected entities is made using a complex network. A dynamical system is called {\it complex} if its final state does not depend on the existence of a central controller, i.e., the cooperative behavior depends only on the interactions among entities \cite{boccara2010modeling}. In this chapter, it is introduced some concepts of the graphs theory, which is used in the network study. The way that the  connections are distributed characterizes the network connection topology. In this thesis, all the networks were generated using a Python language package called ``NetworkX'' \cite{SciPyProceedings_11} which was developed for exploration and analysis of networks and network algorithms.

\section{Graphs theory}

\begin{figure}[t]
\centering
\includegraphics[width=0.75\columnwidth]{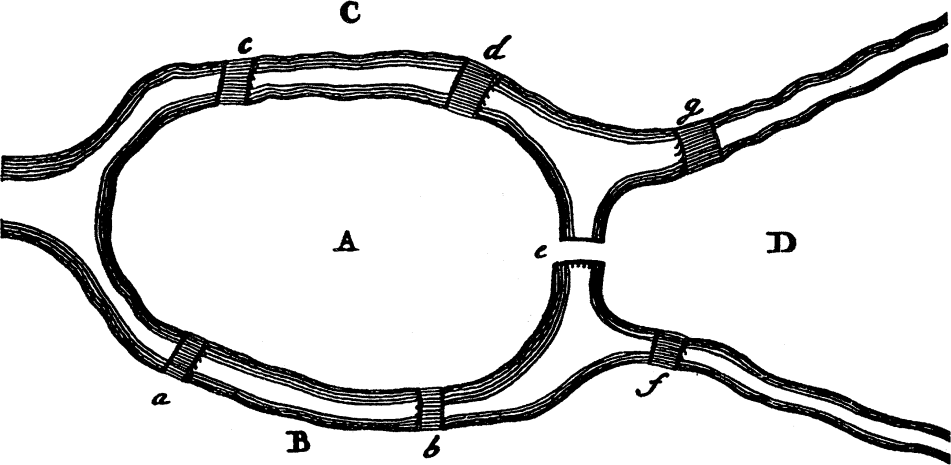}
\caption[Illustration of the Königsberg bridges.]{\textbf{Illustration of the Königsberg bridges.} The Pregel River ramifications and form two islands $A$  and $D$, which are connected to the city ($B$ and $C$) by the $7$ bridges $a$, $b$, $c$, $d$, $e$, $f$ and $g$. Figure taken from \cite{alexanderson2006cover}, an adaptation of the original work of Euler \cite{euler1741solutio}.}
\label{fig:bridges}
\end{figure}

The graph theory is a section of discrete mathematics that was developed in part by Leonard Euler to solve the challenge of the \textit{Königsberg} bridges. The \textit{Pregel} river, when cross the \textit{Königsberg} city is ramified creating two islands, which were connected to the city by seven bridges. The challenge consisted at to cross all the bridges without repetition, independently where the journey was started or finished. Figure~\ref{fig:bridges} depicts one adaptation of the original illustration of the challenge made by Euler \cite{euler1741solutio,alexanderson2006cover}.

The idea used to solve this challenge was to represent each earth region and the bridges as sites connected by edges. It was found that, since all the earth regions have an odd number of bridges, there is no solution for this challenge. The challenge could be solved if any of the bridges were removed. The way in that Euler obtained his results gives a start to the study of graph theory  \cite{west2001introduction}.

A graph $\mathcal G$ is an ordered pair of distinct sets $\mathcal V$ and $\mathcal E$, where $\mathcal V$ is a non-empty discrete set of elements called sites or nodes, and $\mathcal E$ a subset constituted of ordered pair of $\mathcal V$ elements, called edges or connections. For two given elements of $\mathcal V$, the pair  $(v_i,\,v_j) \in \mathcal E$ is a line which connects $v_i$ and $v_j$. The connections among sites can be binary or weighted: for binary connections, $(v_i,\,v_j) = 1$ if the sites are connected or $0$, otherwise; when weighted, some connections are more relevant than others $(v_i,\,v_j) \in \mathbb R$. On the other hand, the connections can be directed and non-directed: when non-directed the connections are reciprocal which means that, if $v_i$ is connected to $v_j$, implies that $v_j$ is connected to $v_i$; when directed, the connections are not reciprocal. The notation $\mathcal G(N,n)$ represents a graph with $N$ sites and $n$ connections. The graph $\mathcal G(N,n)$ can be represented as a squared matrix $G$, called connection matrix, where $N$ is the number of rows and columns, and $n$ is the number of non-zero elements. In this sense, the element $e_{ij}$ of the matrix $G$ represents the connection between the site $i$ and $j$. 

\section{Complex networks}

The graph theory can be used in the study of networks, in particular, neural networks, where each node of the network is a neuron and their connections are the synapses. The way of the connections are distributed in the network determines its topology which plays a role in the global dynamics of the network \cite{strogatz2001exploring}. In this sense, besides the size of the network $N$ and the number of connections $n$, it is possible to classify the topology of the network using some properties.

The average shortest path length $\mathcal{L}$ is a quantity that refers to the shortest path between the $i$-th site and $j$-th site. This path relates to the number of sites that the information needs to pass to reach the target \cite{boccara2010modeling}. If $\mathcal L_{ij}$ is the shortest path between $i$ and $j$, the average shortest path length of the network is
\begin{equation}
\mathcal{L}=\dfrac{1}{N(N-1)}\sum_{i=1}^{N} \sum_{\substack{j=1\\ j\neq i}}^{N} \mathcal{L}_{ij}.
\label{caminhomedio}
\end{equation}

The clustering coefficient $\mathcal C$ is a quantity of how the sites tend to cluster. The evaluation is based on the number of sites which groups in trios  \cite{boccara2010modeling,saramaki2007generalizations}. Considering three sites, a closed trio is when all sites are connected while an open trio one of the connections is missing. The clustering coefficient is defined by the ratio between the number of closed trios divided by the total number of trios (open and closed),
\begin{equation}
    \mathcal{C} =\frac{\mathcal{T}_\mathrm{\triangle}}{\mathcal{T}_\mathrm{\triangle}+\mathcal{T}_{\land}}, 
\end{equation}
where $\mathcal{T}_\mathrm{\triangle}$ and $\mathcal{T}_{\land}$ are the number of closed trios and open trios, respectively. If $\mathcal T_{\triangle} \gg \mathcal T_{\land}$, $\mathcal C \approx 1$, which means in a clustered network \cite{boccara2010modeling}. 

\begin{figure}[t]
\begin{center}
\includegraphics[width=.9\columnwidth]{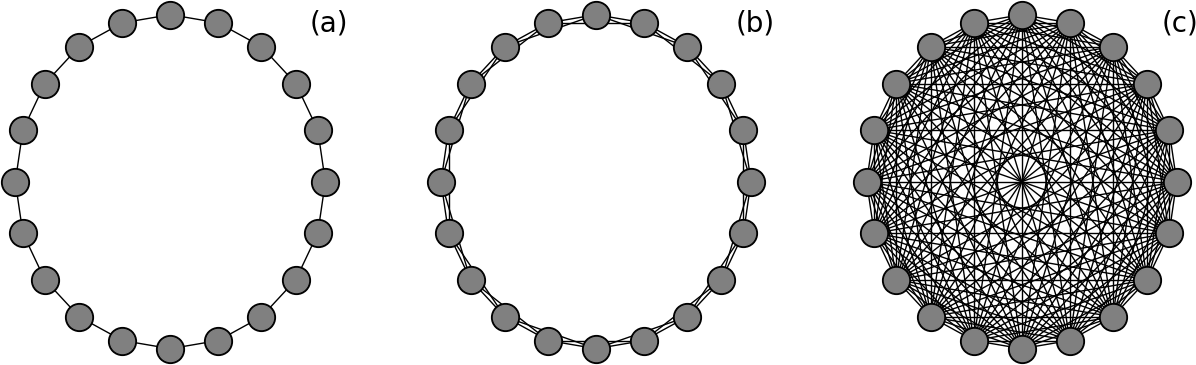}
\caption[Examples of regular networks.]{\textbf{Examples of regular networks.} Using $20$ sites. (a) First-neighborhood network. (b) Second-neighborhood network. (c) Global network (complete network).}\label{fig:regular}
\end{center} 
\end{figure}

The simplest network topology is called a regular network, where all the sites depict the same number of connections, as is the case of a first-neighborhood network, where the $i$-th site is connected to the $i+1$ and $i-1$ sites (closest neighbors), or the second-neighborhood network, and global network where all the possible connections exist. Figure~\ref{fig:regular} depicts an example of these three regular networks with $N=20$ sites. 

\begin{figure}[t]
\begin{center}
\includegraphics[width=.9\columnwidth]{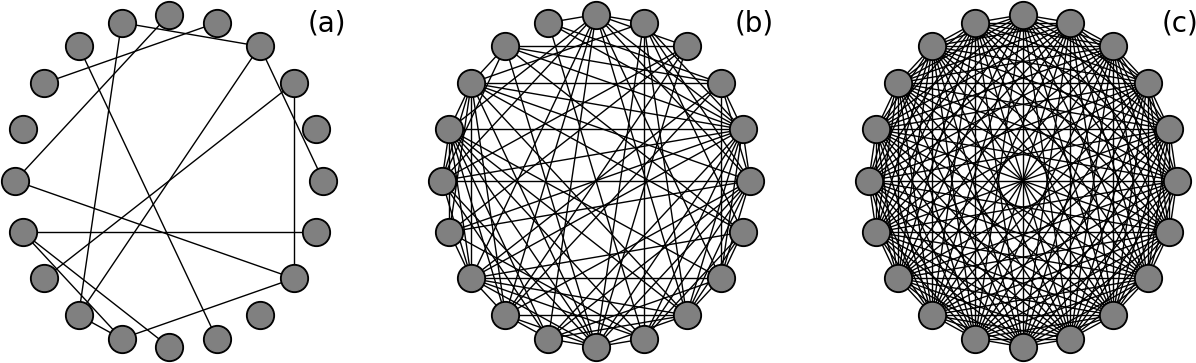}
\caption[Examples of random networks.]{\textbf{Examples of random graphs.} Using $20$ sites. (a) For $p_\mathrm{rand}=0.05$ there are $20$ connections. (b) For $p_\mathrm{rand}=0.5$ there are $190$ connections. (c) For $p_\mathrm{rand}=1$ there are $n(n-1)$ connections.}
\label{fig:rand}
\end{center} 
\end{figure}

Another topology often used in the literature is the random topology, where their connections are randomly distributed \cite{boccara2010modeling}. One of the possibilities to build such network is using the Erdos-Rényi algorithm \cite{erdos1959random}, which consider a probability of connection $0\leq p_\mathrm{rand} \leq 1$, where the number of connections in the network is given by $n_\mathrm{rand}=p_\mathrm{rand}N(N-1)$. Figure~\ref{fig:rand} depicts examples of random networks with $N=20$ and (a) $p_\mathrm{rand}=0.05$ and $n=20$ connections, in this case the small number of connections makes possible to find isolated sites from the network, (b) $p_\mathrm{rand}=0.5$ and $n=190$ connections, and (c) $p_\mathrm{rand}=1$ all possible connections.

However, regular and random networks are idealization cases, real networks are believed to be between these extremes of order and randomness \cite{strogatz2001exploring}. Watts and Strogatz have found that with the substitution of local connections for random connections in a regular network, it is possible to create a network with shortcuts that presents the following properties: low average shortest path length and a high clustering coefficient (in comparison with random networks). These networks were called small-world networks \cite{watts1998collective}. In addition to being networks optimized from the point of view of information, propagation \cite{watts1998collective}, it was found that some real networks have the same topological characteristics of a small-world network, like social networks such as the world-wide-web \cite{latora2001efficient, watts2004small, ebel2002dynamics}, electric power grids \cite{watts2004small}, and even neural networks, as in the case of the nematode nervous system {\it C. Elegans} \cite {watts1998collective, achacoso1991ay}, and in the anatomy of the human brain and other mammals \cite{bassett2006small}.

\begin{figure}[t]
\begin{center}
\includegraphics[width=.9\columnwidth]{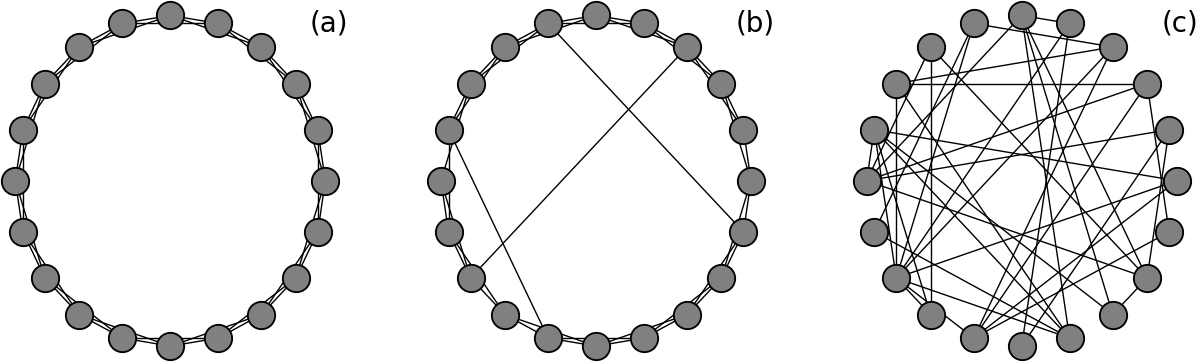}
\caption[Example of the Watts-Strogatz route.]{\textbf{Example of the Watts-Strogatz route.} (a) Regular network $p_\mathrm{ws}=0$, (b) Small-world network $p_\mathrm{ws}=0.05$, (c) Random network $p_\mathrm{ws}=1$. The small-world regime is achieved between the extremes of order and randomness.}
\label{fig:ws}
\end{center} 
\end{figure}

One of the ways to generate a small-world network is using the Watts-Strogatz algorithm \cite{watts1998collective}. Starting with a regular network with $N$ sites and $n_0$ connections, the idea is to replace $n_\mathrm{ws}$ connections randomly. Controlling $n_\mathrm{ws}$ with a probability  $p_\mathrm{ws}$, where $n_\mathrm{ws}=p_\mathrm{ws}n$. For lower values of $p_\mathrm{ws}\approx 0$ the network depicts the same features of a regular one, and higher values $p_\mathrm{ws}\approx 1$ of a random one. But for intermediate values of $p_\mathrm{ws}$ the number of shortcuts can put the network in the small-world regime, presenting a high cluster coefficient and low average shortest path length. Figure~\ref{fig:ws} presents an example of the Watts-Strogatz route, at panel (a) a second-neighborhood regular network ($p_\mathrm{ws}=0$), (b) a small-world network ($p_\mathrm{ws}=0.05$), and (c) a random network ($p_\mathrm{ws}=1$).

\begin{figure}[t]
\begin{center}
\includegraphics[width=.9\columnwidth]{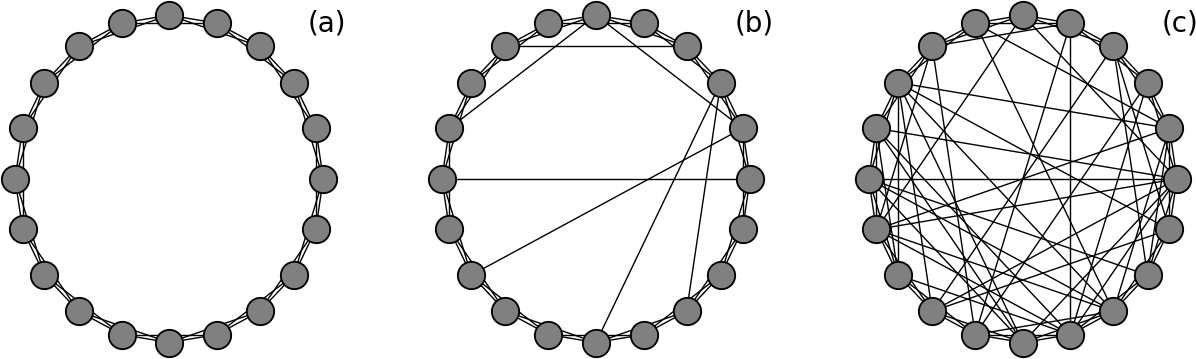}
\caption[Example of the Newman-Watts route.]{\textbf{Example of the Newman-Watts route.} (a) Regular network $p_\mathrm{nw}=0$, (b) Small-world network $p_\mathrm{nw}=0.2$, (c) Network with $p_\mathrm{nw}=1$ in this case the network is no longer at the small-world regime.}
\label{fig:nw}
\end{center} 
\end{figure}

However, in the Watts-Strogatz route, since the connections are randomly replaced, it is possible to disconnect a neuron from the network. To avoid this, Newmann and Watts proposed an alternative algorithm to achieve the small-world regime \cite{newman1999renormalization}, the idea is to add random connections instead of replacing \cite{boccara2010modeling}. For this route, the number of connections depends on a probability $p_\mathrm{nw}$, if $n_0$ is the number of connections of the regular network, the new number of connections is given by $n_\mathrm{nw} = (1+p_\mathrm{nw})n_0$. This means that for each existing connection ($n_0$), a new random connection can be added with a probability $p_\mathrm{nw}$. The Fig.~\ref{fig:nw} represents an example of the Newman-Watts route, at the panel (a) a regular network ($p_\mathrm{nw}=0$), (b) a small-world network ($p_\mathrm{nw}=0.1$) and (c) a network which is no longer in the small-world regime.
\begin{figure}[htb!]
\begin{center}
\includegraphics[width=\columnwidth]{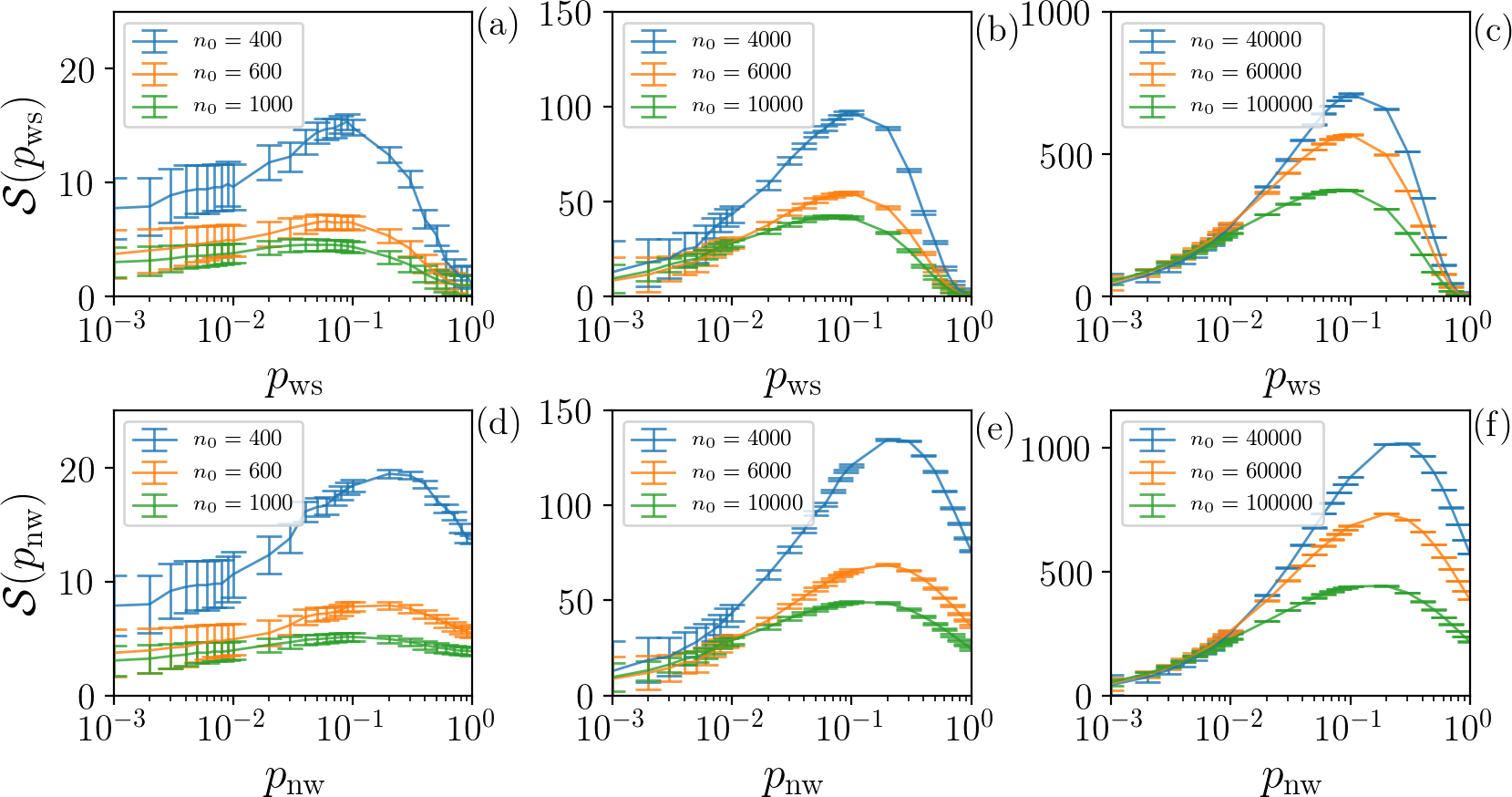}
\caption[Small-worldness coefficient.]{\textbf{Small-worldness coefficient.} $\mathcal S$ as a function of the probability $p_\mathrm{ws}$ (first row), $p_\mathrm{nw}$ (second row), for networks with $N=100$ (a) and (d), $N=1000$ (b) and (e), $N=10000$ (c) and (f). Each color refers to an initial number of connections on a regular network. The error bars are the standard deviation over $10$ different initialization.}
\label{fig:merito}
\end{center} 
\end{figure}

Despite the fact that there is no ideal way to characterize the small-world regime \cite{latora2001efficient,telesford2011ubiquity}, it is possible to define a quantity to quantify the {\it small-worldness} of a generic network using
\begin{equation}
    \mathcal S(p) = \frac{(\mathcal C(p)/\mathcal C_\mathrm r)}{(\mathcal L(p)/\mathcal L_\mathrm r)}
\end{equation}
where $\mathcal L$ and $\mathcal C$ are the average shortest path length and the clustering coefficient of a generic network, and  $\mathcal L_\mathrm r$ and $\mathcal C_\mathrm r$ for a random network. A network is commonly classified as a small-world if $\mathcal S \gg 1$, which happens when $\mathcal C > \mathcal C_\mathrm r$ and $\mathcal L < \mathcal L_\mathrm r$. Figure~\ref{fig:merito} presents $\mathcal S$ as a function of the probability  $p_\mathrm{ws}$ (at the first row) and $p_\mathrm{nw}$ (at the second row) for networks with different sizes $N=100$ (a) and (d), $N=1000$ (b) and (e), $N=10000$ (c) and (f). The Watts-Strogatz route has a maximum of $\mathcal S$ at $p_\mathrm{ws} \approx 0.1$, and for the Newman-Watts the maximum occurs $p_\mathrm{nw}\approx 0.3$. Therefore, the network is in a small-world regime for values close to these maxima.

\chapter{Phase synchronization}\label{chap:ferramentas}

\initial{S}{ynchronization}, this basic nonlinear phenomena, detected in the $17$th century when the scientist Christiaan Huygens reported his observations that two very weakly coupled pendulum clocks become synchronized in phase \cite{huygens1980horologium} (English translation \cite{huygens1986pendulum}). In the context of dynamical systems, many different synchronization states have been studied: complete synchronization, also called identical synchronization, is the simplest form of synchronization, it consists of the exact convergence of all the trajectories of the system to a unique synchronization manifold; frequency synchronization consists in the frequency locking of the system, where all elements of the system evolve with the same periodicity; and phase synchronization were beyond the frequencies the phases are also locked \cite{boccaletti2002synchronization}. In this chapter, it is introduced some concepts of the synchronization of dynamical systems, based on the Kuramoto oscillator model. In particular, it shows a powerful tool to quantify phase synchronization, called the Kuramoto order parameter, which can be generalized to measure phase synchronization of distinct oscillator models \cite{pikovsky2003synchronization}, and in special, to neuronal models.


The Kuramoto model consists of an ensemble of $N$ oscillators with a nonlinear coupling where the phase of the $i$-th oscillator evolves as 
\begin{equation}
    \dot\theta_i = \omega_i + \frac \varepsilon N \sum_{j=1}^N \sin (\theta_j - \theta_i)\,\,\,\,\,\,\,\,\,\, i=1,\cdots,N,
\end{equation}
where $\omega_i$ is the natural frequency of the oscillator and $\varepsilon$ the coupling strength. For $\varepsilon=0$ each oscillator evolves linearly which his own frequency $\theta_i(t)=\omega_i t + \theta_i(0)$,  where $\theta_i(0)$ is the initial condition of the $i$-th oscillator.

For the particular case of identical oscillators $\omega_i = \omega_0, \; \forall\; i$, and, without losing generality $\omega_0=0$ (since for $\omega_0 \neq 0$, it is always possible to choose a frame that rotates at the same frequency $\omega_0$), therefore
\begin{equation} \label{eq:identical_kura}
\dot\theta_i =\frac \varepsilon N \sum_{j=1}^N \sin (\theta_j - \theta_i)\,\,\,\,\,\,\,\,\,\, i=1,\cdots,N.
\end{equation}

With the definition of a vector ${\bm\theta} = (\theta_1,\theta_2,\cdots,\theta_N)$, it is shown that Eq. (\ref{eq:identical_kura}) is a gradient equation $\dot{\bm \theta}=-\nabla \mathcal U$ \cite{wiley2006size}, where the potential function is given by
\begin{equation}
    \mathcal U = -\frac{\varepsilon}{2N}\sum_{i=1}^N\sum_{j=1}^N \cos(\theta_j-\theta_i),
\end{equation}
which means that the trajectories of such systems flow monotonically at the potential surface and asymptotically approaching the equilibrium point $\dot{\bm \theta} = 0$. For $\varepsilon>0$, any symmetric initial condition around the circle $[0,2\pi]$ will lead the system to a steady-state solution, called twisted states \cite{wiley2006size}, since the sinusoidal sum among all oscillators mutually cancel. For non-symmetric initial conditions, the coupling drives the system to the phase-locking stable state, where $\theta_1 = \theta_2 = \cdots = \theta_N$, also called complete phase synchronized state. 

\begin{figure}[htb!]
     \centering
     \includegraphics[width=\columnwidth]{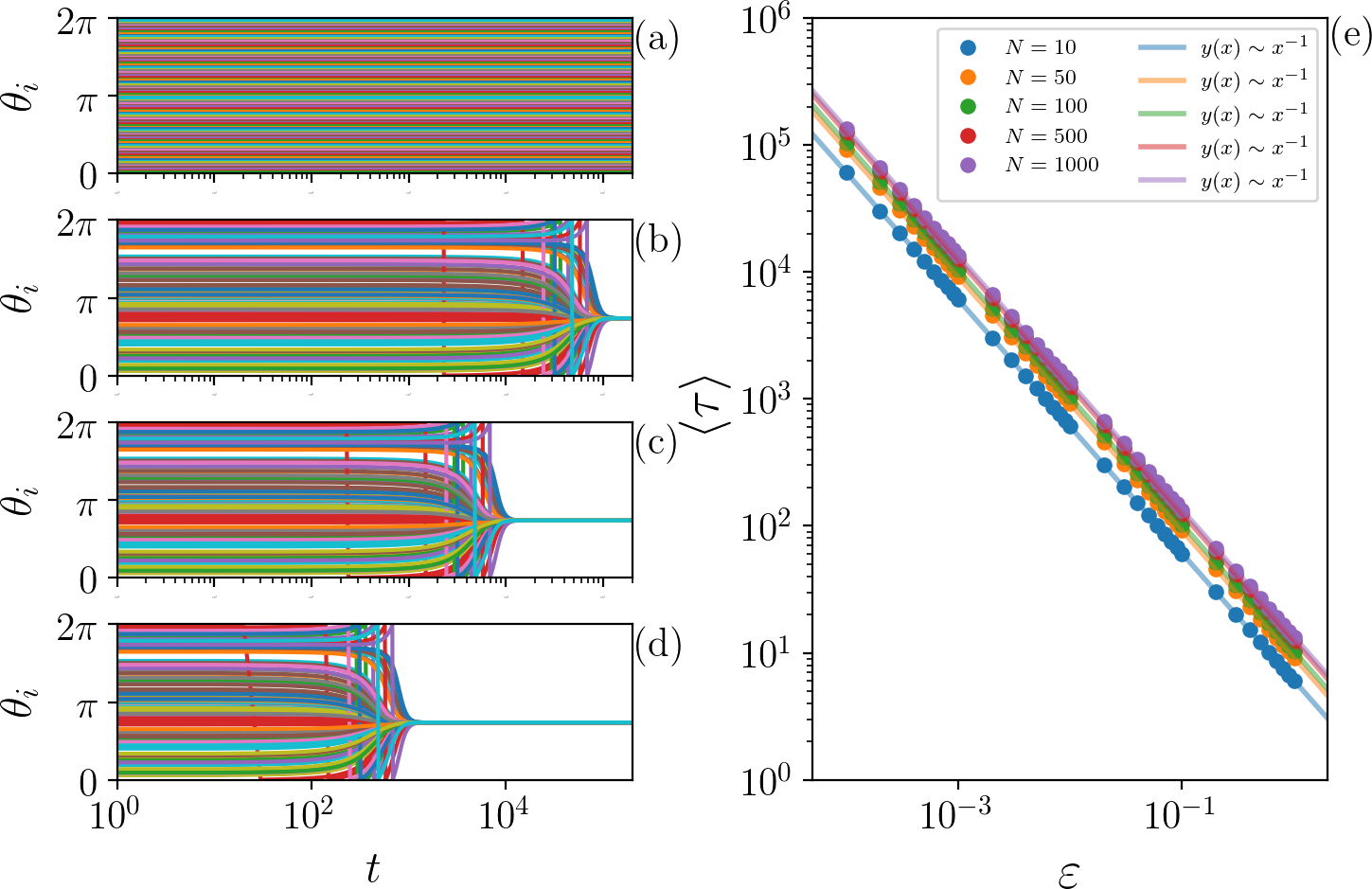}
     \caption[Time evolution of the phases of identical Kuramoto oscillators and the relaxation time of the network.]{\textbf{Time evolution of the phases of identical Kuramoto oscillators and the relaxation time of the network.} (a) For a initial condition which follows a uniform distribution, i.e., $\theta_i(0) = 2\pi i/N$, the symmetry of the system cancels the network dynamics, which is kept in a steady state called twisted state. (b -- d) For a random initial condition ($\theta_i(0) \in [0,2\pi]$), the phases evolve asymptotically to the phase-locking state (synchronized state), where $\theta_1 = \theta_2 = \cdots = \theta_N$, with  $\varepsilon=10^{-4},\;10^{-3},\;10^{-2}$, respectively. Greater the coupling strength $\varepsilon$, fastest is the time to reach the synchronized state $\tau$ (relaxation time). (e) $\langle \tau\rangle$ as a function of $\varepsilon$ for different $N$ values (number of oscillators), where a clear inverse relation is observed $\tau (\varepsilon) \propto \varepsilon^{-1}$ . The $\langle \cdot \rangle$  denotes the average over 1000 realizations.}
     \label{fig:identical_kura}
 \end{figure}

Figure \ref{fig:identical_kura} depicts the time evolution of the phases of $N=100$ identical oscillators, at panel (a) a symmetric initial condition given by a uniform distribution $\theta_i(0) = 2\pi i/N$, where even for $\varepsilon\neq 0$ the solution is an unstable equilibrium state. At panels (b -- d) a random initial condition is studied ($\theta_i(0) \in [0,2\pi]$), the phases asymptotically approach to the phase-synchronized state, for $\varepsilon=10^{-4},\;10^{-3},\;10^{-2}$, respectively. The time required for the system to reach the synchronized state is called the relaxation time $\tau$ \cite{rodrigues2016kuramoto,son2008relaxation}. $\tau$ is presented in panel (e) using different values of $N$, it is observed that $\tau$ decreases with the increase of the coupling parameter in such a way that $\tau (\varepsilon) \propto \varepsilon^{-1}$. Hence, despite of the particular cases of $\varepsilon=0$ and the twisted states, a network of identical coupled oscillators always reach the synchronized state. 
 
On the other hand, for non-identical oscillators, the natural frequency of the oscillators can be described according to a probability density $\varkappa(\omega)$ which is defined to be symmetric as a function of a mean frequency $\Omega$,
\begin{equation}
    \varkappa(\Omega-\omega) =\varkappa(\Omega+\omega). 
\end{equation}
With this exact configuration,
Kuramoto had defined a complex mean-field $\mathcal Z$, defined as
\begin{equation}\label{eq:Rtime}
    \mathcal Z = R(t)e^{i\psi(t)} = \frac{1}{N}\sum_{j=1}^N e^{i\theta_j(t)},
\end{equation}
where $\psi$ is the circular average frequency. The absolute value of $\mathcal Z$, called Kuramoto order parameter $R$, quantifies the degree of phase coherence in the oscillators and can be used as an indicator of phase synchronization.
%
%

Considering $N\rightarrow \infty$ oscillators, the phase transition from the non-synchronized to the phase-synchronized state occurs in a critical value  $\varepsilon^*$ given by
\begin{equation} \label{eq_criticos}
    \varepsilon^* = \frac{2}{\pi \varkappa(0)},
\end{equation}
and the modulus of the Kuramoto order parameter is equal to
\begin{equation} \label{eq_mod}
    R = \begin{dcases}
    0,\;\;\; &  \mathrm{for} \;\; \varepsilon\leq \varepsilon^*,\\
 \sqrt{\frac{-16}{\pi(\varepsilon^*)^{3}\varkappa''(0)}\left(1-\frac{\varepsilon^*}{\varepsilon}\right)}, \; & \mathrm{for}  \;\; \varepsilon>\varepsilon^*,
    \end{dcases}
\end{equation}
which means that, if $\varepsilon$ is lower than a critical value $\varepsilon^*$, $R\approx 0$ (desynchronization) and for $\varepsilon> \varepsilon^*$, $R(\varepsilon)$ approaches asymptotically to $1$, and is equal to $1$ at the limit of $\varepsilon\rightarrow\infty$. The results of the  Eqs. (\ref{eq_criticos}) and (\ref{eq_mod}) are deduced in the appendix \ref{cap_a}.

Considering the case where natural frequency of the ensemble of oscillators that follows a zero-centered Cauchy–Lorentz distribution,
\begin{equation} \label{eq_cl}
    \varkappa(\omega) = \frac{\zeta}{\pi(\zeta^2+\omega^2)},
\end{equation}
where $\zeta$ is related to the width of the distribution. In this sense,
\begin{equation*}
    \varkappa(0) = \frac{1}{\pi\zeta}, \;\; \varkappa''(\omega) = -\frac{2\zeta(\zeta^2-3\omega^2)}{\pi(\zeta^2+\omega^2)^3} \Rightarrow \varkappa''(0) =  -\frac{2}{\pi\zeta^3}, 
\end{equation*}
therefore the critical coupling for the Cauchy-Lorentz distribution
\begin{equation}
    \varepsilon^* = 2\zeta, 
\end{equation}
and the order parameter, for $\varepsilon> \varepsilon^*$,
\begin{equation}\label{eq:kurapar}
    R = 
    \sqrt{\frac{-16}{\pi (8\zeta^3)(-2/\pi\zeta^3)} \left(1-\frac{\varepsilon^*}{\varepsilon}\right)} = \sqrt{1-\frac{\varepsilon^*}{\varepsilon}}.
\end{equation}

\begin{figure}
    \centering
    \includegraphics[width=0.9\columnwidth]{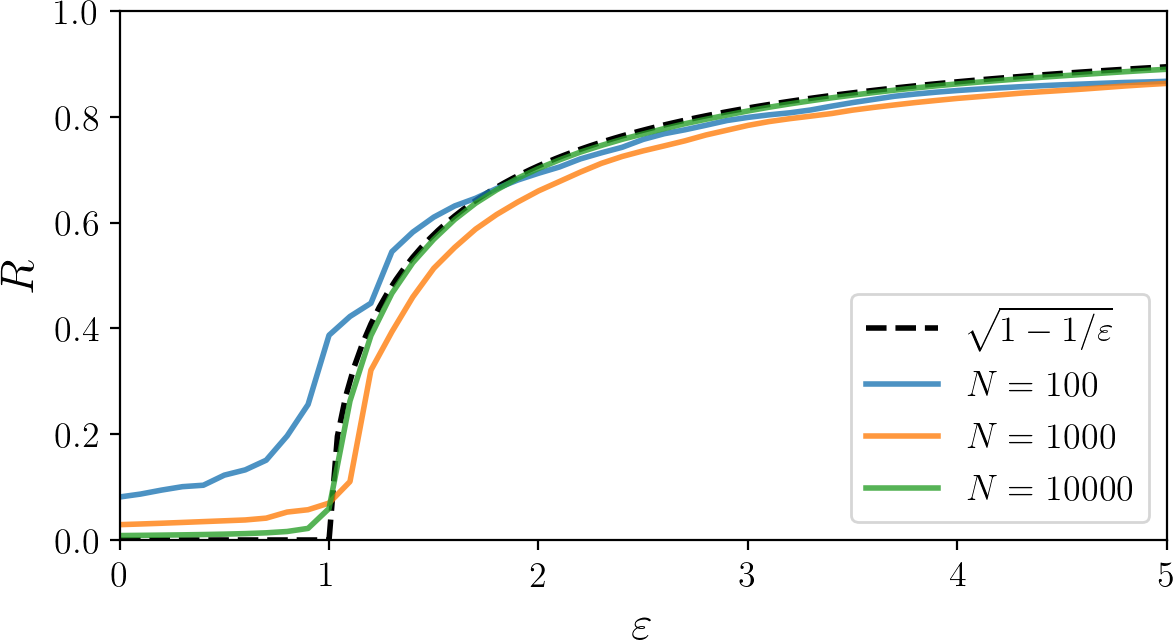}
    \caption[Phase synchronization of the Kuramoto model.]{\textbf{Phase synchronization of the Kuramoto model.} Kuramoto order parameter $R$ as a function of the coupling $\varepsilon$ using a Cauchy-Lorenz distribution with $\zeta = 0.5$, which gives $\varepsilon^*=1$.}
    \label{fig:kuramoto}
\end{figure}

Figure \ref{fig:kuramoto} presents the Kuramoto order parameter $R$ as a function of the coupling $\varepsilon$ considering a Cauchy-Lorenz distribution (Eq. (\ref{eq_cl})) with $\zeta=0.5$ which gives a critical coupling $\varepsilon^*=1$. The dashed line represents the exact solution of $R$ (Eq. \ref{eq:kurapar}) and the solid lines show the numerical results for different values of $N$. The greater the value of $N$, closer is the numerical result to the exact solution. 

Despite the order parameter being created to calculate the phase synchronization of Kuramoto oscillators, this method can be effectively applied to distinct oscillators and also neurons regardless of periodicity presence. To do this, it is necessary to associate a phase to the dynamical model. The phase can be obtained with a projection of the dynamical system $\theta (t) = h(\mathbf x)$ and also with the definition of an appropriate Poincare section where the orbit crosses once for each $2\pi$ rotation. In this sense, the phase can be defined
\begin{equation}
\theta_{i}(t)=2\pi k_{i} +2\pi\frac{ t-t_{{k,i}}}{t_{k+1,i}-t_{{k,i}}}, \hspace{0.5cm} t_{k,i}\leq t<t_{k+1,i},
\label{eq:phase}
\end{equation}
where $t_{k,i}$ represents the $k$th time where the $i$-th system crosses the Poincare surface. Therefore, for every $t=t_k$, the second term of the equation vanishes and the phase is equal to $2\pi k_i$, for the other instants of time, the second term interpolates these times until the next cross occurs in $ t = t_{k + 1} $. 
The phase is increased by a factor of $2\pi$ every $t_{k+1,i}- t_{k,i}$ \cite{ivanchenko2004phase}. After this association, it is possible to quantify the phase synchronization of the system using the Eq. (\ref{eq:Rtime}) for the order parameter. Different from the Kuramoto model, a dynamical system might not present an stationary state of $R(t) = R$. To measure phase synchronization in distinct models, the average order parameter is defined, which consists of the temporal average of the order parameter module
\begin{equation}\label{eq:parmed}
    \langle R\rangle = \frac{1}{t_\mathrm f-t_\mathrm i} \sum_{t=t_\mathrm i}^{t_\mathrm f} R(t),
\end{equation}
where $t_\mathrm i$ and $t_\mathrm f$ are the start and end times of the order parameter computation, respectively. If $\langle R\rangle =1$ represents a completely phase-synchronized state, in which all elements start to cross the surface at the same time. On the other hand, $\langle R\rangle = 0$ means that each element in the network has a corresponding pair that is completely out-of-phase. This can correspond to a completely incoherent state (completely unsynchronized) or a state with clusters of in-phase neurons that are anti-phase between themselves. If the $N$ phases were to be randomly distributed, the result would be $\langle R\rangle  \sim \sqrt{1/N}$ \cite{arenas2008synchronization}.

\chapter{The role of the individual dynamics in the synchronization processes of neural networks}\label{chap:role}

\initial{M}{ost} of the neuronal behaviors produced by the brain are directly related to the collective patterns of activation of groups of neurons, generating (partially) synchronized dynamics of these groups \cite{kandel2013principles}. However, the brain activity can be disturbed by pathological states caused by some neuropathies, as in the case of epilepsy, where the increase in neuron synchronization generates seizure episodes \cite{mormann2000mean}, and in Parkinson's disease, where an excessive synchronization in the basal ganglia is evidenced \cite{hammond2007pathological,popovych2014control,popovych2012desynchronizing}. In this sense, drugs and other treatments are used to normalize brain activity, allowing the neurons to depolarize and repolarize in a healthy way \cite{loscher2002new,loscher2013new,perlmutter2006deep}. In this scenario, the use of simulated neural networks have proven to be very useful to understand the synchronization mechanism to optimize or suppress the synchronization of neurons \cite{boaretto2017suppression,batista2010delayed,louzada2012suppress}.

This chapter is devoted to studying the influence of the dynamics of the neurons in their synchronization processes of a neural network with HH$\ell$ neurons. It is shown that exists a strong correlation between the individual dynamics of the neuron with the type of synchronization of the network. For neurons that individually exhibit chaotic behavior, phase synchronization occurs analogously to the transition of chaotic oscillators known in the literature \cite{kuramoto1975self}. However, when non-chaotic neurons are coupled, the network may present phase synchronization to weak coupling regimes (a coupling parameter close to zero) due to the influence of the regularity of the individual dynamics, and the increase of the coupling parameter can decrease the phase synchronization of the network \cite{prado2014synchronization,budzinski2017detection,boaretto2017suppression}, characterizing a {\it non-monotonic} evolution of synchronization as a function of coupling. 

This phenomenon has already been explored in other works using different approaches: network of networks \cite{prado2014synchronization,budzinski2019synchronization}, detection of nonstationarity \cite{budzinski2017detection}, suppression of phase synchronization \cite{boaretto2017suppression,boaretto2018neuron,boaretto2019protocol,boaretto2019suppression} and by temperature changes \cite{budzinski2019temperature,rossi2021phase}. Here, the dynamics of the neurons are changed by variations in the ion conductances of the model, and with the application of an external pulsed current to the neurons, both situations show similar results. To confirm the generality of this behavior, a verification test is successfully performed with another neuron model in the same conditions. In conclusion, the occurrence of this non-monotonic phase transition occurs due to an interplay between the individual-regular behavior of the neurons and the influence of the synaptic current. Hence, it is possible to change the phase synchronization in the weak coupling regime just with small variations to achieve a chaotic transition in the neuron dynamics. Most of the results of this chapter are published in the article ``The role of individual neuron ion conductances in the synchronization processes of neuron networks" Neural Networks 137 (2021) 97–105 \cite{boaretto2021role}. 

\section{Network properties}

It is considered a network with $N=1000$ HH$\ell$ neurons coupled in a small-world network, generated with the Watts-Strogatz route with $n_0=6000$ connections and $p_\mathrm{ws}=0.1$. The membrane potential of the $i$-th neuron is described by
\begin{equation} \label{eq_rede_1}
    C_\mathrm{M} \frac{dV_i}{dt} = -I_{i,\mathrm{d}} - I_{i,\mathrm{r}} - I_{i,\mathrm{sd}} - I_{i,\mathrm{sr}} - I_{i,\mathrm{l}}+I_{i,\mathrm{syn}},
\end{equation}
where the ionic currents are given by the Eqs. (\ref{eq_hb_1}-\ref{eq_hb_2}). For this specific case, the neurons are coupled with a chemical synapse
\begin{equation}
   I_{i,\mathrm{syn}} = \frac{\varepsilon}{\bar{n}}( E_\mathrm{syn} - V_i) \sum\limits_{j=1}^N e_{i,j}r_j(t),
   \label{syn_current}
\end{equation}
with $\varepsilon$ being the coupling strength parameter, $\bar{n}$ is the average of the number of connections in the network, $E_\mathrm{syn}$ is the synaptic reversal potential. It is used $E_\mathrm{syn}=20$ $\SI{}{\milli\volt}$, since $ E_\mathrm{syn}>V_i$, the synaptic term $I_\mathrm{syn}$ is always positive featuring an excitatory synapse. $e_{i,j}$ gives the element of the connection matrix, so that if $i$ and $j$ are connected $e_{i,j}=1$, otherwise $e_{i,j}=0$. The variable $r_i$ represents the fraction of bound receptors in the synapse where the kinetics model is described by \cite{destexhe1994efficient}
\begin{equation}
\label{r_ij}
\frac{dr_{i}}{dt}=\left(\frac{1}{\tau_{r}}-\frac{1}{\tau_{d}}\right)\frac{1-r_{i}}{1+\exp[-s_0(V_{i}-V_{0})]}-\frac{r_{i}}{\tau_{d}},
\end{equation}
where $s_{0}$ is an unitary constant $\SI{1}{\milli\volt^{-1}}$, $V_0=-20$ $\SI{}{\milli\volt}$, $\tau_r=0.5$ $\SI{}{\milli\second}$, and $\tau_d=8$ $\SI{}{\milli\second}$ are constants.

\begin{figure}[htb!]
    \centering
    \includegraphics[width=.9\columnwidth]{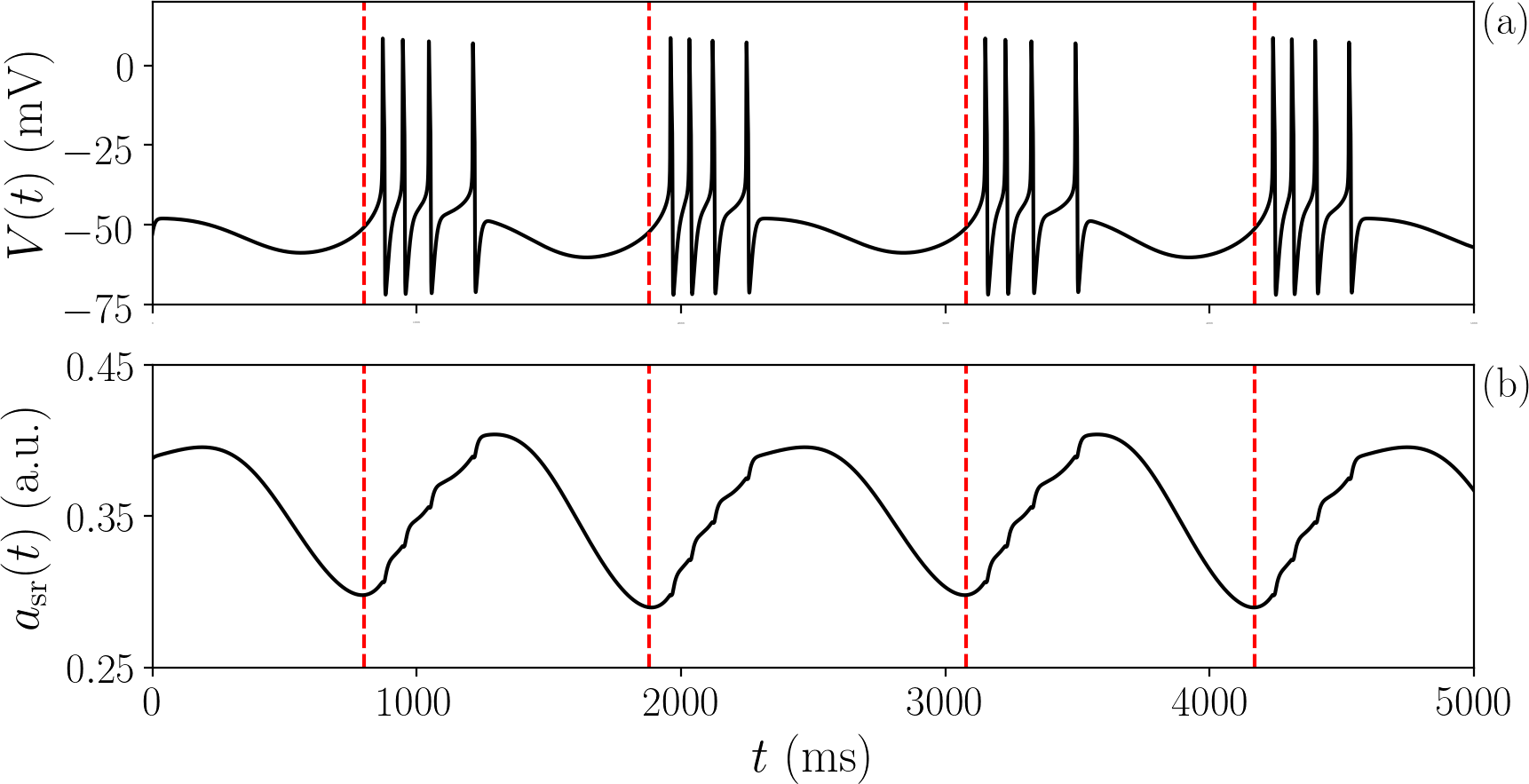}
    \caption[Phase association of the HH$\ell$ model.]{\textbf{Phase association of the HH$\ell$ model.} Time evolution of the: (a) membrane potential of the neuron, (b) $a_\mathrm{sr}$ variable, using the parameters of Table \ref{tabeladeconstantes}. The local minimums of $a_\mathrm{sr}$ (dashed line) can be used as an auxiliary to compute the beginning of a burst.}
    \label{fig:hb_phases}
\end{figure}

Figure \ref{fig:hb_phases} presents the dynamics of the membrane potential (a) and the variable $a_\mathrm{sd}$ (b) for one neuron simulated using the values of Table \ref{tabeladeconstantes}. The temporal dynamic is composed of a set of (four) spikes followed by a resting time, characterizing bursting dynamics. The bursting dynamic is maintained for all parameters used in this Chapter. The $a_\mathrm{sr}$ variable, described in Eq. (\ref{a_sr_eq}), is used as an auxiliary variable to compute the bursting time since the local minimum of $a_\mathrm{sr}$ coincides with the beginning of a burst. After compute all the bursts of all the neurons, using Eq. (\ref{eq:phase}) it is possible to evaluate $\theta_i$ and the phase synchronization $\langle R \rangle$ with Eq. (\ref{eq:parmed}), here, considering $t_\mathrm i=150$ $\SI{}{\second}$ (avoiding transient effects) and $t_\mathrm f=250$ $\SI{}{\second}$ as a final time of computation. The initial conditions for each neuron are randomly distributed to avoid any initial synchronization bias. Each synchronization result ($\langle R\rangle$) is a mean value over $10$ distinct initial conditions and $10$ different networks preserving the same connection properties.

\section{Results}

\begin{figure}[htb!]
    \centering
    \includegraphics[width=\columnwidth]{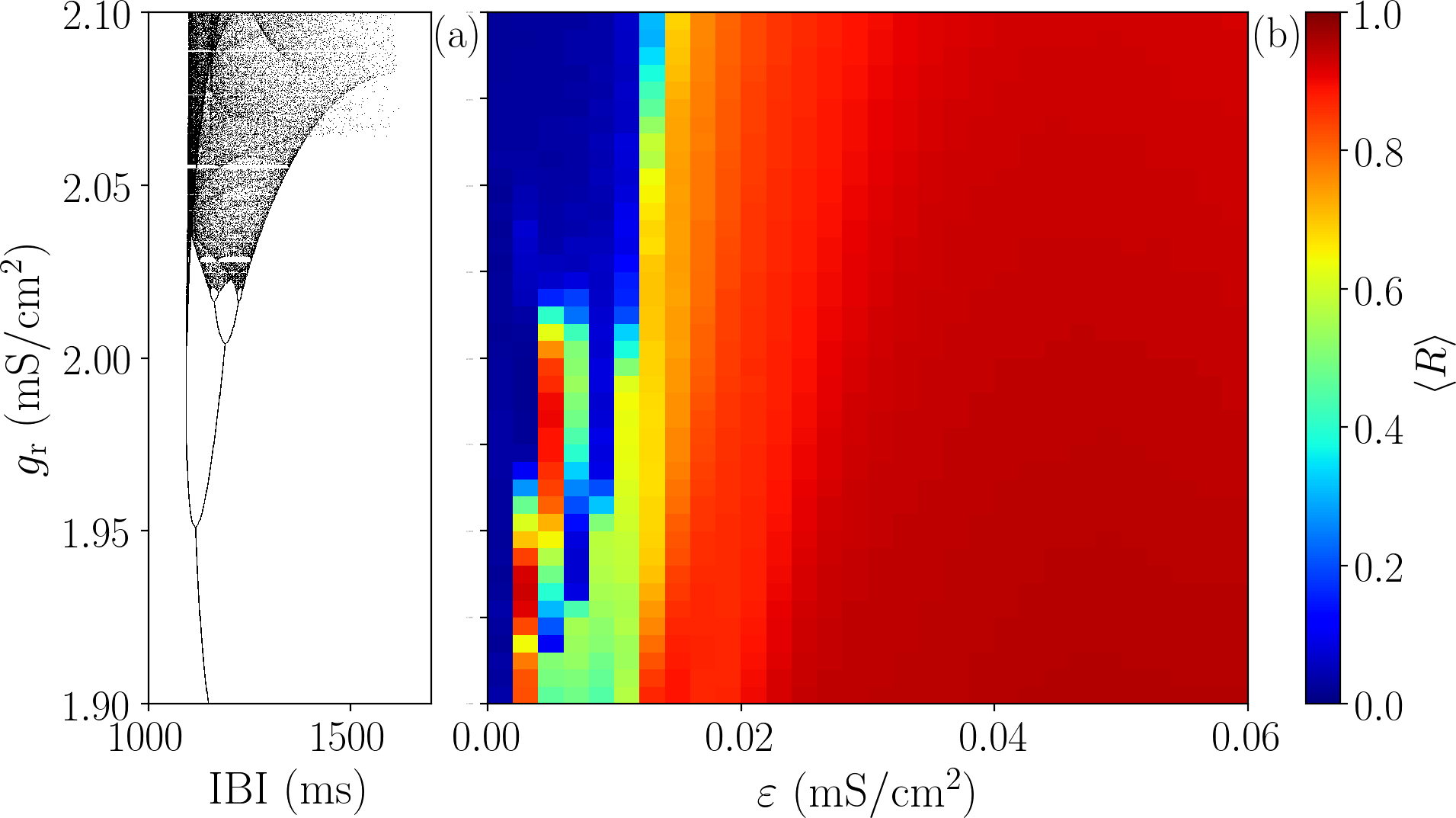}
    \caption[Neural dynamics of HH$\ell$ model as a function of ${g_\mathrm r}$.]{\textbf{Neural dynamics of HH$\ell$ model as a function of $\bm {g_\mathrm r}$.} (a) Bifurcation diagram of the inter-burst-interval (IBI) for the individual neuron as a function of the maximum conductance $g_\mathrm{r}$. (b) The Kuramoto order parameter of the network $\langle R\rangle$ as a function of the coupling parameter ($\varepsilon$) and the conductance $g_\mathrm r$.}
    \label{fig:hb_din_syn}
\end{figure}

The effects of variations in neuron dynamics due to changes in the ion conductance can be illustrated by a bifurcation diagram of the Inter-Burst-Interval (time between two successive bursts, IBI). Figure \ref{fig:hb_din_syn} (a) depicts the IBI (x-axis) for one single isolated neuron ($\varepsilon=0$) as a function of $g_\mathrm r$ (y-axis). It is observed for $1.90<g_\mathrm{r}\lesssim 2.03$ that the neuron dynamics has a periodic behavior. When $g_\mathrm{r}$ is increased above $1.95$ a cascade of period-doubling bifurcations occurs producing infinite orbits and finally, for $g_\mathrm{r}\gtrsim 2.03$, the large variability of IBI illustrates that the neuron dynamics is chaotic \cite{ott2002chaos}. At panel (b), it is computed the Kuramoto order parameter $\langle R\rangle$ in color codes from blue tones ($\langle R\rangle = 0$, non-synchronized) to red tones ($\langle R\rangle = 1$, complete phase synchronized), as function of $\varepsilon$ and $g_\mathrm r$. It is noted that for values of $g_\mathrm r>2.03$, the transition for the non-synchronized state to the synchronized state is very similar to the transition seen in chaotic oscillators \cite{kuramoto1975self,batista2010delayed}, which means that for $\varepsilon\lesssim 0.01$, $R\approx 0$, and for $\varepsilon > 0.01$ the network gains phase synchronization, $R\approx 1$. On the other hand, for lower values of $g_\mathrm r$, the network starts to synchronize for very low values of $\varepsilon$, this behavior is associated with the fact that the neurons, which individually present regular dynamics, are susceptible to synchronize since the synaptic current is not strong enough to disrupt the periodic features of the neurons. Hence, due to the frequency-locked phenomenon, the synaptic current acts to aligning the neurons in phase generating the partial phase synchronization behavior. As the coupling strength increases, the synaptic current gains relevance, inducing chaoticity in the neurons, resulting in a sharp decrease in the phase synchronization level, characterizing the non-monotonic evolution of the synchronization.

\begin{figure}[htb!]
    \centering
    \includegraphics[width=\columnwidth]{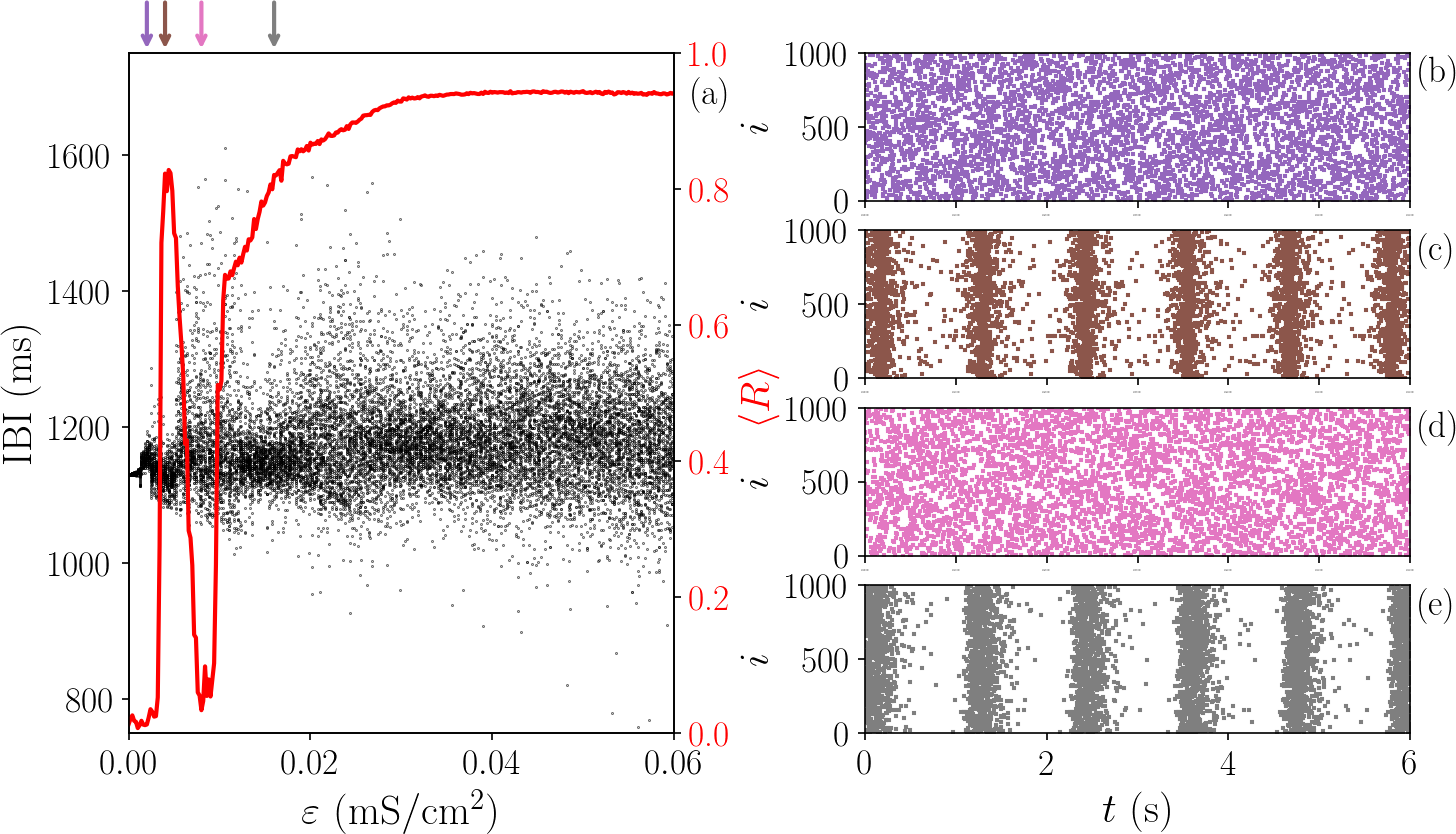}
    \caption[Synchronization transition of individual regular neurons.]{\textbf{Synchronization transition of individual regular neurons.} (a) Bifurcation diagram of the IBI (left scale) for an $i$-th coupled neuron in the network, as a function of the coupling parameter ($\varepsilon$), the red curve, represents the Kuramoto order parameter of the network (right scale) using $g_\mathrm{r}=1.92$. (b -- e) Raster plots of the network where each dot corresponds to the beginning of a burst ($t_{k,i}$). The colored arrows indicate the coupling values when the raster plots are obtained, which match with the colors of the dots.}
    \label{fig:coupled_ibi_1}
\end{figure}

\begin{figure}[htb!]
    \centering
    \includegraphics[width=\columnwidth]{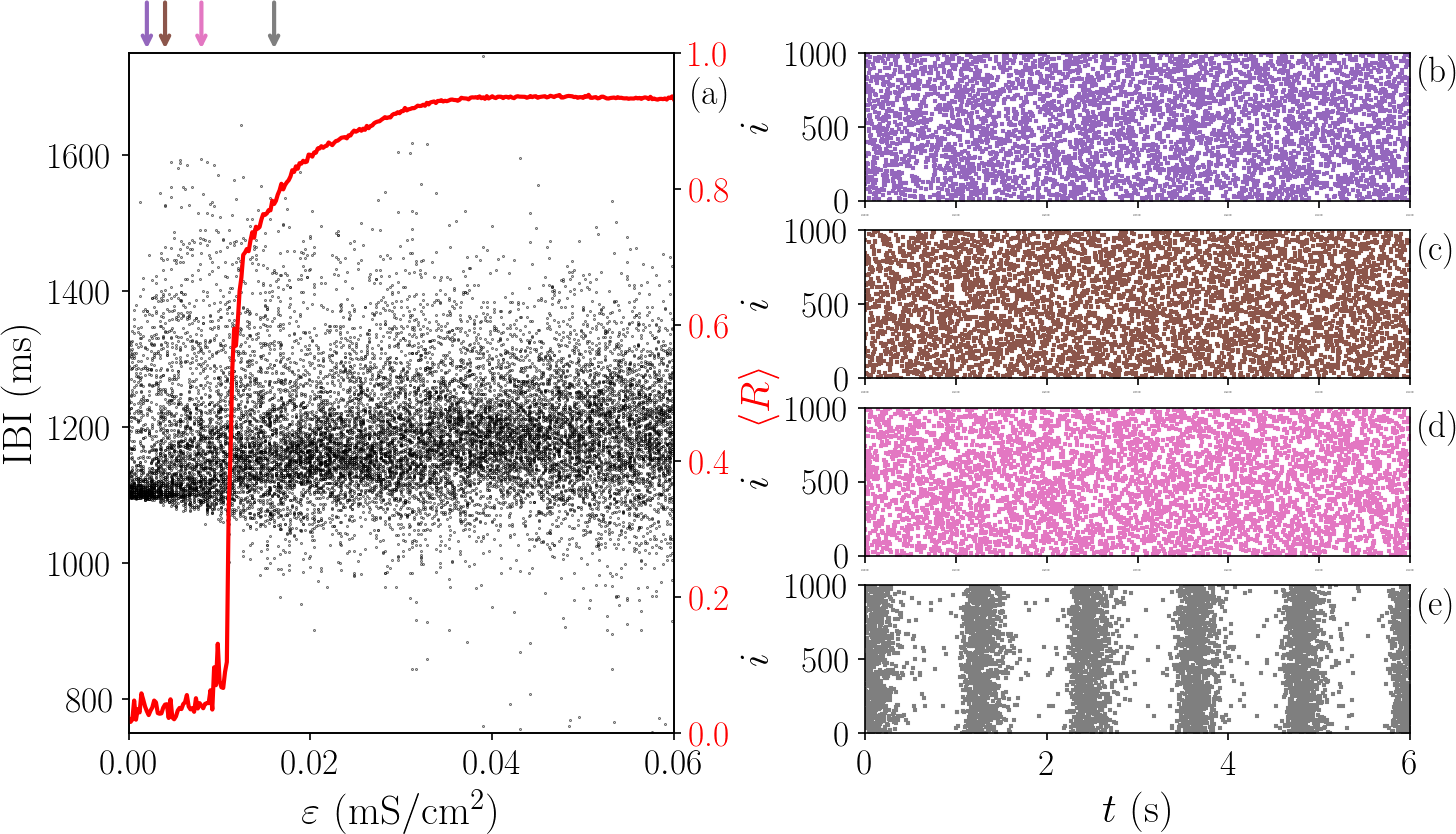}
    \caption[Synchronization transition of individual chaotic neurons.]{\textbf{Synchronization transition of individual chaotic neurons.} (a) Bifurcation diagram of the IBI (left scale) for an $i$-th coupled neuron in the network, as a function of the coupling parameter ($\varepsilon$), and the red curve, represents the Kuramoto order parameter of the network (right scale) using $g_\mathrm{r}=2.05$. (b -- e) Raster plots of the network where each dot corresponds to the beginning of a burst ($t_{k,i}$). The colored arrows in panel (a) indicate the coupling values when the raster plots are obtained, which match with the colors of the dots.}
    \label{fig:coupled_ibi_2}
\end{figure}

Figure \ref{fig:coupled_ibi_1} shows how this transition occurs for $g_\mathrm r=1.92$ (regular individual dynamics), where at the panel (a), the IBI for one arbitrary neuron of the network is plotted as a function of $\varepsilon$ (left scale) with $\langle R \rangle$ which is presented in red (right scale). In this case, a weak coupling initially leads the network to a partial phase-synchronized state $\langle R \rangle \approx 0.8$, but as the coupling is increased the phase synchronization is decreased and for even stronger coupling a chaotic-phase-synchronized behavior is achieved. It is important to mention that a weak chaotic synchronized dynamic is still preserved for a moderate dispersion of the IBI and some intervals of the coupling parameter. Panels (b -- e) present raster plots of the network where each dot corresponds to the beginning of a burst. In panel (b), $\varepsilon=0.002$, an incoherent behavior is observed. For panel (c) $\varepsilon=0.004$, some vertical structures are noticed in the raster plot, this characterizes a partial phase synchronization behavior, where bursts of neurons occur at close time instants. In panel (d), $\varepsilon=0.008$ it shows the decrease of the phase synchronization, and (e) $\varepsilon=0.020$ the chaotic-phase-synchronized behavior. On the other hand, Fig. \ref{fig:coupled_ibi_2} depicts the same configuration, but with $g_\mathrm{r}=2.05$, which confirms that once the individual dynamics is already chaotic only the final transition from desynchronized state to the chaotic-phase-synchronized state is observed. This can be seen at the monotonic evolution of $R$ (panel (a)), and at the raster plots (panels (b -- e)). 

\begin{figure}[htb!]
    \centering
    \includegraphics[width=\columnwidth]{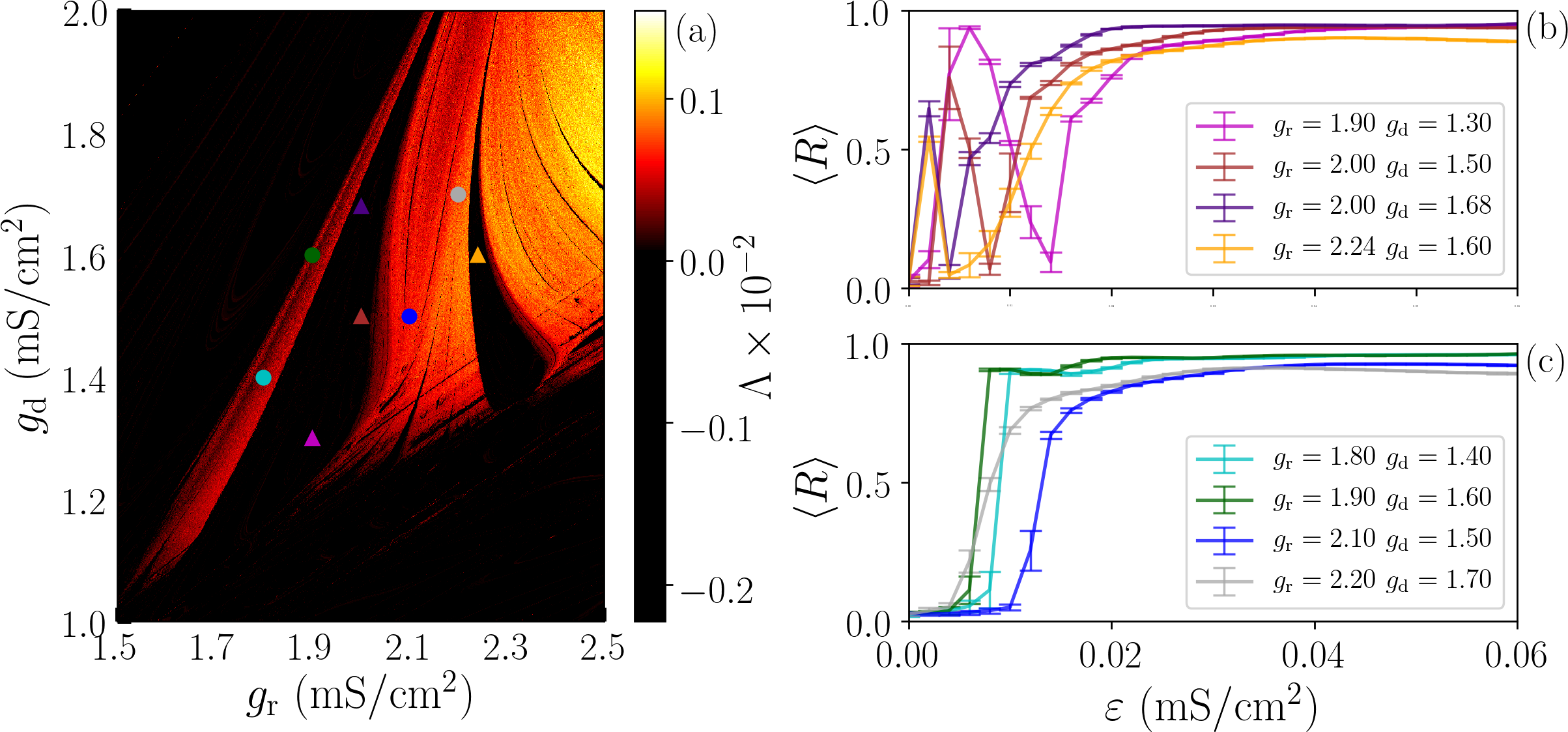}
    \caption[Correlation between the individual dynamics and the phase synchronization varying the parameter space $g_\mathrm{r} \times  g_\mathrm{d}$.]{\textbf{Correlation between the individual dynamics and the phase synchronization varying the parameter space $\bm{g_\mathrm{r} \times  g_\mathrm{d}}$.} (a) Largest Lyapunov Exponent ($\Lambda$) obtained for one isolated neuron in the parameter space $g_\mathrm{r} \times  g_\mathrm{d}$. Two sets of $4$ points are selected inside chaotic $\Lambda >0$ (colored dots) and regular $\Lambda\approx 0$ (colored triangles) parameter space regions. (b, c) $\langle R\rangle$ as a function of $\varepsilon$ for the two sets of $4$ points of the parameter space, (b) regular dynamics (triangles in (a))  and (c) chaotic dynamics (dots in (a)).}
    \label{fig:hb_surf_1}
\end{figure}

Another option to study the individual dynamics of the neuron, and correlates to the type of transition of the phase synchronization, is comparing the largest Lyapunov exponent ($\Lambda$) of the individual neuron with the synchronization features of the network. The panel (a) of Fig. \ref{fig:hb_surf_1}, presents the largest Lyapunov exponent ($\Lambda$) presented in Chapter \ref{chap:sistemas} of one isolated neuron for a parameter space $g_\mathrm{r} \times g_\mathrm{d}$ using a grid of 800 $\times$ 800 parameter-pairs. The color bar codifies the value of $\Lambda$ going from black $\Lambda \leq 0$ to hot colors $\Lambda > 0$. It is selected two sets of 4 points each in the parameter space where the colored triangles represent points where $\Lambda \leq 0$, $g_\mathrm{r} \times g_\mathrm{d} =\{(1.90, 1.30)$, $(2.00, 1.50)$, $(2.00, 1.68)$, $(2.24, 1.60)\}$ and the colored dots represent points where $\Lambda>0$, $g_\mathrm{r} \times g_\mathrm{d}=\{(1.80, 1.40)$, $(1.90, 1.60)$, $(2.10, 1.50)$, $(2.20, 1.70)\}$. In panels (b) and (c) it is measured the phase synchronization transition for each selected point, plotting $\langle R\rangle$ as a function of $\varepsilon$ for each triangle (b) and dot (c) point (with the respective colors). It is noted that at the panel (b), the $4$ curves depict a non-monotonic transition, which means that for lower coupling values there is a local maximum of phase synchronization, for the magenta line it happens at $\varepsilon=0.006$, brown $\varepsilon=0.004$, purple and yellow $\varepsilon=0.002$. On the other hand, at the panel (c), all $4$ curves the intrinsic chaoticity of the neurons is sufficient to avoid phase synchronization for the weak coupling regime, and the network presents phase synchronization only for couplings greater than a critical coupling $\varepsilon^{*}$ which its value depends on the parameter space. It is noted that a synchronization transition can be easily achieved for weaker values of $\varepsilon$ with the change of more than one parameter since it is necessary to achieve a bifurcation point in the parameter space to change the individual dynamics. In these cases, small changes in one specific ion-conductance may act as a catalyst to the second conductance change \cite{boaretto2021role}. 

\begin{figure}[htb!]
    \centering
    \includegraphics[width=\columnwidth]{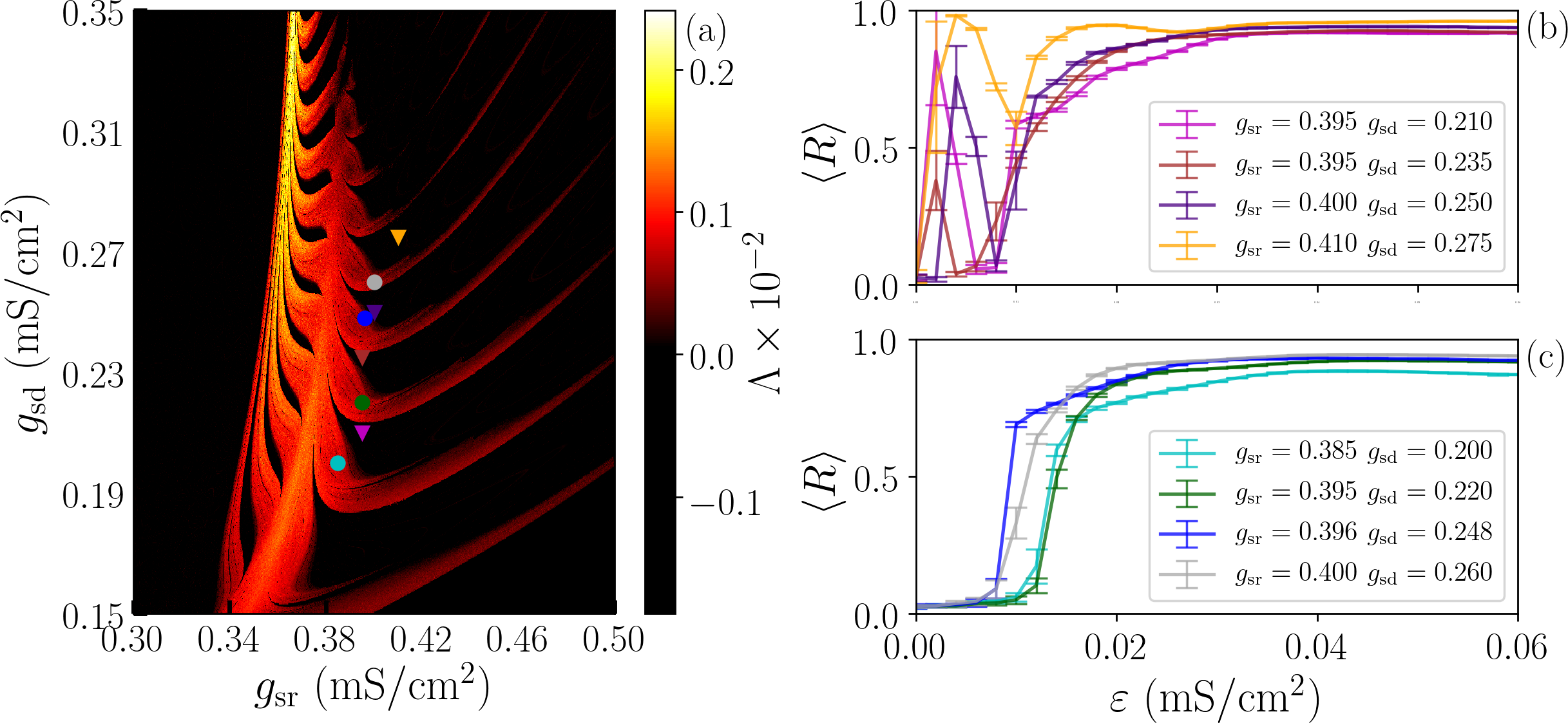}
   \caption[Correlation between the individual dynamics and the phase synchronization varying the parameter space $g_\mathrm{sr} \times  g_\mathrm{sd}$.]{\textbf{Correlation between the individual dynamics and the phase synchronization varying the parameter space $\bm{g_\mathrm{sr} \times  g_\mathrm{sd}}$.} (a) $\Lambda$ obtained for one isolated neuron in the parameter space $g_\mathrm{sr} \times  g_\mathrm{sd}$. Two sets of $4$ points are selected inside chaotic $\Lambda >0$ (colored dots) and regular $\Lambda\approx 0$ (colored triangles) parameter space regions. (b, c) $\langle R\rangle$ as a function of $\varepsilon$ for the two sets of $4$ points of the parameter space, (b) regular dynamics (triangles in (a))  and (c) chaotic dynamics (dots in (a)).}
    \label{fig:hb_surf_2}
\end{figure}

To investigate further the relation between the individual dynamics with the phase synchronization, assuming that other conductances also can play a similar role in the network, it is fixed $g_\mathrm{r}$ and $g_\mathrm{d}$ in $2.00$ and $1.50$ (original values of Table \ref{tabeladeconstantes} \cite{postnova2010computational}), but with the variations of the slow currents of the model $g_\mathrm{sr}$ and $g_\mathrm{sd}$. The results are depicted in Fig. \ref{fig:hb_surf_2} (the same methodology of Fig. \ref{fig:hb_surf_1} is used). Panel (a) $\Lambda$ is plotted as a function of $g_\mathrm{sr}$ and $g_\mathrm{sd}$, and, once again panels (b) and (c) depict the $\langle R\rangle$ as a function of $\varepsilon$ for two sets of $4$ points selected in the parameters space where (b) corresponds to the colored triangles (regular individual dynamics) $g_\mathrm{sr} \times g_\mathrm{sd} = \{(0.395,0.210)$, $(0.395,0.235)$, $(0.400,0.250)$, $(0.410,0.275)\}$ and (c) the colored dots (chaotic individual dynamics) $g_\mathrm{sr} \times g_\mathrm{sd} = \{(0.385,0.200)$, $(0.395,0.220)$, $(0.396, 0.248)$,  $(0.400,0.260)\}$. Even for very similar parameters, as the purple triangle ($0.400,0.250$) and the the blue dot ($0.396,0.248$) the phase synchronization evolves in each case in completely different ways when the lower coupling regime is considered.  
%
%

\subsection{External perturbation by electrical pulses}

In the last section, it was shown a correlation between the individual dynamics of the neuron with the transition of the phase synchronization of the network. Precisely, the synchronization transition of periodic neurons is different from chaotic ones. In this sense, instead of change the parameters of the neuron to achieve a bifurcation transition, it is possible to induce chaoticity with the application of external perturbations. To do this, it is made the addition of an external current $\xi(t)$ in Eq. (\ref{eq_rede_1}), which describes a sequence of electrical pulses
\begin{equation}
    \xi(t) = \frac{\xi_0}{2} + \sum_{m=1,3,5,\cdots}^\infty \frac{2\xi_0}{m\pi}\sin(2m\pi \nu t),
\end{equation}
where $\xi_0$ is the amplitude of the pulse measured in $\SI{}{\micro\ampere/\centi\meter^2}$, $\nu$ is the frequency measured in $\SI{}{\hertz}$, and $m$ is an odd integer, and for the simulations it is used the $m<1000$ terms. The choice for a pulsed current is inspired in the {\it Deep Brain Stimulation} procedure, where electrodes are placed deep in the brain and are connected to a stimulator device. When turned on, the stimulator emits electrical pulses, which in high frequencies ($>100$ $\SI{}{\hertz}$ \cite{blumenfeld2015high} in comparison with the frequency of the inter-bursts $\sim 1$ $\SI{}{\hertz}$) suppresses tremor symptoms associated with essential tremor or Parkinson disease \cite{perlmutter2006deep}. 

\begin{figure}[t]
    \centering
    \includegraphics[width=\columnwidth]{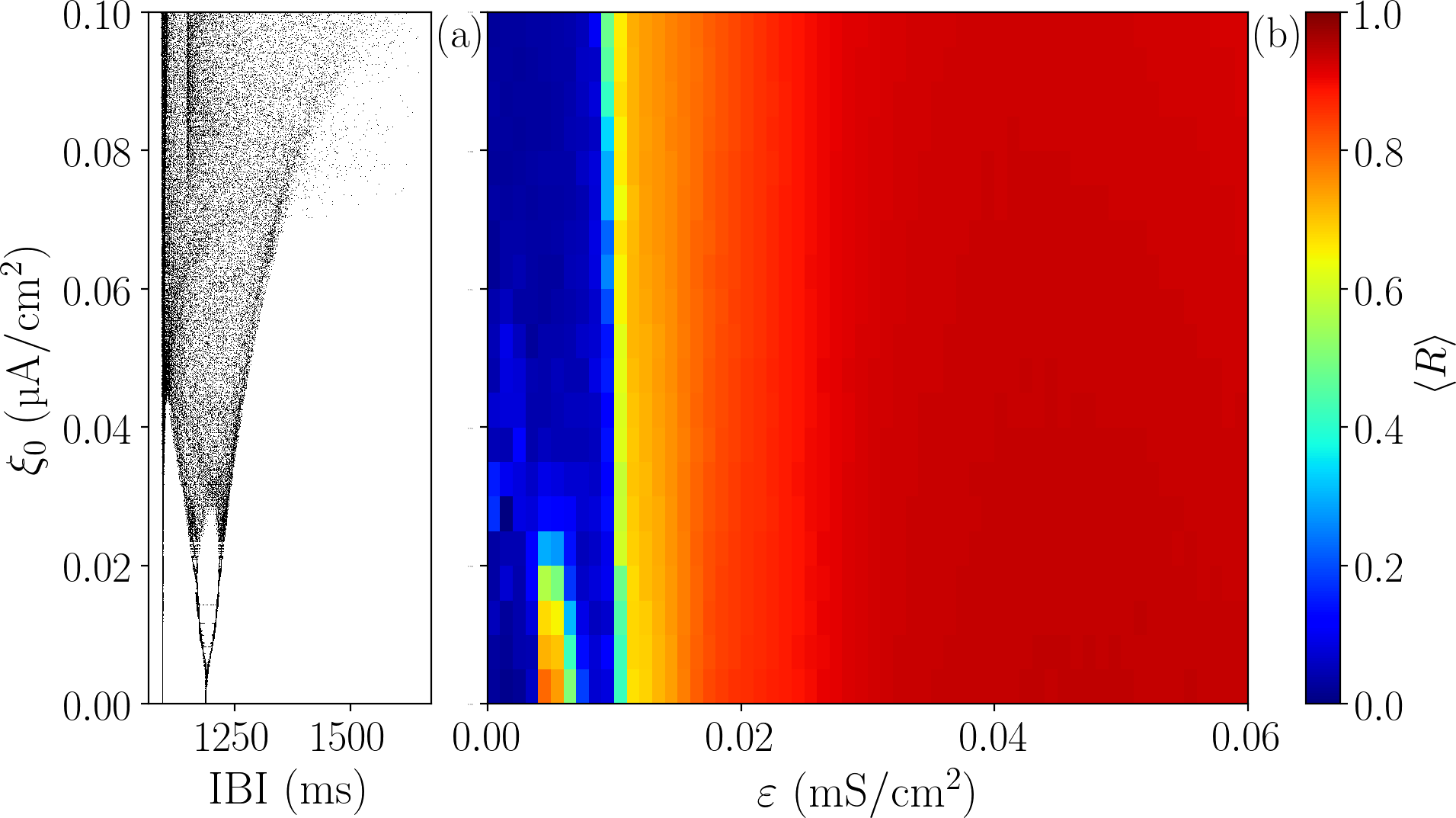}
    \caption[Chaos induced by the application of external pulses.]{\textbf{Chaos induced by the application of external pulses.} (a) Bifurcation diagram of IBI for an individual neuron as a function of the amplitude of the current $\xi_0$. (b) The Kuramoto order parameter of the network $\langle R\rangle$ as a function of $\varepsilon$ and $\xi_0$.}
    \label{fig:surf_pulso}
\end{figure}

The results are depicted in Fig.~\ref{fig:surf_pulso}, where panel (a) presents the IBI of the individual neuron as a function of the amplitude of the applied pulse $\xi_0$ with a fixed frequency of $\nu = 100$ $\SI{}{\hertz}$, for fixed parameters as in Table $\ref{tabeladeconstantes}$. It is seen that the regular dynamic exhibited by the neuron is kept until $\xi_0 \approx 0.02$, where a great variability of IBI appears in the dynamics. At panel (b), it is measured the phase synchronization ($\langle R\rangle$) as a function of the coupling parameter $\varepsilon$ and $\xi_0$. Similarly, the individual dynamics of the neuron rule the synchronization transition, since for $\xi_0<0.02$ there is a local maximum of synchronization for $\varepsilon\approx 0.006$, which is associated with the non-monotonic evolution of $\langle R\rangle$. On the other hand, for $\xi_0>0.02$, the individual neuron presents chaotic dynamics, and the phase synchronization is reached only for coupling values greater than a critical value $\varepsilon^*\approx 0.01$, corroborating with the results of the previous section.

\subsection{Verification test with the Hindmarsh-Rolse model}

\begin{figure}[t]
    \centering
    \includegraphics[width=\columnwidth]{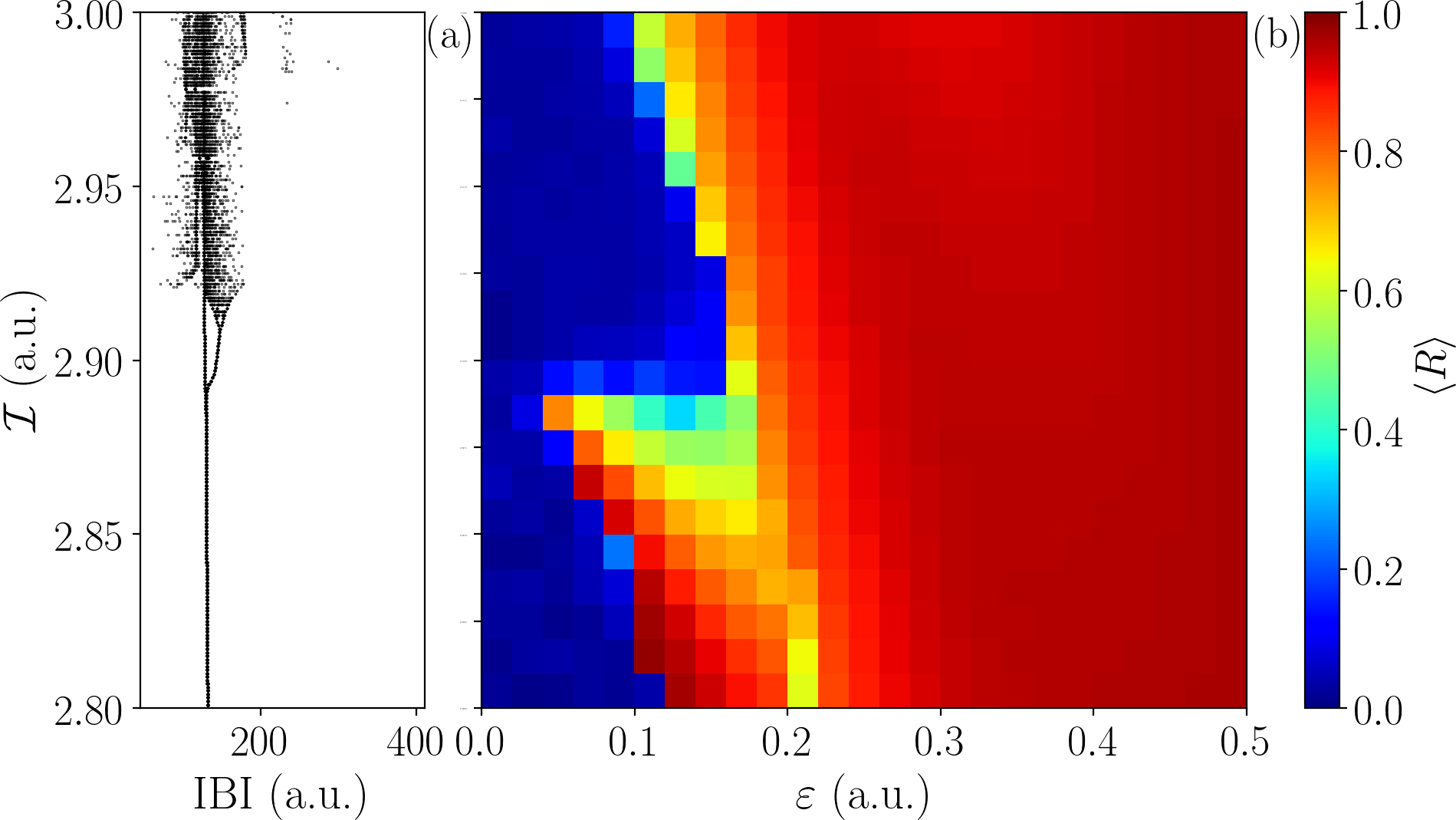}
    \caption[Verification test with the Hindrmarsh-Rose model.]{\textbf{Verification test with the Hindmarsh-Rose model.} (a) Bifurcation diagram of IBI for an individual Hindrmarsh-Rose neuron as a function of the parameter $\mathcal I$. The Kuramoto order parameter of the network $\langle R\rangle$ as a function of $\varepsilon$ and $\mathcal I$.}
    \label{fig:hr_surf}
\end{figure}

To generalize the non-monotonic of the synchronization phenomenon, the same situation is studied as a verification test using a second network in the same conditions, i.e. same size, connections, and topology, but simulating the Hindmarsh-Rose neuron model described by Eqs. (\ref{hr_I} -- \ref{hr_III}), which now reads
\begin{eqnarray}
    \frac{dx_i}{dt} &=& y_i-ax_i^3 + bx_i^2-z_i+\mathcal I + \mathcal I_ {i,\mathrm{syn}},\\
    \frac{dy_i}{dt} &=& c-dx_i^2-y_i,\\
    \frac{dz_i}{dt} &=& r[s(x_i-x_\mathrm r)-z_i],
\end{eqnarray}
where $a$, $b$, $c$, $e$, $r$, $s$, and $x_\mathrm r$ are constants whose values are depicted in Table \ref{tablehr}, $\mathcal I$ is a free parameter, and $ \mathcal I_ {i,\mathrm{syn}}$ is the dimensionless ``synaptic'' parameter defined as
\begin{equation}
\mathcal I_{i,\mathrm{syn}}=\frac{\varepsilon}{\bar n}(x_\mathrm s - x_i)\sum_{j=1}^{N}e_{i,j}\frac{1}{1+\exp[-\lambda_\mathrm s (x_j-\beta_\mathrm s)]}.
\label{i_acopla}
\end{equation}
where $\varepsilon$ is the coupling parameter, $\bar n$ is the average of connections, and $e_{i,j}$ is the element of the connection matrix. $x_s$, $\lambda_s$ and $\beta_s$ are constants related to the chemical synapses, as described in Chapter \ref{chap:redes} where it is used $x_s = 2$, $\lambda_s = 30$, and $\beta_s = 0$. To evaluate the bursting times $t_{k,i}$ to compute the phase of the $i$-th neuron ($\theta_i$), it is used the local minimum values of $z_i(t)$ which correlates with the beginning of a burst, as shown in Fig. \ref{fig:hr_2} (analogously to the $a_\mathrm{sr}$ variable of the HH$\ell$ model). The dynamic of the individual neuron is depicted in panel (a) of Fig. \ref{fig:hr_surf} where is computed the IBI as a function of $\mathcal I$. The neuron presents a periodic behavior for $\mathcal I<2.92$, and vast variability of IBI (chaotic dynamics) $\mathcal > 2.92$. At panel (b) it is evaluated the $\langle R\rangle$ in color codes as a function of $\varepsilon$ and $\mathcal I$. In a similar way to the results of the previous sections, the transition from non-synchronized state to the synchronized state varies with the individual neuron dynamics: non-monotonic transition for $\mathcal I<2.92$ and monotonic for $\mathcal I>2.92$.

\subsection{Topological effects}

\begin{figure}[t]
    \centering
    \includegraphics[width=\columnwidth]{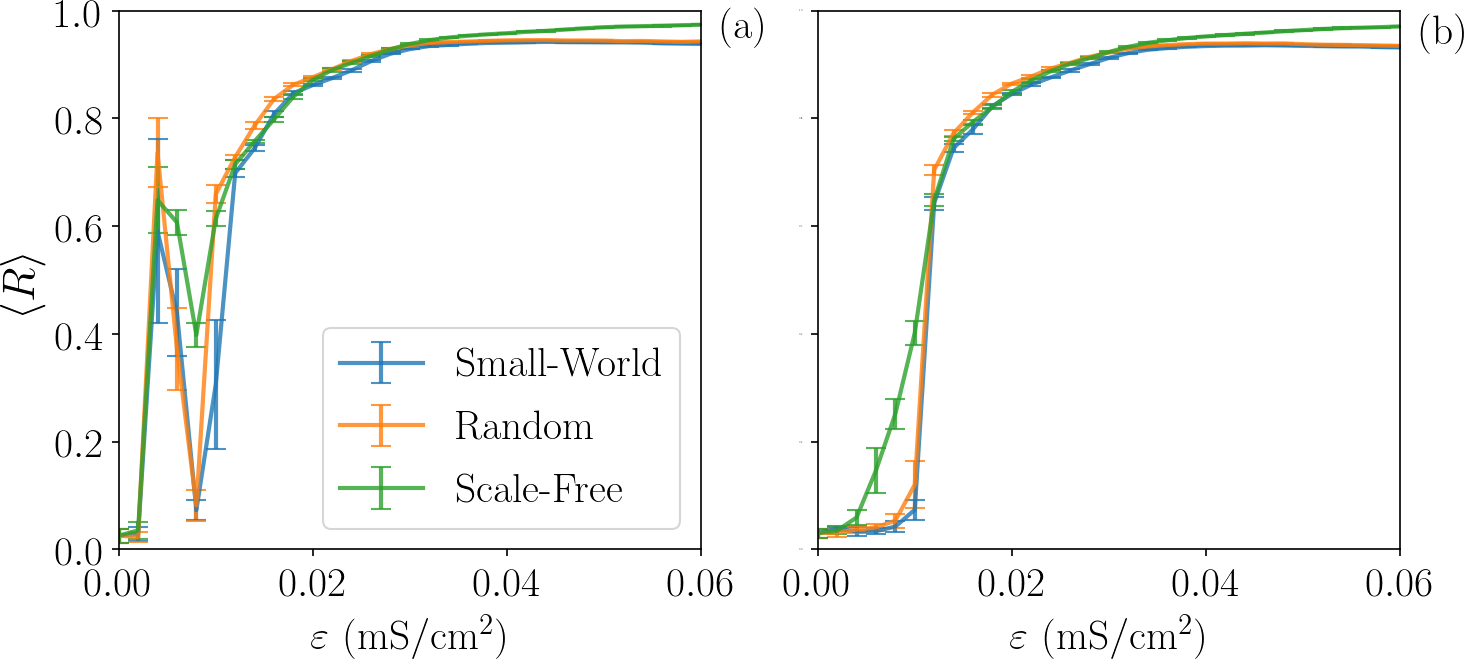}
    \caption[Synchronization  transition  for different networks' topologies.]{\textbf{Synchronization  transition for different networks' topologies.} $\langle R\rangle$ as a function of $\varepsilon$ (a) for $g_\mathrm r = 1.92$, and (b) $g_\mathrm r=2.05$ comparing different topologies the same number of connections. The error bars are the standard deviation over $10$ different realizations.}
    \label{fig:topological}
\end{figure}

In this project, it is opted to couple the neurons in a complex network with a small-world topology. This choice is justified since, as discussed in Chapter \ref{chap:redes}, the small-world features are detected in real neural networks. However, one can believe that the non-monotonic evolution of the network can arise due to the small-world topology. To disprove this hypothesis, in Fig. \ref{fig:topological}, it is studied $\langle R\rangle$ as a function of the coupling $\varepsilon$ for different networks' topologies generated with approximately the same number of connections ($n_\mathrm{sw} = n_\mathrm{random} = 6000$ and $n_\mathrm{sf} = 5982$). In panel (a) $g_\mathrm r =1.92$ (individually periodic) is used and the phase synchronization transitions are non-monotonic and panel (b) $g_\mathrm r=2.05$ (individually chaotic) is used and the transitions are monotonic. These results confirm that the same effects of the individual dynamics can be detected in different complex networks' topologies. 

\section{Discussions}

Throughout this chapter, the phenomenon of non-monotonic synchronization was analyzed, which arises in the weak coupling regime when neurons that individually exhibit non-chaotic behaviors are coupled. A network composed of $1000$ neurons was simulated, firstly with the HH$\ell$ model, secondly with the HR model. The bifurcation diagram of a coupled neuron shows that in the lower coupling regime, the phase synchronization behavior occurs due to the partial-regular behavior of the neurons. This phase synchronization signal is lost with the increase of the coupling parameter since the synaptic current gains relevance inducing chaoticity in the neurons. After that, for greater values of coupling strength, the partial-regular behavior no longer exists, and the neurons phase synchronize in a chaotic regime.

This phenomenon is explored with different approaches, by varying intrinsic parameters of the model with the change of ionic conductances and disturbing the neuron dynamics with external disturbances by applying an external pulsed current. In both cases, it is possible to transition the neuronal dynamics of one isolated neuron from periodic to chaotic, and consequently, change the phase synchronization transition of a network composed of such neurons. In this context, this bifurcation point can be achieved more easily with the mutual change of more than one ion conductance, where small changes in one specific ion-conductance may act as a catalyst to the second conductance change \cite{boaretto2021role}. On the other hand, the direct application of external currents in the neuron can induce chaoticity in the regular dynamics, suppressing the non-monotonic transition. It is important to mention that, for the parameters used in this work, it was not possible to find a situation in which the pulsed current applied to individually chaotic neurons induces a regular behavior in neurons. Nevertheless, this possibility should not be discarded, as several works use the application of external disturbances as a chaotic control mechanism \cite{ott1990controlling,shinbrot1993using,schiff1994controlling}.
 
Also, it is important to notice that similar results can be obtained using the other networks' topologies of connections and other neuron models with the same conditions, i.e., bursting behavior, and chemical coupling. Hence, this correlation between the individual neuron dynamics and the collective behavior of the network can be directly applied to the suppression (or optimization) of phase synchronization of the network. 

\chapter{Bistability in the synchronization of identical neurons}\label{chap:bistability}

\initial{A}{s} shown in Chapter \ref{chap:sistemas}, nonlinear dynamical systems are known to exhibit a complex behavior with the variation of a control parameter, going from a stable periodic orbit to a chaotic state. Another interesting dynamical behavior is called {\it multistability}, which is the coexistence of different states for a given set of parameters \cite{feudel2008complex,ott2002chaos,strogatz2001nonlinear}. This dynamical feature has been studied for several years and it is observed in areas such as Physics \cite{arecchi1982experimental}, Chemistry \cite{ganapathisubramanian1984bistability}, climatology \cite{paillard1998timing}, and also neuroscience  \cite{hertz2018introduction,canavier1993nonlinear}. In this sense, multistability can be associated with different states of the brain and was proposed as a possible mechanism for memory storage and pattern recognition \cite{hertz2018introduction,canavier1993nonlinear}. On the other hand, in simulated neuronal systems, this coexistence can mean that a neuron may depict distinct stable states, with different firing patterns, frequencies, regularity, and chaoticity \cite{feudel2008complex,foss1996multistability,sainz2004influence,ma2007multistability}. 

In this Chapter, it is studied how a bistable state can affect the synchronization of a neural network. Firstly, it shows the existence of a parameter region where the neuron exhibits \textit{bistable} behavior, that is, a neuron initialized with different initial conditions can present two different stable states. Using the Lyapunov spectrum, it is detected that one state, namely state I, is always periodic while the other state, namely state II, depending on the parameters used, can exhibit periodic behavior or a chaotic one. Furthermore, it is shown that state II is more sensitive to noise than the state I, and the meantime that the system spends before escape from state II to state I follow a Kramers law \cite{hanggi1990reaction,kraut1999preference}. After that, a network of identical neurons is constructed using a generic mean-field electrical coupling. If all the neurons are initialized in the periodic state, as shown in Chapter \ref{chap:ferramentas} for identical oscillators, the network always reaches phase synchronization, with more weakly-coupled networks needing more time to reach the phase-synchronized state. But if there is bistability in the network, for high coupling values the synchronization state is reached, but, it takes longer compared to the case without bistability. On the other hand, in the situations where both states are periodic, smaller values of coupling strength are not able to make the network reach the phase synchronized state, and the system depicts two groups with different synchronization features. In both cases, interesting dynamic phenomena are observed such as \textit{chimera} states and anti-phase synchronization \cite{abrams2004chimera}. The results of this Chapter are published in the article ``\textit{Bistability in the synchronization of identical neurons.}" Physical Review E 104.2 (2021): 024204 \cite{boaretto2021bistability}.

\section{Bistability on the HH$\ell$ model}

\begin{figure}[t]
    \centering
    \includegraphics[width=\columnwidth]{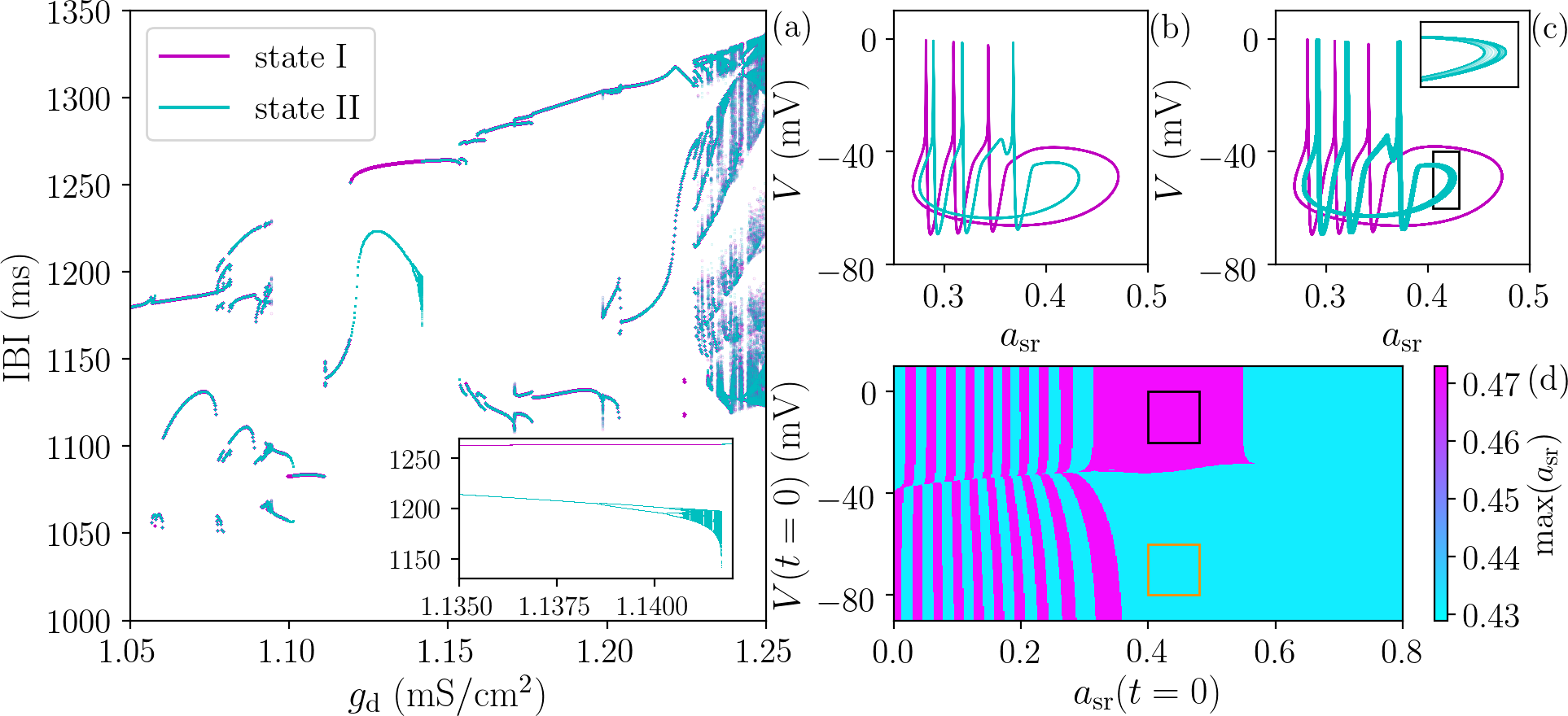}
    \caption[Bistability on the HH$\ell$ model.]{\textbf{Bistability on the HH$\ell$ model.}  (a) Bifurcation diagram of the Inter-Burst-Intervals (IBI) as a function of $g_\mathrm{d}$, where two different initial conditions are shown - cyan and magenta points. The magnification in the inner panel exhibits the bistability region where state I (magenta) and state II (cyan) are defined. The former is always periodic, while the latter can depict periodic and chaotic behavior depending on the $g_\mathrm{d}$ value (see inner panel). (b -- c) Two-dimensional projection $V(t)\times a_\mathrm{sr}(t)$ of the system's phase portrait, at (b) $g_\mathrm{d}=1.1350$ resulting in two periodic states, and (c) $g_\mathrm{d}=1.1415$ a periodic (magenta) and chaotic (cyan) states are detected. Each color corresponds to a distinct initial condition: magenta (cyan) line with IC-1 (IC-2). The inner panel exhibits a magnification of the chaotic attractor. (d) Maximum values of $a_\mathrm{sr}$ as a function of $V(t=0)$ and $a_\mathrm{sr}(t=0)$ for $a_\mathrm{d}(0) = a_\mathrm{r}(0) = a_\mathrm{sd}(0) = 0$ using $g_\mathrm{d}=1.1350$. The black (yellow) rectangle delimits each initial condition regions to obtain the initialization at each state I (state II).}
    \label{fig:bistability}
\end{figure}

The bistability in the HH$\ell$ model can be seen in Fig. \ref{fig:bistability}. In panel (a) it presents the bifurcation diagram of the IBI of one single neuron, using Eqs. (\ref{eq_hb_1}-\ref{eq_hb_2}) for the same parameters of the Table \ref{tabeladeconstantes}, but varying the sodium maximum conductance $g_\mathrm d$. The different colors refers to different initial conditions $\{V(0),a_\mathrm{d}(0),a_\mathrm{r}(0),a_\mathrm{sd}(0),a_\mathrm{sr}(0)\}$: where the magenta dots represent $\{-10,0,0,0,0.45\}$, named IC-1, and the cyan dots represent $\{-70,0,0,0,0.45\}$, named IC-2. It is noted that the region $1.120 \lesssim g_\mathrm{d}\lesssim 1.142$, the neuron is bistable, since IC-1 (IC-2) leads to an upper (bottom) state called state I (state II). The inner panel exhibits a magnification of the two states: while state I is always periodic, the state II goes through a sequence of period doubling bifurcation at $g_\mathrm{d}\gtrsim 1.1385$ and becomes chaotic due to mechanisms discussed in Chapter \ref{chap:sistemas}.  

To proceed with the study of bistability, two values of conductances are selected, $g_\mathrm{d}=1.1350$ (before the period doubling) and $g_\mathrm{d}=1.1415$ (after the period doubling). Figure \ref{fig:bistability}(b) depicts the evolution of a two-dimensional projection of the phase portrait $V\times a_\mathrm{sr}$ for $g_\mathrm d=1.1350$ where the magenta (cyan) line represents the IC-1 (IC-2). The variations in the amplitude of $a_\mathrm{sr}$ distinction the two states, while the state I $\mathrm{max}(a_\mathrm{sr})\approx 0.47$, for state II $\mathrm{max}(a_\mathrm{sr})\approx 0.43$. Figure \ref{fig:bistability}(c) $g_\mathrm d=1.1415$, it is observed that state II (cyan) depicts a greater thickness, indicating chaotic dynamics. The inner panel exhibits a magnification of this orbit, showing a projection of the chaotic attractor \cite{ott2002chaos}. Figure \ref{fig:bistability}(d) displays the maximum values of $a_\mathrm{sr}$ in color tones (from cyan $\approx 0.43$ to magenta $\approx 0.47$) ($g_\mathrm{d}=1.1350$) as a function of different initial conditions $\{V(0), 0, 0, 0, a_\mathrm{sr}(0)\}$, varying $V(0)$ and $a_\mathrm{sr}(0)$. It is seen that the $\max(a_\mathrm{sr})$ can be used to characterize each state, since the initial conditions leading to state I (state II) are represented in magenta (cyan). The black and yellow rectangles delimit the initial conditions used in this work to initialize the neurons in each state: $(V(0),a_\mathrm{sr}(0)) \in ([-20,0],[0.40,0.48])$, for state I and $(V(0),a_\mathrm{sr}(0)) \in ([-80,-60],[0.40,0.48])$ for II. It has to mention that the results of panel (d) changes subtly with $g_\mathrm{d}=1.1415$, but the rectangles still lead to their respective states.

\begin{figure}[htb!]
    \centering
    \includegraphics[width=\columnwidth]{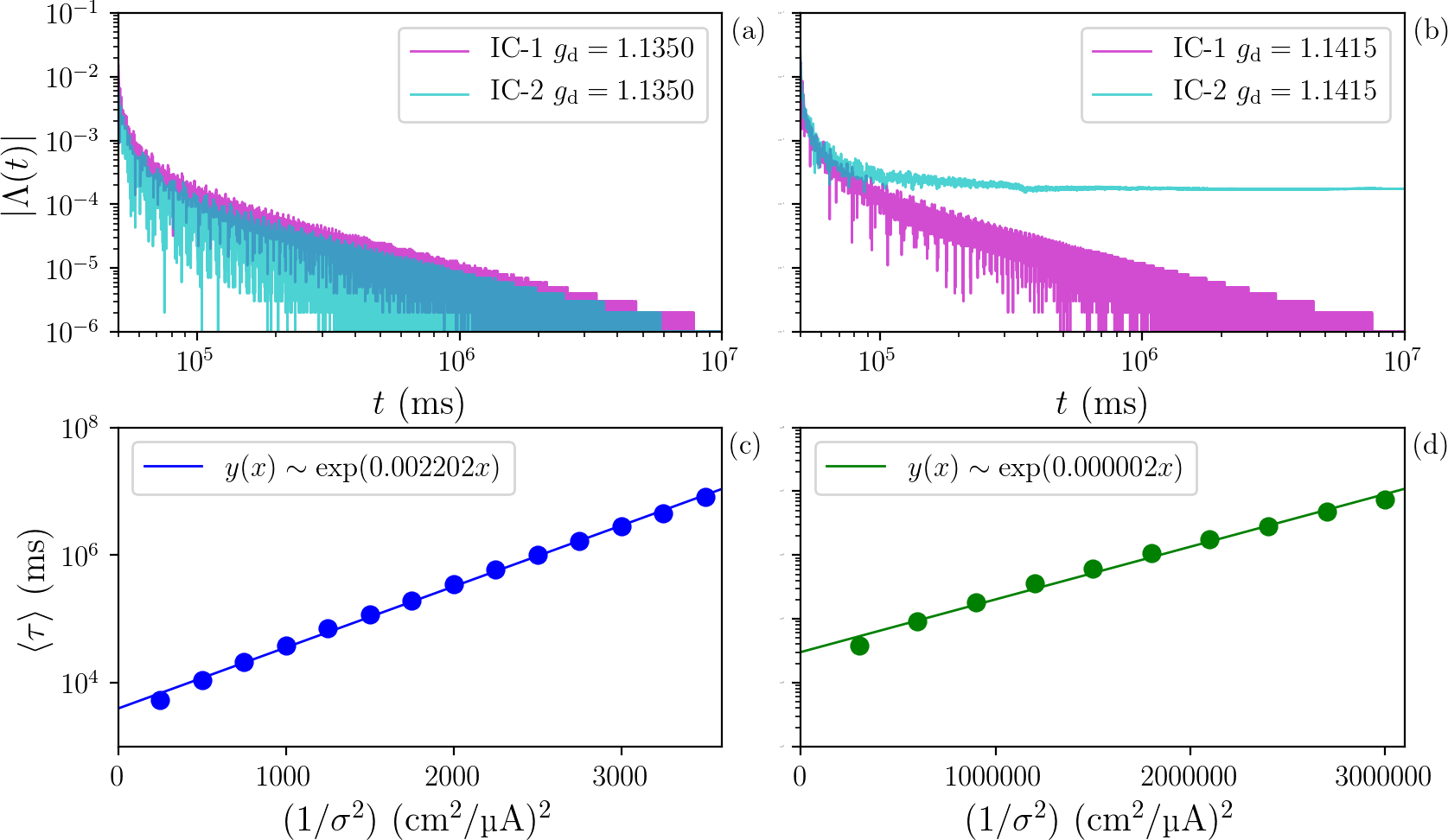}
    \caption[Stability of each state.]{\textbf{Stability of each state.} Time evolution of the Largest Lyapunov exponent for IC-1 and IC-2 (a) $g_\mathrm d = 1.1350$ and (b) $g_\mathrm d = 1.1415$. Mean time that the system spends in state II before escaping to the state I, as a function of the $1/\sigma^2$ ($\sigma$ is the noise strength) considering $1000$ simulations and the threshold time of $10^{8}$ $\SI{}{\milli\second}$. The slope $U$ correspond to the height of the potential barrier of a Kramers law $\langle \tau(\sigma)\rangle \sim \exp (U/\sigma^2)$. (c) $g_\mathrm d = 1.1350$ (periodic state)  $U \approx 2.2 \times 10^{-3}$. (d) $g_\mathrm d = 1.1415$ (chaotic state) $U \approx 2.0 \times 10^{-6}$.}
    \label{fig:stability_hb}
\end{figure}

To verify the chaotic features of the states, Table \ref{table_lyap} depicts the Lyapunov spectrum for the two values of $g_\mathrm d$ and computed for both IC-1 and IC-2. As mentioned in Chapter \ref{chap:sistemas} the Lyapunov exponent computes the divergences of orbits arbitrarily closed \cite{wolf1985determining}. If at least one of the exponents $\lambda_i > 0$, it is an indication of chaos. For $g_\mathrm d=1.1350$ both initial conditions depict a non-chaotic behavior since the largest Lyapunov exponent $\Lambda = \max \{\lambda_1,\cdots,\lambda_5\} \approx 0$ while for $g_\mathrm d=1.1415$, the IC-2 produces a state with $\Lambda \approx 10^{-4} > 0$, evidencing its chaoticity. Besides this, the smaller Lyapunov exponents indicate a difference in the stability of each state, thereby showing that state I is more stable than state II, and state II periodic is more stable than state II chaotic. In order to demonstrate the convergence of the exponents, panels (a) and (b) of Fig. \ref{fig:stability_hb} present the time evolution of the absolute value of the largest Lyapunov exponent ($\Lambda$) for IC-1 (magenta) and IC-2 (cyan) after discarding $10^5$ $\SI{}{\milli\second}$ to avoid transient effects. It is seen that, despite the IC-2 with $g_\mathrm d = 1.1415$ where $\Lambda$ is stable $\sim 10^{-4} $ (positive), for the other three cases $\Lambda$ converges  asymptotically to zero, exactly to zero if $t\rightarrow \infty$, denouncing the regular dynamics. 
\begin{table}[htb!]
    \centering
    \caption{Lyapunov spectra for IC-1 and IC-2 computed for two values of $g_\mathrm{d}$.}
    \begin{tabular}{l r r r r r}
         \hline \hline
        $g_\mathrm{d}=1.1350$ & $\lambda_1$& $\lambda_2$& $\lambda_3$& $\lambda_4$& $\lambda_5$\\ \hline
         IC-1 & -0.000007 & -0.001657 & -0.102236 & -0.197086 & -5.466736\\
         IC-2 & -0.000002 & -0.001059 & -0.122687 & -0.213510 & -5.420790\\
         \hline \hline
        $g_\mathrm{d}=1.1415$ & $\lambda_1$& $\lambda_2$& $\lambda_3$& $\lambda_4$& $\lambda_5$\\ \hline
         IC-1 & -0.000006 & -0.001818 & -0.099646 & -0.195736 & -5.471647 \\
         IC-2 & 0.000173 & -0.000017 & -0.122036 & -0.217536 & -5.418144\\
         \hline \hline
    \end{tabular}
    \label{table_lyap}
\end{table}

It is known that multistable systems are generally sensitive to noise \cite{feudel2008complex,kraut1999preference,sharma2013controlling,pisarchik2014control}. The mean time to perform a noise-induced transition from one state to another gives an insight into the stability of the system. In this sense, a current term is summed to Eq. (\ref{eq_1}), characterized by a noisy signal which follows a random normal distribution with average $0$ and standard deviation $\sigma$. It is noted that for the values of $\sigma$ and considering a maximum simulation time of $10^{8}$ ms, state II switches to state I only, and never the other way around. Panels (c) and (d) of Fig. \ref{fig:stability_hb} show the meantime that the system spends in state II before escaping to state I as a function of the inverse of the noise strength $1/\sigma^2$ considering a $1000$ simulations. The time when a transition from state II to state I occur is evaluated when $a_\mathrm{sr}$ crosses the threshold $0.45$. The results show that the meantime to perform a transition follows a Kramers law $\langle \tau(\sigma)\rangle \sim \exp (U/\sigma^2)$ where $U$ corresponds to the height of the potential barrier \cite{hanggi1990reaction,kraut1999preference}.  In panel (c) $g_\mathrm{d}=1.1350$ (periodic state) $U \approx 2.2\times 10^{-3}$. In panel (d) $g_\mathrm{d}=1.1415$ (chaotic state) $U \approx 2.0\times 10^{-6}$. Based on this test, the results indicate that the chaotic state II switches more easily than the periodic state II. 

\section{Network properties}

To explore the role of bistability in synchronization, is considered a neural network composed of $N = 100$ HH$\ell$ identical neurons. The evolution of the membrane potential for the $i$-th neuron is given by
\begin{equation} \label{eq_rede}
    C_\mathrm{M} \frac{dV_i}{dt} = -I_{i,\mathrm{d}} - I_{i,\mathrm{r}} - I_{i,\mathrm{sd}} - I_{i,\mathrm{sr}} - I_{i,\mathrm{l}}+I_{i,\mathrm{syn}},
\end{equation}
where the ionic currents are given by the Eqs. (\ref{eq_hb_1}-\ref{eq_hb_2}) and the synaptic current is characterized by a mean-field coupling (all-to-all) 
\begin{equation}\label{eq_global}
    I_{i,\mathrm{syn}} = \frac{\varepsilon}{N}\sum_{j=1}^N(V_j-V_i) = \varepsilon(\langle V\rangle - V_i), 
\end{equation}
where $\varepsilon$ is the coupling (synaptic) strength measured in $\SI{}{\milli\siemens/\centi\meter^{2}}$ which, for simplicity, is subsequently omitted. And
\begin{equation}
   \langle V\rangle = \frac{1}{N}\sum_{i=1}^N V_i 
\end{equation}
is the mean-field of the network. This configuration is chosen because allows the isolation of the bistability effect since $\varepsilon$ and $\langle V\rangle$ are the same for all neurons, the only difference between them is the initial conditions. 

Moreover, the network is artificially subdivided in two groups  $\Omega_1=\{1,\cdots,N_1\}$ and  $\Omega_2=\{N_1+1,\cdots,N\}$ with $N_1$ and $N_2$ neurons, respectively. With this definition, it is possible to adapt the Kuramoto order parameter (defined to the whole network) to analyze the phase synchronization level of each group separately. To do this, Eq. (\ref{eq:Rtime}), for the $\ell$-th group is
\begin{equation}
    R_\ell = \left |\frac{1}{N_\ell}\sum_{j \in \Omega_\ell}e^{i\theta_j(t)} \right|, \;\;\;\;\; \ell=1,2.
\end{equation}

%
%
%
%
Finally, to measure if there is a difference in the synchronization of each group is defined the absolute difference between $R_1$ and $R_2$:
\begin{equation}
    \Delta R = |R_2-R_1|.
\end{equation}

The initial conditions for the network simulations were selected to be in a random position on the attractor of each state. To do so, an uncoupled neuron is simulated for $10^{6}$ $\SI{}{\milli\second}$ with random initial conditions according to the rectangles of Fig. \ref{fig:bistability}(d) (for each state). To select the initial conditions for each neuron it is selected a random time instant of the last $5\times10^{5}$ $\SI{}{\milli\second}$ of the simulation. This approach avoids any initial synchronization bias ($R(t=0)\approx 1$) and allows the initialization of each desired state. All simulations are performed using a time limit of $10^8$ $\SI{}{\milli\second}$.

\section{Network results}

The investigation of the role of the bistability in the network is made by considering three different situations:
\begin{itemize}
        \item[] (i) - All neurons are in the state I (periodic) with $g_{\mathrm{d}} = 1.1350$ (absence of bistability);
        \item[] (ii) - Half of neurons of the network in each group ($N_1=N_2=N/2$), where neurons in $\Omega_1$ are initialized in state I (periodic) and $\Omega_2$ in the state II (chaotic) with $g_{\mathrm{d}} = 1.1415$;
        \item[] (iii) - Half of neurons of the network in each group ($N_1=N_2=N/2$), where neurons $\Omega_1$ are initialized in the state I (periodic) and $\Omega_2$ in the state II (periodic) with $g_{\mathrm{d}} = 1.1350$.
\end{itemize}

\begin{figure}[t]
    \centering
    \includegraphics[width=.9\columnwidth]{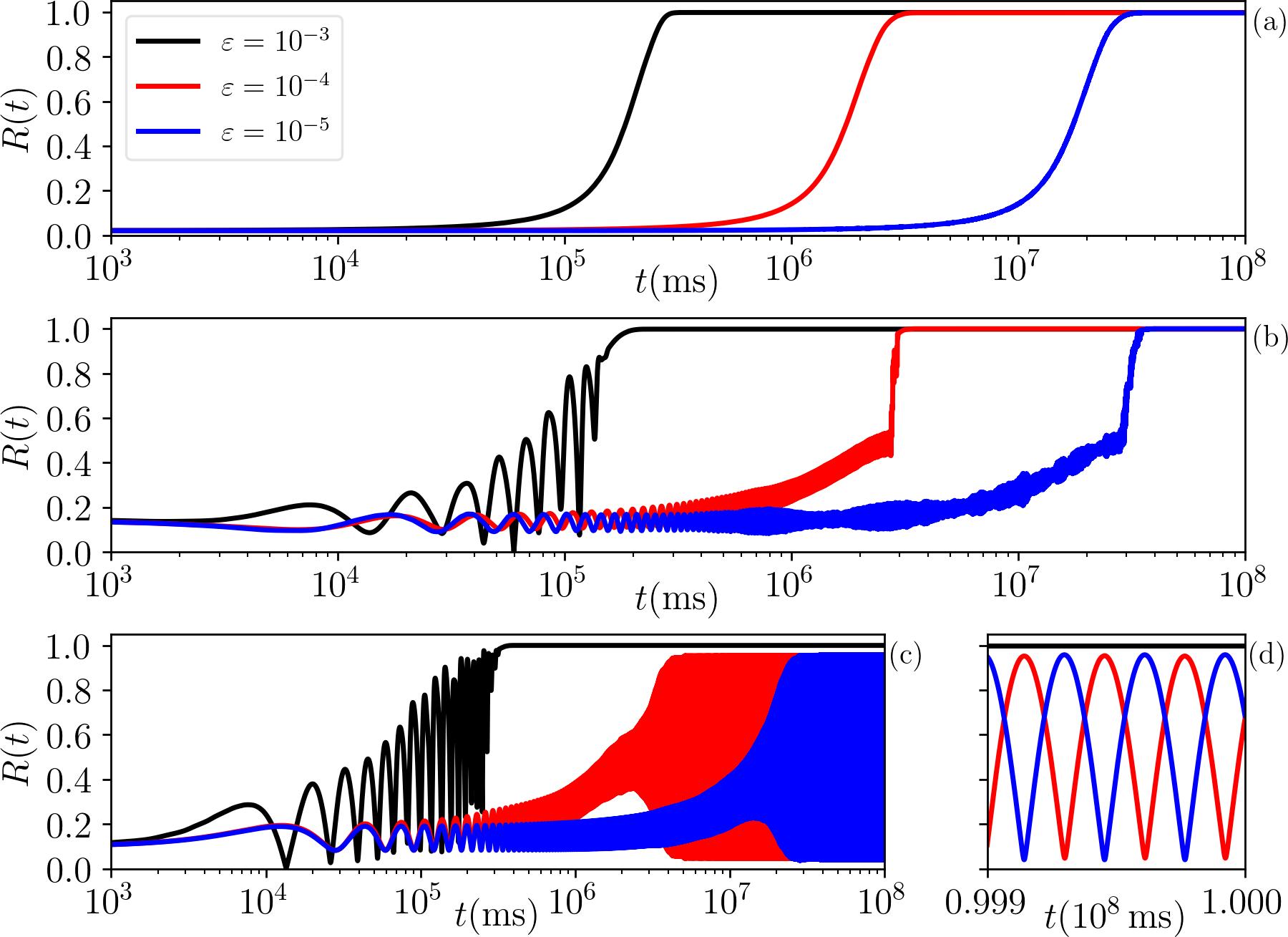}
    \caption[The role of the bistability at the synchronization in a network of identical HH$\ell$ neurons.]{\textbf{The role of the bistability at the synchronization in a network of identical HH$\ell$ neurons.} Kuramoto order parameter $R(t)$ for three values of coupling $\varepsilon=10^{-3}$ (black lines), $\varepsilon = 10^{-4}$ (red lines), and $\varepsilon = 10^{-5}$ (blue lines). Three network configurations are used: (a) - a network with all neurons in state $\mathrm{I}$; (b) - a mixed network with half of neurons in state $\mathrm{I}$ (periodic) and half in state $\mathrm{II}$ (chaotic); (c) - a mixed network with half of neurons in state $\mathrm{I}$ (periodic) and half in state $\mathrm{II}$ (periodic). (d) Magnification of the last $10^{5}$ times of panel (c).}
    \label{fig:r_bistability}
\end{figure}

In Fig. \ref{fig:r_bistability}, it is evaluated the time-evolution of the Kuramoto order parameter $R(t)$ for the three different values of coupling $\varepsilon$, $10^{-3}$ (black line), $10^{-4}$ (red line), and $10^{-5}$ (blue line). Panel (a) shows the result for the situation (i) (absence of bistability), for the three values of $\varepsilon$ the system monotonically evolves from a non-synchronized state $R\approx 0$ to a complete phase synchronized state $R=1$, with the time needed to reach phase synchronization decreasing as the coupling strength increases. A similar scenario is observed in panel (b), where situation (ii) is studied, but the transition is not monotonic as in panel (a). Panel (c) depicts the result for situation (iii) where the stable phase synchronized state is achieved only for $\varepsilon = 10^{-3}$, otherwise, for $\varepsilon=10^{-5},\,10^{-4}$ the system stays in an oscillatory state where $R$ vary between $0$ and $1$. This behavior is maintained for a long period as observed in panel (d), which shows the last $10^{5}$ of the simulation. The difference between the panels highlights that the bistability can influence the final state of the system. While situation (i) the network depicts the same behavior of periodic coupled oscillators, the existence of two different states in the network makes things more complicated. To reach the synchronized state, the coupling first needs to induce a transition in the neurons leading them to a final unique state, with the same frequency, and only then, it can align their phases. The details of these processes are analyzed in the next sections.

\subsection{Bistability with periodic-chaotic configuration (situation (ii))}

\begin{figure}[t]
    \centering
    \includegraphics[width=.9\columnwidth]{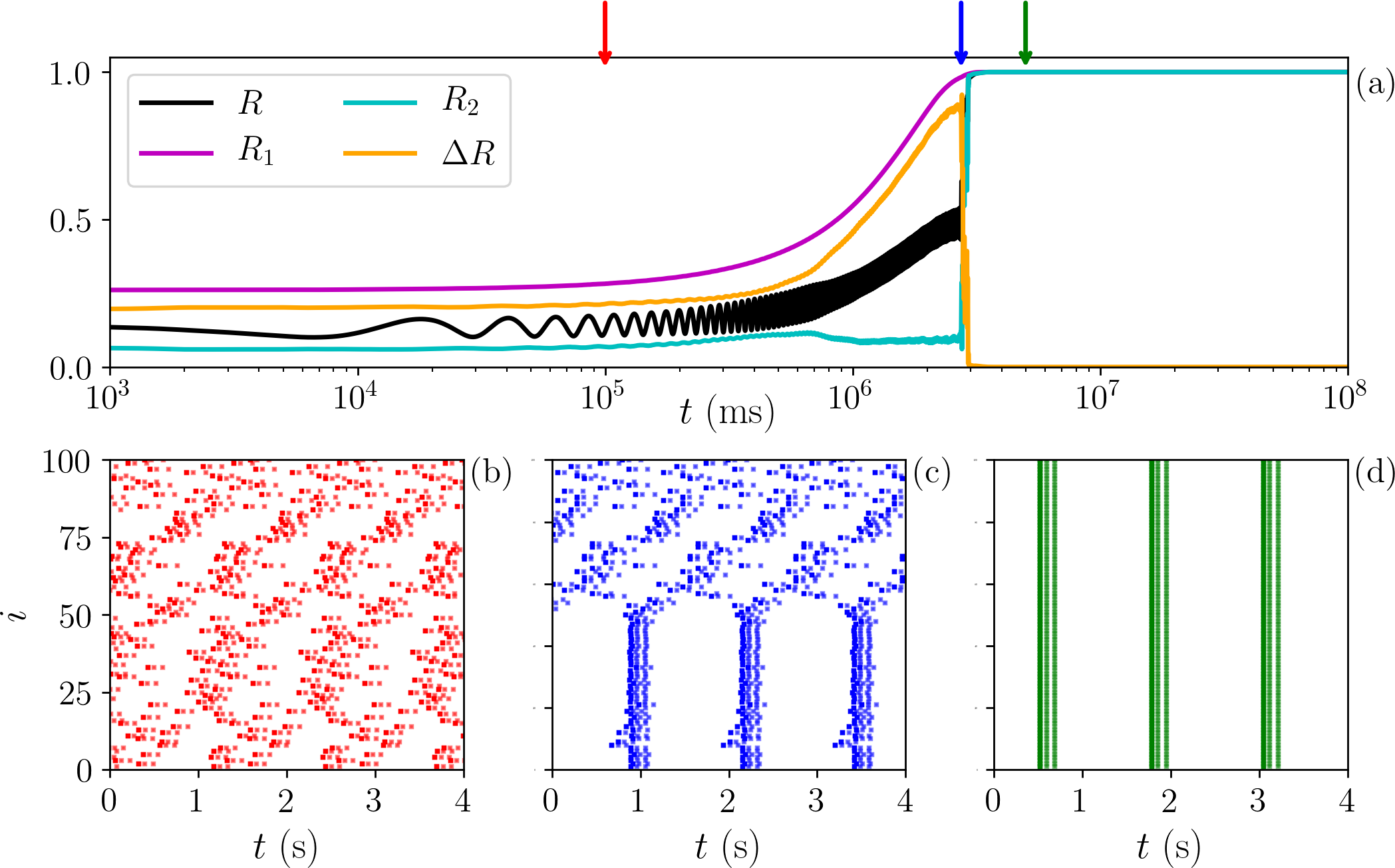}
    \caption[Temporal evolution of the synchronization in the network at the situation (ii) with $\varepsilon = 10^{-4}$.]{\textbf{Temporal evolution of the synchronization in the network at the situation (ii) with $\mathbf{\bm{\varepsilon} = 10^{-4}}$.} (a) $R$, $R_1$, $R_2$, and $\Delta R$ (black, magenta, cyan, and orange lines, respectively). (b - d) Raster plots of the network where each dot corresponds to the beginning of a spike. The colored arrows indicate the times when the raster plots are obtained, which match with the dots' colors.}
    \label{fig:loc_1}
\end{figure}

In this section, it is considered a network with $g_\mathrm d=1.1415$ with neurons $\Omega_1$ initialized in state I (periodic), and neurons of $\Omega_2$ in state II (chaotic). As shown in Fig. \ref{fig:r_bistability} (b) the network presents the same behavior for the three $\varepsilon$ values. In order to study this situation it is used $\varepsilon=10^{-4}$ (red line). The results are presented in Fig. \ref{fig:loc_1} wherein panel (a), the black line in panel corresponds to $R$ (whole network), the magenta and cyan lines represent the local order parameters $R_1$ and $R_2$, respectively, and the orange line presents $\Delta R$. As expected the network starts from the non-syncrhronized state and, as the system evolves, $R_1$ increases slowly while $R_2$ remains close to $0$, leading to a local maximum of $\Delta R$. After the first group reach the synchronized state $R_1=1$, $R_2$ quickly rises to $1$. The colored arrows above panel (a) are three instants to be analyzed separately in raster plots in panels (b -- d). In this sense, each dot corresponds to the beginning of a spike for each neuron, evaluated when $V_i(t)$ reaches $-20$ $\SI{}{\milli\volt}$ (with a positive derivative). Panel (b) shows the raster plot for a non-synchronized state of the network. Panel (c) presents an interesting behavior where the first group is synchronized ($R_1\approx 1$) while the second one does not ($R_2 \approx 0$). This behavior is called \textit{chimera}, where the system displays the coexistence of one coherent-phase-locked group with an incoherent-non-synchronized one \cite{abrams2004chimera}. In this case, the chimera is transient \cite{zakharova2014chimera,kemeth2016classification} and disappears when the second group gains synchronization, shown in panel (d), where the network as a whole is synchronized.

\subsection{Bistability with periodic-periodic configuration (situation (iii))}

\begin{figure}[t]
    \centering
    \includegraphics[width=.9\columnwidth]{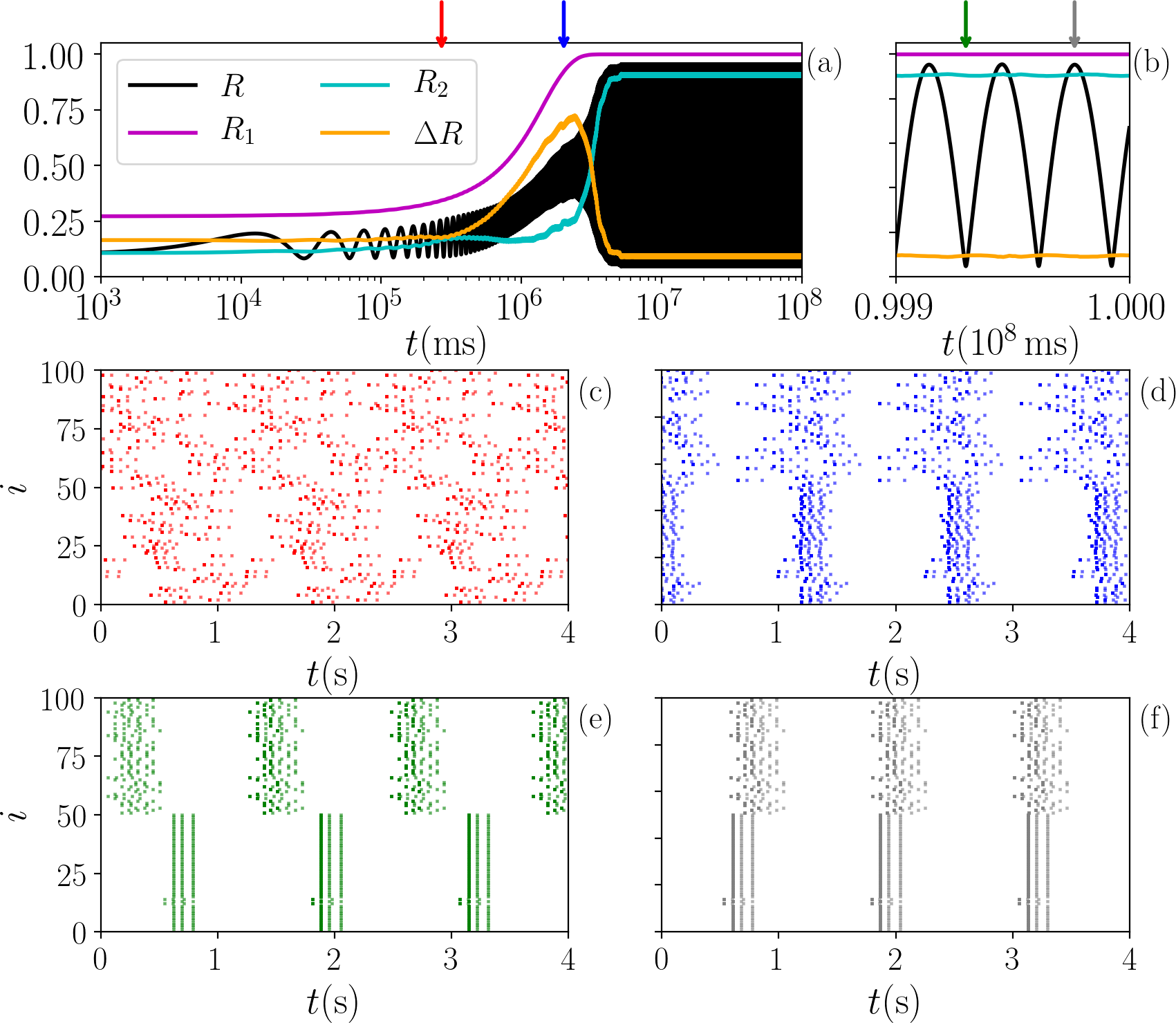}
    \caption[Detailed analysis in the synchronization of situation (iii) with $\varepsilon=10^{-4}$.]{\textbf{Detailed analysis in the synchronization of situation (iii) with $\bm{\varepsilon=10^{-4}}$.}
    (a) $R$, $R_1$, $R_2$ and $\Delta R$ (black, magenta, cyan, and orange lines) as a function of time. (b) Last $10^5$ $\SI{}{\milli\second}$ of the simulation, where an oscillatory behavior of $R$ is evidenced while each group assumes phase synchronization separately. (c - f) Raster plots obtained for the network in the times indicated by the colored arrows.}
    \label{fig:loc_2}
\end{figure}

In this section, it is consider a network with $g_\mathrm d=1.1350$ with neurons $\Omega_1$ initialized in the state I (periodic), and neurons of $\Omega_2$ in the state II (periodic). As shown in panels (c) and (d) of Fig.~\ref{fig:r_bistability}, the final state of the network depends on the $\varepsilon$. Firstly it is studied the non-synchronizaed state where $\varepsilon=10^{-4}$ (red line of Fig.~\ref{fig:r_bistability} (c - d)). Figure \ref{fig:loc_2} (a) depicts the synchronization features of the network, where $R$, $R_1$, $R_2$, and $\Delta R$ are represented by the black, magenta, cyan, and orange lines, respectively. The network goes from the non-synchronized state ($R \approx 0$) and, as time evolves, the $R$ oscillates with increasing amplitude. On the other hand, $R_1$ and $R_2$ grows, but $R_1$ increases more quickly than $R_2$, leading to a local maximum of $\Delta R$. After that, $R_2$ also increases, leading both $R_1$ and $R_2$ close to $1$, but $R_{1} > R_{2}$. Panel (b) depicts the last $10^5$ $\SI{}{\milli\second}$ of the simulation, suggesting a beating process of $R$.

In order to understand this oscillatory behavior of $R$ in Fig. \ref{fig:loc_2}, $4$ time instants of the simulation are selected, represented by the colored arrows above panels (a) and (b). Panels (c - f) represent the raster plots corresponding to the colors of the arrows. First, at the panel (c) an incoherent behavior is observed in the red dots. As time evolves, in panel (d) the network depicts a transient chimera state since $\Delta R$ increases. Eventually, each group depicts phase synchronization separately, since $R_1$ and $R_2$ show values close to one. However, each group evolves following its frequency, which leads to momentary non-synchronization (e) and phase-synchronization (f) (the oscillatory behavior of $R$). 

\begin{figure}[t]
    \centering
    \includegraphics[width=.9\columnwidth]{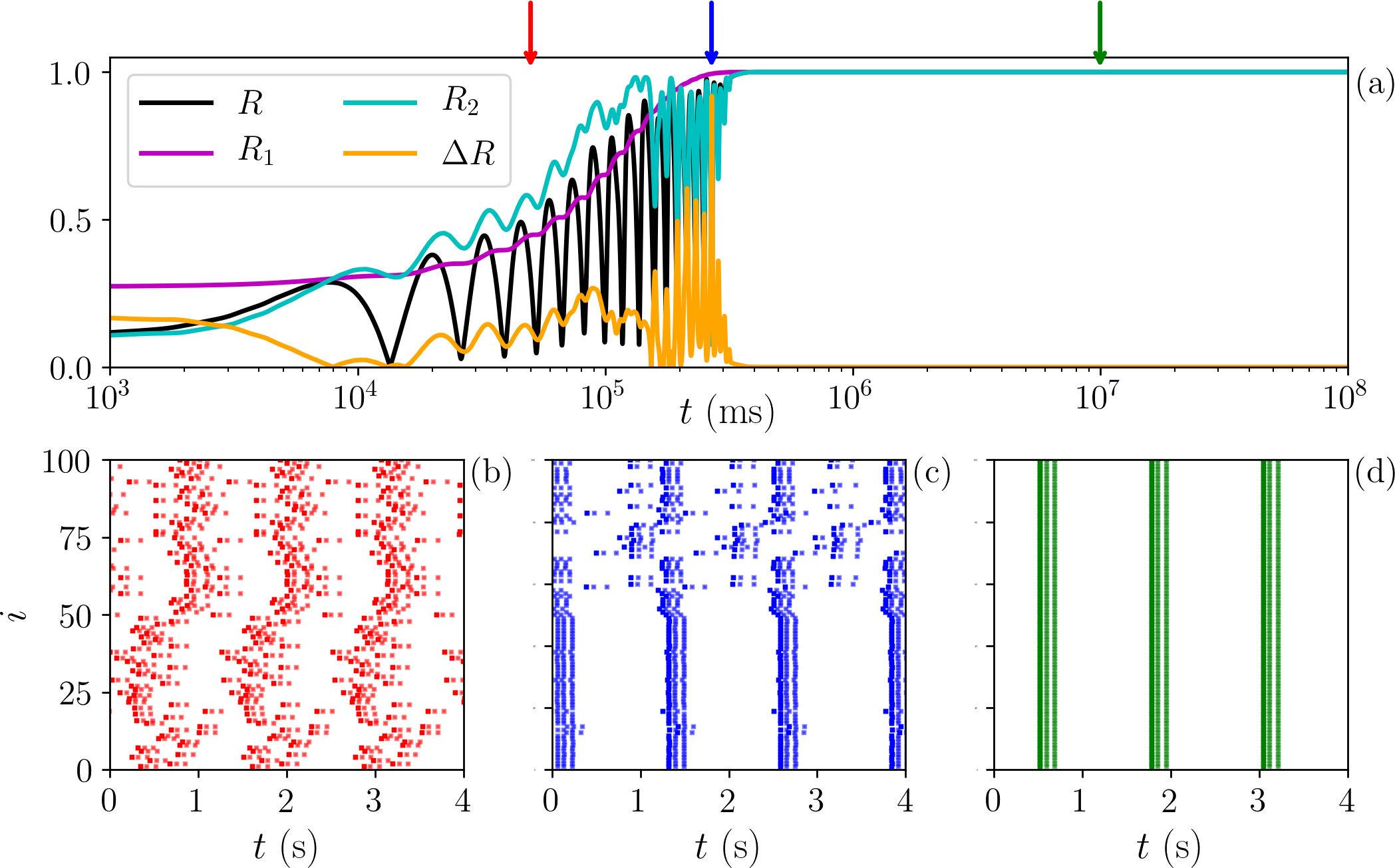}
    \caption[Detailed analysis in the synchronization of situation (iii) with $\varepsilon=10^{-3}$.]{ \textbf{Detailed analysis in the synchronization of situation (iii) with $\bm{\varepsilon=10^{-3}}$.} (a) $R$, $R_1$, $R_2$, and $\Delta R$ (black, magenta, cyan, and orange lines) as a function of time. (b - d) Raster plots for the network where the dots' colors match the arrows' colors, which indicate the analyzed times.}
    \label{fig:loc_3}
\end{figure}

The detailed analysis for the situation (iii) with $\varepsilon=10^{-3}$ (black line of Fig.~\ref{fig:r_bistability} (c - d)) is presented in Fig.~\ref{fig:loc_3}. The color scheme follows the same one of the previous figure. The network starts in a non-synchronized case and $R$ gains amplitude in an oscillatory way, reaching the phase synchronized asymptotic state ($R = 1$) as time evolves. While $R_1$ approaches monotonically the state of phase synchronization, $R_2$ approaches in an oscillatory manner. In addition, for $t\sim 2 \times 10^{5}$ $\SI{}{\milli\second}$, the $R_2$ momentarily loses synchronization and a peak can be observed in $\Delta R \approx 0.9$. Three-time instants (colored arrows) are selected to evaluate the raster plots. Panel (b) depicts the non-synchronized case. Panel (c) depicts the raster plot for the network in the time instant where the $R_2$ loses synchronization and $\Delta R$ assumes a maximum characterizing a transient chimera state \cite{zakharova2014chimera}. This behavior occurs for a short period. Panel (d) shows the raster plot for the network when phase synchronization is reached and maintained until the end of the simulation. Different from the previous case, with $\varepsilon=10^{-4}$, here, the coupling is strong enough to induce a transition in the network which yields a phase synchronized state, the oscillatory behavior of the neurons in $\Omega_2$ indicates that it undergoes a series of transitions to the state of the $\Omega_1$.

\begin{figure}[t]
    \centering
    \includegraphics[width=.9\columnwidth]{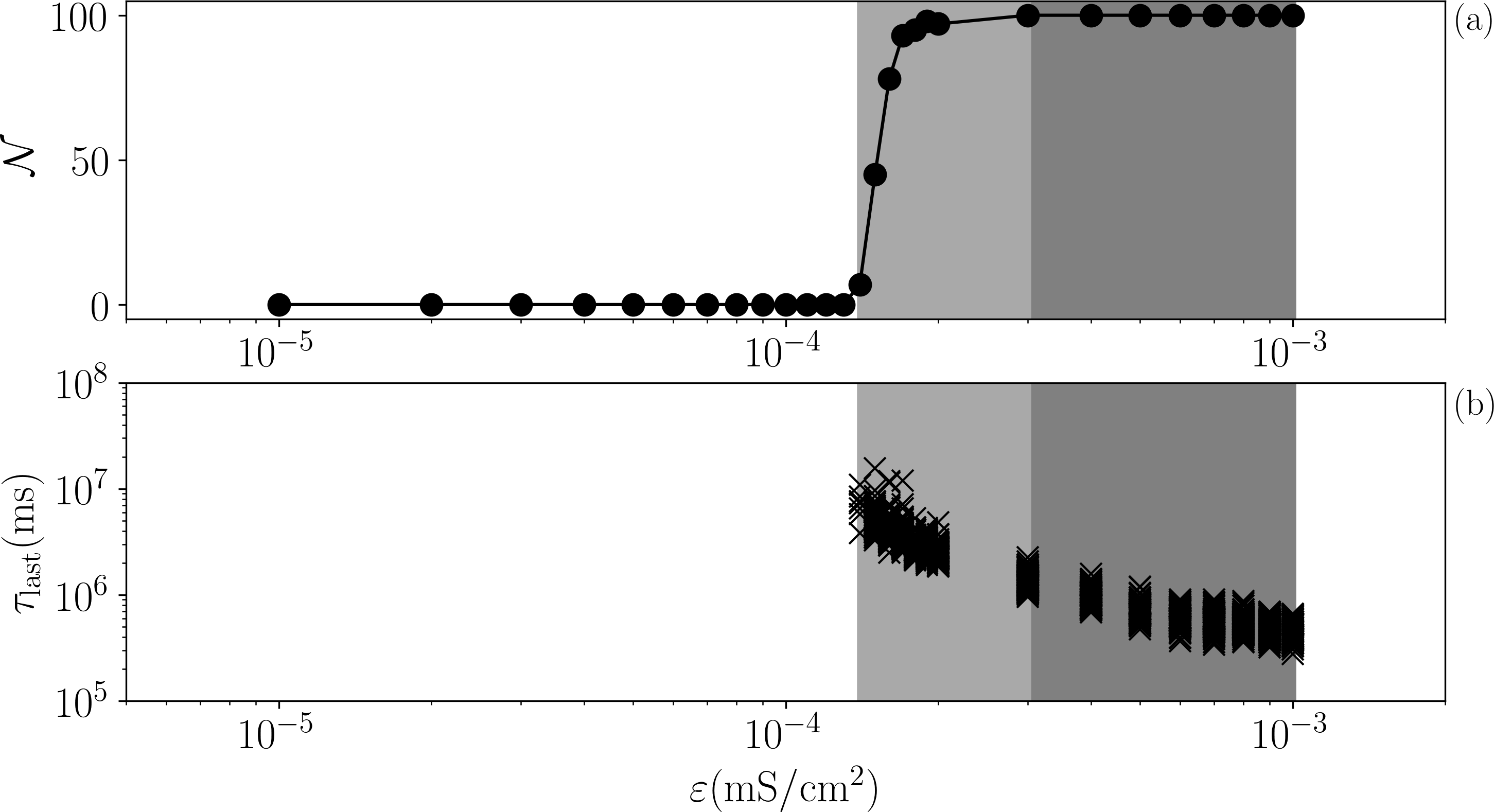}
    \caption[A complete transition to state I depends on the initial conditions.] {\textbf{A complete transition to state I depends on the initial conditions.} (a) Number of simulations where the network as a whole depicts phase synchronization, meaning all neurons transition to state I, as a function of $\varepsilon$. For $\varepsilon < 0.00014 $, no transitions are observed. For $0.00014  < \varepsilon < 0.00030$ (light-gray area), a fraction of simulations induce the network to present phase synchronization, and, at last, for $\varepsilon > 0.00030$ (dark-gray area), the network depicts phase synchronization for all simulations. (b) The time when the last neuron transitions to state I occur ($\tau_\mathrm{last}$) as a function of $\varepsilon$. Here, $100$ simulations with different initial conditions are performed.}
    \label{fig:gray_fig}
\end{figure}

To understand the influence of the coupling strength $\varepsilon$ in the final state of the network in Fig. \ref{fig:gray_fig} it is simulated $100$ different initials conditions for the situation (iii). At the panel (a) is computed the number of simulations $\mathcal N$ where there is a complete transition from state II to state I and the entire network reaches phase synchronization. The transition from state II to state I is recorded when $a_\mathrm{sr}$ crosses the threshold 0.45 for the first time. For $\varepsilon > 0.00030$ (dark-gray area) the network depicts phase synchronization for all simulations $\mathcal N=100$, for $0.00014 <\varepsilon < 0.00030$ (light-gray area), a fraction of simulations induce the network to present phase synchronization $0<\mathcal N<100$, and, at last, for $\varepsilon < 0.00014 $, no transitions are observed $\mathcal N=0$. At panel (b) it is computed the time when the last neuron initialized in state II transitions to state I, named $\tau_\mathrm{last}$. In this case, the higher the coupling, the smaller the time at which this happens.

\begin{figure}[t]
    \centering
    \includegraphics[width=.9\columnwidth]{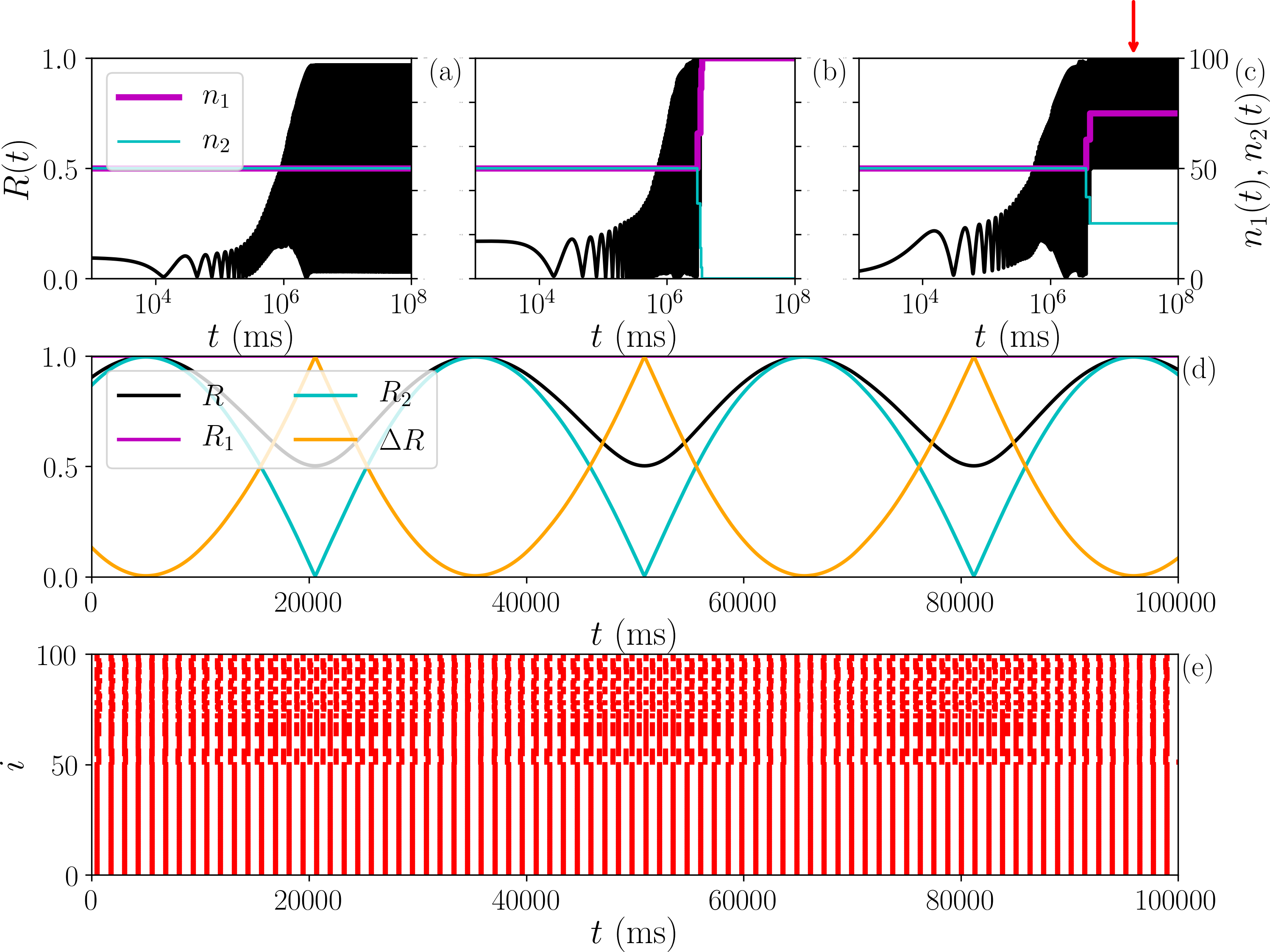}
    \caption[The dependence of synchronization features on the initial conditions analyzed for $\varepsilon = 0.00016$.]{\textbf{The dependence of synchronization features on the initial conditions analyzed for $\bm{\varepsilon = 0.00016}$.} (a), (b), and (c) present the Kuramoto order parameter for the entire network (left scale) and the number of neurons in each state as a function of $t$ (right scale). Here, different initial conditions are considered, in which no transition (a), complete transition (b), and a partial transition (c) from state II to state I are observed. (d) Synchronization features of the network for the case of partial transition considering the time region indicated by the red arrow. (e) Raster plot of the network (bursts only) for the same condition.}
    \label{fig:sync_ic}
\end{figure}

An interesting behavior can be seen at the light-gray-area $0.00014  < \varepsilon < 0.00030$, where the final state of the network depends on the initial condition even for the same value of $\varepsilon$. In this sense, in Fig. \ref{fig:sync_ic} it is studied the sensibility of initial conditions for $\varepsilon=0.00016$. Panels (a -- c) depict the time evolution of $R(t)$ (left scale) and the number of neurons in state I ($n_1(t)$, magenta) and state II ($n_2(t)$, cyan) (right scale). Panel (a) presents an initial condition where $R$ evolves to an oscillatory behavior and $n_1 = n_2 = 50$ during all the simulation. Panel (b) depicts an initial condition where $R$ evolves to a phase synchronized state and at the end of simulation $n_1 = 100$ and $n_2 = 0$, a complete transition. An interesting behavior is described in panel (c), for this initial condition, a partial transition to state I is observed. In this case, $R$ oscillates between $0$ and $1$ as $t$ evolves, but after a certain time, the oscillation becomes restricted between $0.5$ and $1$. To understand this behavior it is selected a time instant (red arrow) to analyze the phase synchronization and the spatiotemporal of the network. Panel (d) shows $R$, $R_1$, $R_2$ and $\Delta R$, where it is observed that neurons in $\Omega_1$ are phase synchronization while $R_2$ oscillates between $0$ and $1$. The raster plot of the network (bursts only), depicted in panel (e), shows the coexistence of two frequencies in the $\Omega_2$, explaining the oscillations in $R_2$ and $R$. In this situation, chimera states occur: they appear when the second group is non-synchronized, then disappear when it synchronizes and reappears later again.

\begin{figure}[t]
    \centering
    \includegraphics[width=.9\columnwidth]{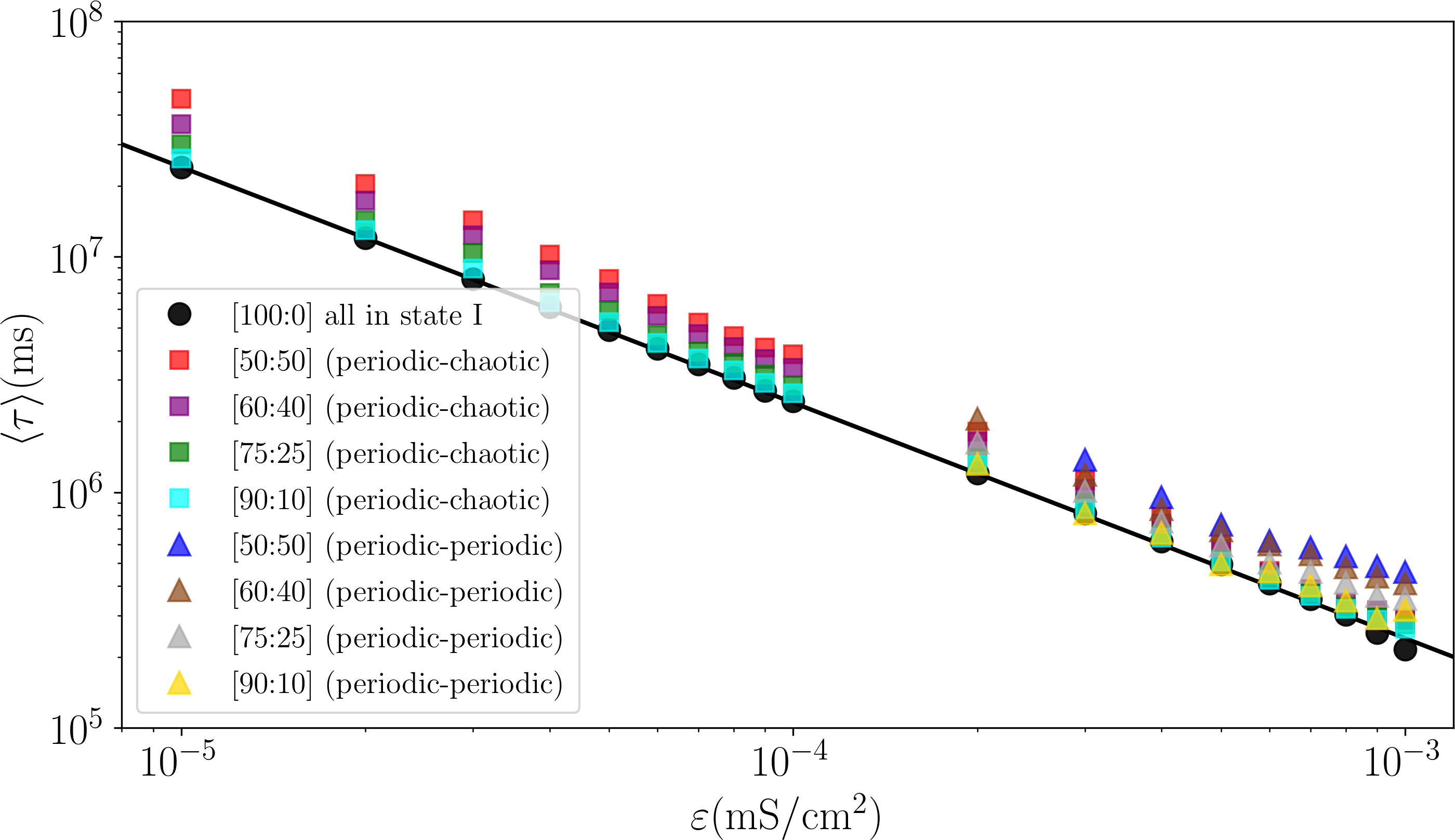}
    \caption[Mean time $\langle \tau\rangle$ needed for the network to reach phase synchronization as a function of $\varepsilon$ for $100$ initial conditions.]{ \textbf{Mean characteristic time $\bm{\langle \tau\rangle}$ needed for the network to reach phase synchronization as a function of $\bm \varepsilon$ for $\bm{100}$ initial conditions.} The initialization of the network is made in different configurations of states $[N_1: N_2]$. The black dots correspond to all neurons in state I (situation (i)), following a power-law curve (solid black line) $f(\varepsilon) \sim \varepsilon^{-1}$. The squares represent configurations where state II is chaotic, in red [50:50] (situation (ii)), purple [60:40], green [75:25], and cyan [90:10]. For the triangles, state II is periodic, in blue [50:50] (situation (iii)), brown [60:40], gray [75:25], and yellow [90:10].}
    \label{fig:bi_mean_time}
\end{figure}

In Fig.~\ref{fig:bi_mean_time} it is studied the meantime to reach a synchronized state ($R\geq 0.99$) $\langle \tau\rangle$ as a function of $\varepsilon$ considering $100$ initial conditions where different network configurations are considered, varying the number of neurons initialized in each state $[N_1, N_2]$. The black dots correspond to the situation (i) (all neurons in the state I with $g_\mathrm d=1.1350$) where the mean time decays with a power-law $f(\varepsilon) \sim \varepsilon^{-1}$ represented by the black solid line. It has to be mentioned that this is the same situation of periodic oscillators studied in Chapter \ref{chap:ferramentas}, the relaxation time is inversely proportional to the coupling parameter. The squares represent configurations where state I is periodic and state II is chaotic ($g_\mathrm d=1.1415$) in red [50:50] (situation (ii)), purple [60:40], green [75,25], and cyan [90:10], and the triangles both states are periodic ($g_\mathrm d=1.1350$) in in blue [50:50] (situation (iii)), brown [60:40], gray [75:25], and yellow [90:10]. In a general point of view the existence of bistability in the network increases the time necessary to the system reach phase synchronization, greater is the proportion of neurons in the state I closest is the result to the black line. In particular, the absence of triangles on the left side of the figure can be explained since for coupling values lower than a critical value no transition is observed, consequently there is no phase synchronization to compute. 

\section{Discussions}

Throughout this chapter, the effects of bistability on the synchronization process of neural networks were analyzed. The HH$\ell$ neuron presents a bistability region when the sodium conductance is varied. It is shown that one of the states, namely state I, is always periodic, while the other state, namely state II depending on $g_\mathrm d$ can be periodic or chaotic. By evaluating the Lyapunov spectrum and with noise application it was found that state I is always more stable than state II, and state II periodic is more stable than state II chaotic.

To understand the impact of bistability in the synchronization, it was built a network of identical neurons with a global coupling, where the only difference between neurons is their initial conditions. In this sense, three network configurations have been considered: (i) all neurons in the state I; (ii) half of the network in the state I and half in state II (chaotic); (iii) half of the network in the state I and half in state II (periodic). The existence of bistability in the network increases, on average, the time to reach a synchronized state. This can be explained due to the fact that the phase synchronization is only possible with a unique and final state. Hence, a transition from the initial states to a final and unique state is mandatory. The transitions occur from state II (periodic or chaotic) to state I (periodic). If the state II is periodic small coupling values may not induce the transition, resulting in a local synchronization in each of the groups (not global synchronization), or even in a partial transition of neurons. At last, using different fractions of neurons in each state, the time the network takes to reach phase synchronization decreases, on average, as the coupling strength increases and the behavior gets closer to configuration (i) as the fraction of neurons initialized in the state I increase.

In summary, bistability plays an important role in the synchronization of neural networks. The simple existence of two distinct stable states can lead the network to different states of synchronization, depending on the initialization of the system: from a non-synchronized state to a complete phase synchronized state. Bistability also leads to the existence of a variety of chimera states where the network displays the coexistence of one coherent-phase-locked group with an incoherent-non-synchronized one \cite{abrams2004chimera}. These synchronization and chimera states occur due to the difference in the stability of the states of the uncoupled neurons, thus highlighting the importance of the individual neuronal dynamics.

\chapter{Mechanism for explosive  synchronization of neural networks}\label{chap:mechanism}

\initial{T}{he} phenomenon of explosive synchronization in oscillators was first studied in chaotic networks by Gomez \textit{et al.} \cite{gomez2011explosive}, in which, for the particular case of Kuramoto oscillators coupled by a network with scale-free topology \cite{boccara2010modeling}, when the natural frequency $\omega_i$ of each oscillator is given by the number of connections $\omega_i = n_i$, the transition from the non-synchronized state to the synchronized state occurs abruptly. This abrupt phase-variation regarding the synchronization is named explosive synchronization. This behavior occurs due to the existence of hubs (sites with a high degree of connectivity), a characteristic of the topology of the scale-free network \cite{albert2002statistical}. From the point of view of neural networks, this effect is not necessarily reproduced. This chapter explores how explosive synchronization appears in a network of spiking neurons. It is considered an ensemble of Chialvo neurons coupled in a network generated with the Newman-Watts route \cite{newman1999renormalization}. For different values of the connection probability $p_\mathrm{nw}$ (which controls the number of shortcuts added in the network) it is possible to found different synchronization transitions, in particular, the behavior of explosive synchronization \cite{gomez2011explosive,zhang2015explosive,zhou2015stability}. In addition to this abrupt phase transition, a range of coupling strength values depicts a bistability behavior, where for the same parameters there is a coexistence of both synchronized and non-synchronized states. In the end, it is shown that the dynamical mechanisms of this bistability are described by a saddle-node bifurcation and a boundary crisis  \cite{ott2002chaos, grebogi1986critical,grebogi1987critical}. These results are published in the article ``\textit{Mechanism for explosive synchronization of neural networks.}" Physical Review E 100.5 (2019): 052301, \cite{boaretto2019mechanism}.  

\section{Network properties}

A network with $N=10000$ Chialvo neurons is considered. The neurons are coupled using a complex network generated with the Newman-Watts route (with $n_0=40000$ connections for $p_\mathrm{nw} = 0$). The dynamics of the $i$-th neuron is described as
\begin{eqnarray}
x_{i,t+1} &=& x_{i,t}^2\exp(y_{i,t}-x_{i,t})+k_i+\frac{\varepsilon}{\bar{n}}\sum_{j=1}^N e_{i,j}x_{j,t}, \label{chialvo1}\\
y_{i,t+1} &=& a y_{i,t}-bx_{i,t}+c,\label{chialvo2}
\end{eqnarray}
where $x_{i,t}$ and $y_{i,t}$ are the activation and recovery variables, respectively as defined in Chapter \ref{chap:modelos}. $a=0.89$, $b=0.6$, and $c=0.28$ are constant parameters and $k_i$ acts like an addictive disturbance in the neuron that affects both the amplitude of oscillation and the frequency of neurons.  $\bar{n}$ is equivalent to the average number of connections on the network, $e_{i,j}$ the element of the connection matrix, and $\varepsilon$ is the coupling parameter. As shown in Chapter \ref{chap:modelos}, for this set of parameters this model depicts periodic spikes, different from chaotic oscillators, the synchronization state of periodic oscillators is achieved for any $\varepsilon> 0$. This synchronization can be avoided with the introduction of a dissimilitude parameter $\sigma=0.001$. In this sense, for each neuron, the parameter $k_i$ is randomly selected between $ [0.03,\,0.03+\sigma]$ imposing a network with non-identical neurons.

\begin{figure}[t]
\begin{center}
\includegraphics[width=\columnwidth]{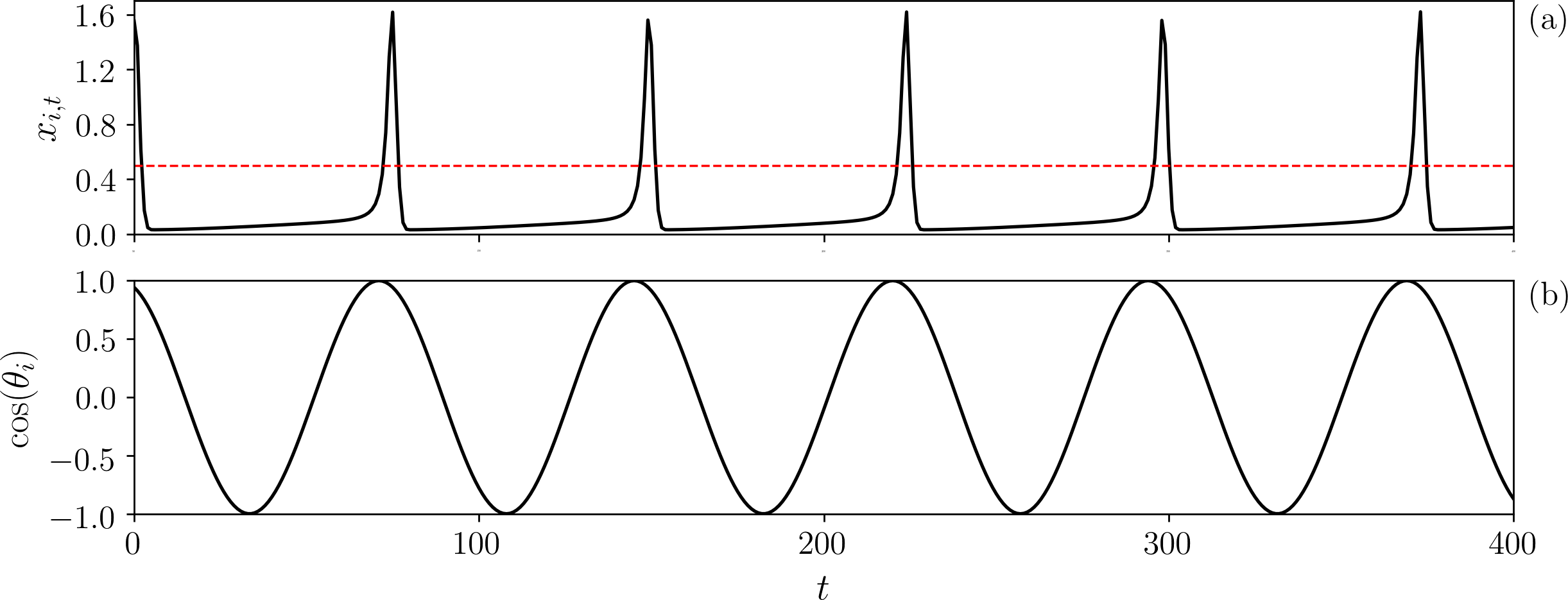}
\caption[Phase association in the Chialvo model.]{\textbf {Phase association in the Chialvo model.} (a) Dynamics of the Chialvo model for $a=0.89$,  $b=0.6$, $c=0.28$, and $k_i=0.03$. The red dashed line delimits $x_0 =0.5$. (b) The cosine of the phase of the neuron, where a spike begins $\theta_i$ is $2\pi$ multiple and $\cos(\theta_i)=1$.}
\label{fig:fase_chialvo}
\end{center} 
\end{figure}

To compute the occurrence of the spikes of each neuron, and further define a phase $\theta_i$ in the Chialvo model, a threshold is defined with a value of $x_0=0.5$, which means that a spike occurs when $x_{i,t}$ crosses $x_0$ (with positive derivative). After computing all the spikes of all the neurons, with Eq. (\ref{eq:phase}), it is possible to evaluate $\theta_i$, and hence, the phase synchronization with Eq. (\ref{eq:parmed}) using $t_\mathrm i=150000$ and $t_\mathrm f=200000$. Figure \ref{fig:fase_chialvo} depicts at panel (a) the time-evolution of $x_{i,t}$ for $\varepsilon=0$, where the red dashed line delimits $x_0$. At panel (b) is computed the $\cos (\theta_i)$, at each spike the phase is $2\pi$ multiple and $\cos(\theta_i)=1$. 

\section{Results}

\begin{figure}[t!]
\centering
\includegraphics[width=.9\columnwidth]{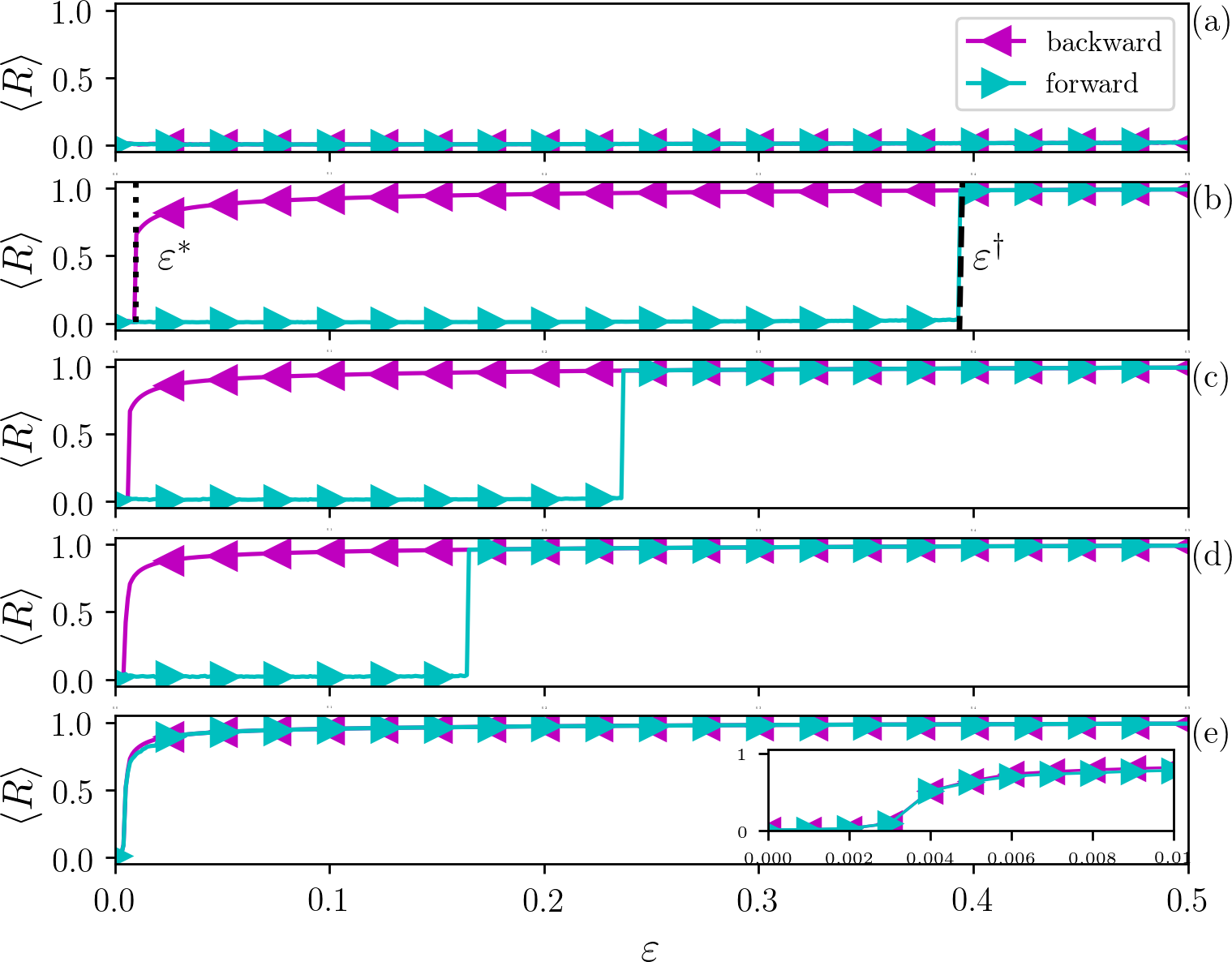}
\caption[The role of the topology in the synchronization behavior.]{\textbf{The role of the topology in the synchronization behavior.} $\langle R\rangle$ as a function of the coupling $\varepsilon$ for a fixed $\sigma=0.001$ and $5$ $p_\mathrm{nw}$ values. The coupling strength is evolved adiabatically ($\delta \varepsilon=0.001$) in two different directions: forward (increment) and backward (decrement). (a) For $p_\mathrm{nw}=0.10$, the network does not depict phase synchronization. (b) For a larger value of $p_\mathrm{nw}=0.15$, a large interval is observed where explosive synchronization is likely to occur, the vertical dotted (dashed) line represents the transition points of $\varepsilon^*$ ($\varepsilon^\dagger$). (c) and (d) for $p_\mathrm{nw}=0.25$ and  $p_\mathrm{nw}=0.35$, explosive synchronization still may exist but for a smaller range of $\varepsilon$. (e) For larger values of $p_\mathrm{nw}=0.45$ the transition from unsynchronized to phase-synchronized state is smooth. The inner panel displays a  magnification of the transition region.}
\label{fig:parmed_p}
\end{figure}

The topological effects of the probability of connection $p_\mathrm{nw}$ are presented in Fig.~\ref{fig:parmed_p}, by evaluating the Kuramoto order parameter $\langle R\rangle$ as a function of $\varepsilon$ for different values of $p_\mathrm{nw}$. To produce this results, it is considered a continuation process as follows: for $\varepsilon=0$ a random initial conditions is selected for the neurons in the network. After evaluating $\langle R\rangle$ for this $\varepsilon$, the coupling is increased adiabatically from $\varepsilon$ to $\varepsilon + \delta\varepsilon$ (with $\delta \varepsilon=0.001$). For this new $\varepsilon$, the initial conditions are the final conditions of the previous $\varepsilon$, without rebooting the system. The increment is made until $\varepsilon=0.5$. After that, it is made the adiabatic decrease ($\delta \varepsilon=-0.001$) until $\varepsilon=0$. Therefore, each $\varepsilon$ value is studied two times, in the forward direction (positive variations) and backward direction (negative variations). If at each $\varepsilon$ value the system was restarted, i.e. without the continuation method previously described, there would be a probability that the chosen initial condition would fall into one of two possible states. In this situation, it has been found that the backward direction is more likely than the forward direction.

At panel (a) $p_\mathrm{nw}=0.1$ the network is kept in the non-synchronized state in both directions due to the small number of non-local connections which forbids the phase synchronization. Panel (b) $p_\mathrm{nw}=0.15$, for the forward direction (cyan triangles), as the coupling is increased there is a non-synchronized region ($\langle R\rangle \approx 0$) which is kept until the critical coupling $\varepsilon=\varepsilon^\dag$ when the network transition abruptly to the synchronized state ($\langle R\rangle \approx 1$). On the other hand, for the backward direction (magenta triangles), the network starts at the synchronized state, which is maintained until $\varepsilon=\varepsilon^* \ne \varepsilon^\dag$. Hence, the forward and backward directions are different, creating a hysteretic behavior that characterizes a bistability region delimited $\varepsilon^*<\varepsilon<\varepsilon^{\dag}$, where at this range it is possible to achieve both synchronized and non-synchronized states depending on the initial conditions. Panels (c) and (d) where $p_\mathrm{nw}=0.25$ and $0.35$, respectively, present the same behavior of panel (b), but for distinct values of critical couplings $\varepsilon^*,\, \varepsilon^\dag$. At panel (e) $p_\mathrm{nw}=0.45$ the transition is not explosive, for this case, the phase transition occurs smoothly in both directions, which can be easily seen in the magnification at the inner panel. 

\begin{figure}[t]
\begin{center}
\includegraphics[width=0.95\columnwidth]{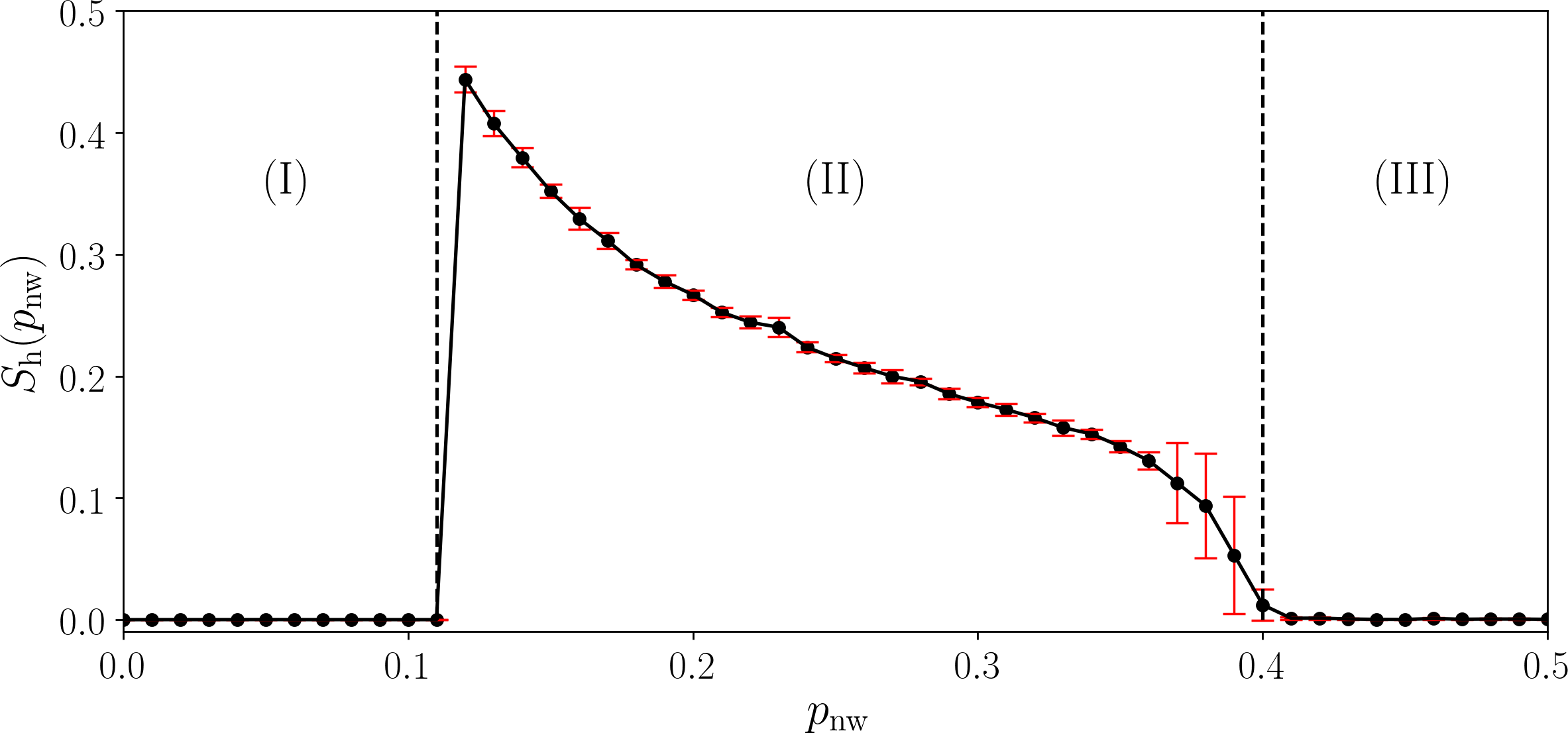}
\caption[Numerical results for ${S_\mathrm{h}}$ as a function of ${p_\mathrm{nw}}$.]{\textbf{Numerical results for $\mathbf{S_\mathrm{h}}$ as a function of $\mathbf{p_\mathrm{nw}}$.} The vertical lines delimit three different scenarios: (I) the network does not show synchronization and only a chaotic-asymptotic state is observed; (II) the network depicts explosive synchronization as the result of attraction basin changes. At the same time, the network displays a hysteretic loop; (III) the network shows a smooth transition from the unsynchronized state to the synchronized state. The error bars are the standard deviation over $50$ samples.}
\label{fig:area}
\end{center} 
\end{figure}

The synchronization features caused by the topological changes can be summarized with the definition of the hysteretic area
\begin{equation}
    S_\mathrm{h}(p_\mathrm{nw})=
    \left | \int_\mathrm{forward}\langle R\rangle (\varepsilon,p_\mathrm{nw})\,d\varepsilon - \int_\mathrm{backward}\langle R\rangle (\varepsilon,p_\mathrm{nw})\,d\varepsilon \right |.
 \label{sXpnl}
\end{equation}
Figure~\ref{fig:area} presents $S_\mathrm h$ as a function of $p_\mathrm{nw}$ considering $0\leq\varepsilon\leq 0.5$, the error bars are the standard deviation over $50$ different simulations. As discussed in Fig.~\ref{fig:parmed_p}, it can be identified three different regions. The region (I) $0\leq p_\mathrm{nw}\leq 0.1$ with $S_\mathrm h \approx 0$ since the network does not synchronize. The region (II) $0.11 \leq p_\mathrm{nw} \lesssim 0.4$ with $S_\mathrm h > 0$ where the network presents its bistability behavior, and $S_\mathrm h$ decreases as $p_\mathrm{nw}$ increases. Region (III) $\varepsilon>0.4$ with $S_\mathrm{h}=0$ different from region (I), $S_\mathrm h\approx 0$ due the smoothly transition in both directions.

\begin{figure}[t]
    \centering
    \includegraphics[width=\columnwidth]{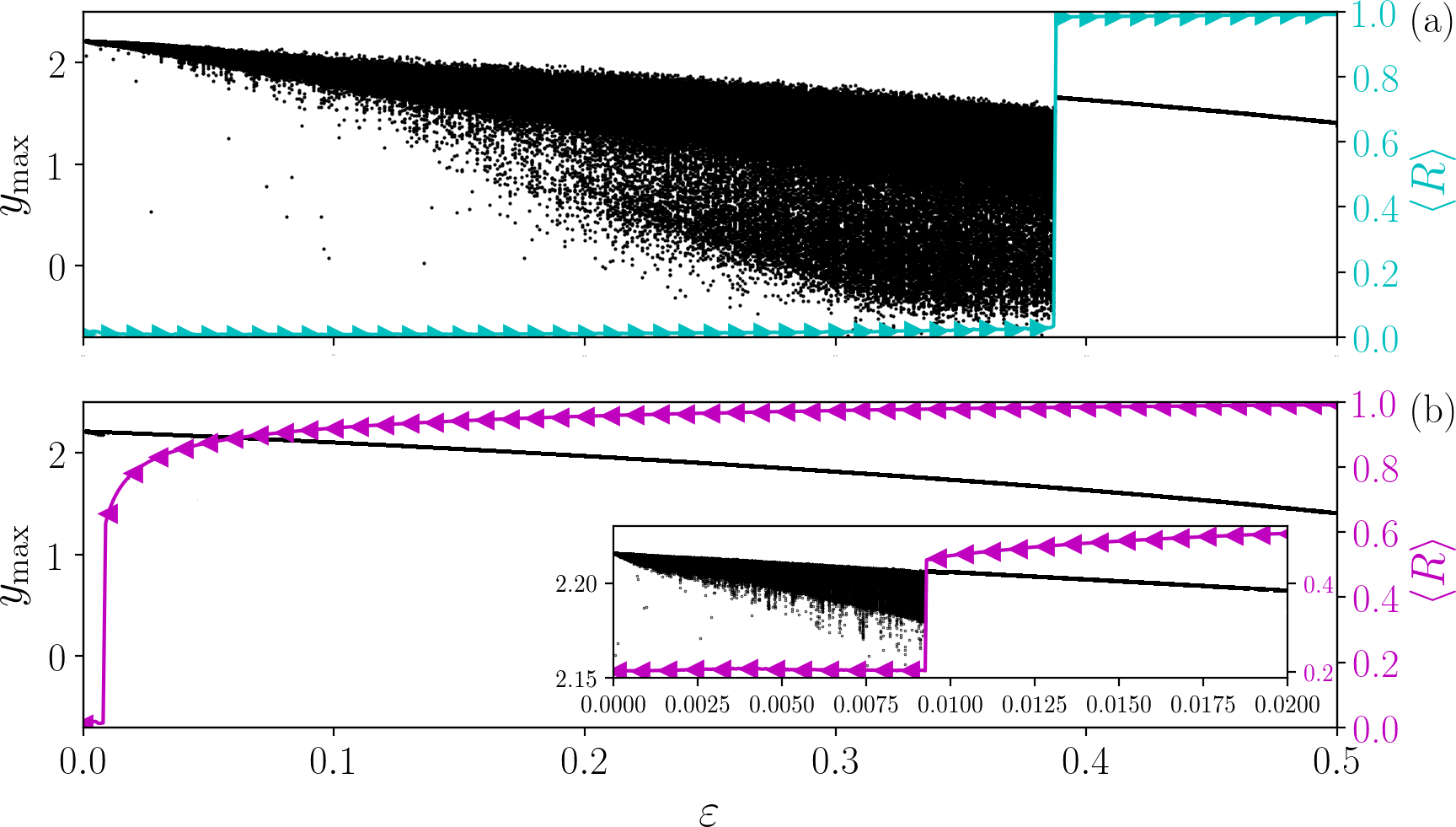}
    \caption[The explosive synchronization occurs due to a chaotic transition.]{\textbf{The explosive synchronization occurs due to a chaotic transition.} Bifurcation diagram of the $y_\mathrm{max}$ of an arbitrary neuron (left scale), $\langle R\rangle$ of the network as a function of $\varepsilon$ (right scale). (a) The coupling is adiabatically increased, the chaotic-non-synchronized state is kept until $\varepsilon=\varepsilon^\dag\approx 0.4$ where the chaotic attractor loses stability to a periodic orbit. (b) The coupling is adiabatically decreased (from $\varepsilon=0.5$ to $0$), the periodic-synchronized state exists for $\varepsilon\approx\varepsilon^*\approx 0.01$ (inner panel) in which the non-synchronized state is reached.}
    \label{fig:chaos_chialvo}
\end{figure}

To investigate the bistability region, Fig. \ref{fig:chaos_chialvo} explores the neuron dynamics and the synchronization behavior of the network for a fixed probability of $p_\mathrm{nw}=0.15$, where it is computed the maxima of $y_\mathrm{max}$ variable of one arbitrary neuron (left scale), comparing with the $\langle R\rangle$ (right scale) of the network. In the forward direction (a), the dissimilitude among neurons allows a chaotic-non-synchronized state, until $\varepsilon=\varepsilon^\dag$, where the chaotic attractor is subtly replaced by a stable periodic orbit for $\varepsilon>\varepsilon^\dag$. In the backward direction (b), the synchronized state is maintained from $\varepsilon = 0.5$ until $\varepsilon=\varepsilon^*$, after that, for $\varepsilon<\varepsilon^*$ (inner panel) the coupling is not sufficiently strong enough to keep the non-identical neurons in the same period, consequently, the synchronized state is lost. Hence, abrupt transitions in the phase synchronization occurs due to the chaotic transitions when the coupling strength $\varepsilon$ reaches a critical value, $\varepsilon^\dag$ in the forward direction, and $\varepsilon^*$ in the backward direction. In the next section, it is investigated the dynamical mechanisms of the chaotic transitions of both directions.

\subsection{The dynamical mechanism of the loss of stability of the chaotic attractor}

\begin{figure}[t]
\begin{center}
\includegraphics[width=0.95\columnwidth]{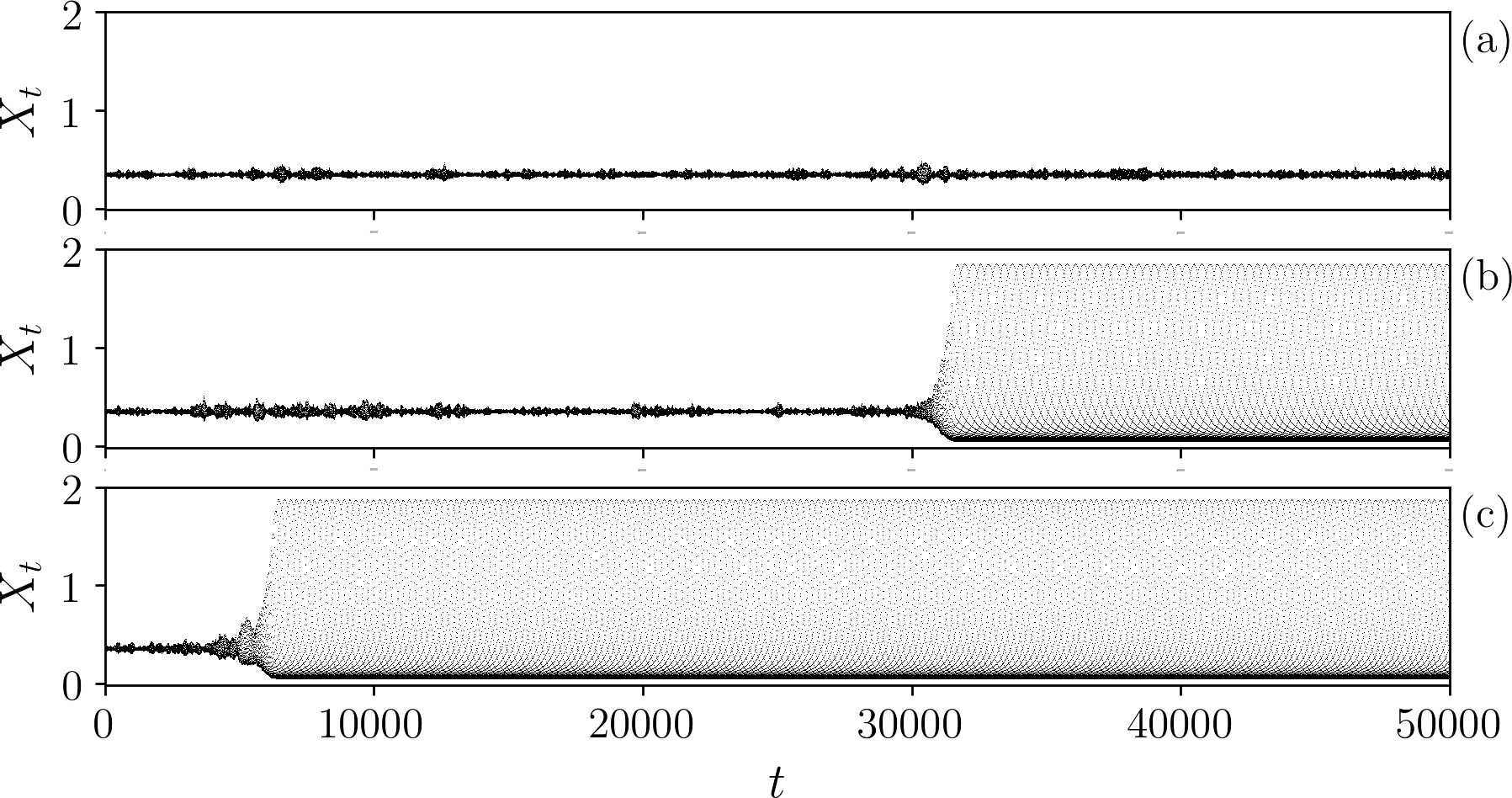}
\caption[Illustration of the chaotic transient in a crisis.]{\textbf{Illustration of the chaotic transient in a crisis.} Time evolution of the mean-field of the network $X_t=(1/N)\sum_{i=1}^N x_{i,t}$ with a initial condition inside the chaotic saddle, where (a) $\varepsilon\lesssim \varepsilon^{\dag}$, (b) $\varepsilon \gtrsim \varepsilon^{\dag}$ and (c) $\varepsilon > \varepsilon^{\dag}$.}
\label{fig_crise_1}
\end{center} 
\end{figure}

The physical mechanism responsible for the abrupt transition in the forward direction, where the network changes abruptally from the non-synchronized to the synchronized state which occurs for $\varepsilon=\varepsilon^\dag(p_\mathrm{nw})$, can be described in terms of a route to chaos called boundary crisis \cite{ott2002chaos,grebogi1986critical,grebogi1987critical}. As described in Chapter \ref{chap:sistemas}, a boundary crisis is characterized by the collision of a chaotic attractor with an unstable periodic orbit \cite{ott2002chaos}. In this sense, the attractor loses stability when the control parameter reaches a critical point $\varepsilon = \varepsilon^{\dag}$. For $\varepsilon>\varepsilon^{\dag}$ the chaotic attractor no longer exists but is replaced by a chaotic transient. If it is considered an initial condition in the basin of attraction (which exists only for $\varepsilon<\varepsilon^{\dag}$) and increase the $\varepsilon \gtrsim \varepsilon^{\dag}$, for a period, the orbit is similar to the chaotic attractor, however, after a chaotic transient, called $\tau_\mathrm{crisis}$, the system loses this behavior \cite{ott2002chaos}. The Fig.~\ref{fig_crise_1} illustrates this chaotic transient where it is calculated the mean-field of the networks, defined as
\begin{equation}
    X_t = \frac{1}{N}\sum_{i=1}^N x_{i,t}.
\end{equation}
With the initial condition of the network being chosen within the chaotic saddle ($\varepsilon\lesssim\varepsilon^{\dag}$). In panel (a) $\varepsilon\lesssim\varepsilon^{\dag}$ the mean-field shows an incoherent low amplitude behavior, related to the out-of-sync state of the network. At panel (b) $\varepsilon\gtrsim\varepsilon^{\dag}$ the chaotic attractor no longer exists, and after a transient time, $\tau_\mathrm{crisis} \approx 3\times 10^4$ the amplitude of the mean-field increases, showing periodic behavior related to the synchronization state. In panel (c) $ \varepsilon>\varepsilon^{\dag}$ (greater than in panel (b)), the coupling value is already far from the critical point, making the transient time extremely short $ \tau_\mathrm{crisis} \approx 0.5 \times 10^4 $.

\begin{figure}[t]
\begin{center}
\includegraphics[width=0.95\columnwidth]{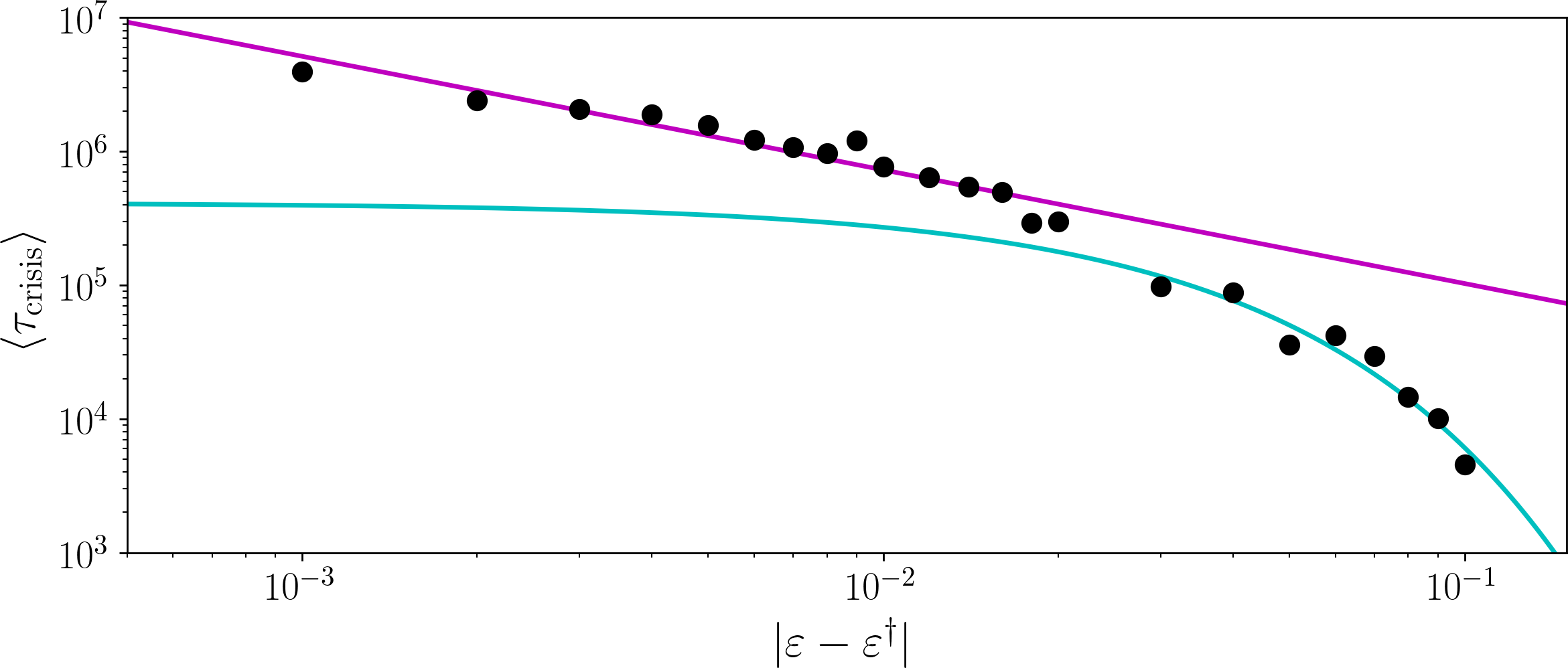}
\caption[Evidence of boundary crisis in the network.]{\textbf{Evidence of boundary crisis in the network.} Average chaotic transient $\langle \tau_\mathrm{crisis}\rangle$ as a function of $|\varepsilon-\varepsilon^\dag|$ for the system initialized at the chaotic saddle $\varepsilon\lesssim\varepsilon^\dag$ (but, iterated with $\varepsilon>\varepsilon^\dag$). The average is evaluated over $50$ distinct initialization. For values sufficient close to the critical point  $\varepsilon^\dag$ the chaotic transient decays with a power-law  $\langle\tau_\mathrm{crisis}\rangle \sim |\varepsilon-\varepsilon^\dag|^{-\kappa}$ with $\kappa=0.85$.}
\label{fig_crise_2}
\end{center} 
\end{figure}

In order to evidence the post-crisis behavior, Fig. \ref{fig_crise_2} presents how this chaotic transient $\langle \tau_\mathrm{crisis} \rangle$ changes as a function of $|\varepsilon-\varepsilon^\dag|$. To do so, firstly an initial condition is evolved from $\varepsilon=0$  until $\varepsilon<\varepsilon^\dag$, to ensure that the initial condition belongs to the chaotic saddle. After that, for this initial condition ($\varepsilon<\varepsilon^\dag$), it is computed how much time the system spends until reach the periodic synchronized state ($R=0.9$) for values $\varepsilon>\varepsilon^\dag$. The $\langle \cdot \rangle$ represents the average over $50$ distinct orbits initialized in the chaotic saddle, using a representative value of $p_\mathrm{nw}=0.15$. In a crisis, it is expected a power-law decay $\langle\tau_\mathrm{crisis}\rangle \sim |\varepsilon-\varepsilon^\dag|^{-\kappa}$, where $\kappa$ is namely the crisis exponent, and $\kappa>0.5$ for multidimensional systems \cite{ott2002chaos,grebogi1986critical,grebogi1987critical,kubo2008crisis}. For sufficient close values $|\varepsilon-\varepsilon^\dag|<2\times10^{-2}$ the power-law decay occurs with $\kappa=0.85$. For $|\varepsilon-\varepsilon^\dag|>2\times10^{-2}$ a power-law decay is still expected, but due the numerical impossibility of start the system exactly in the saddle point $\varepsilon^\dag$ the power-law decay is replaced by an exponential decay.

\subsection{The dynamical mechanism of the loss of stability of the phase-synchronized attractor}

\begin{figure}[t!]
\begin{center}
\includegraphics[width=0.85\columnwidth]{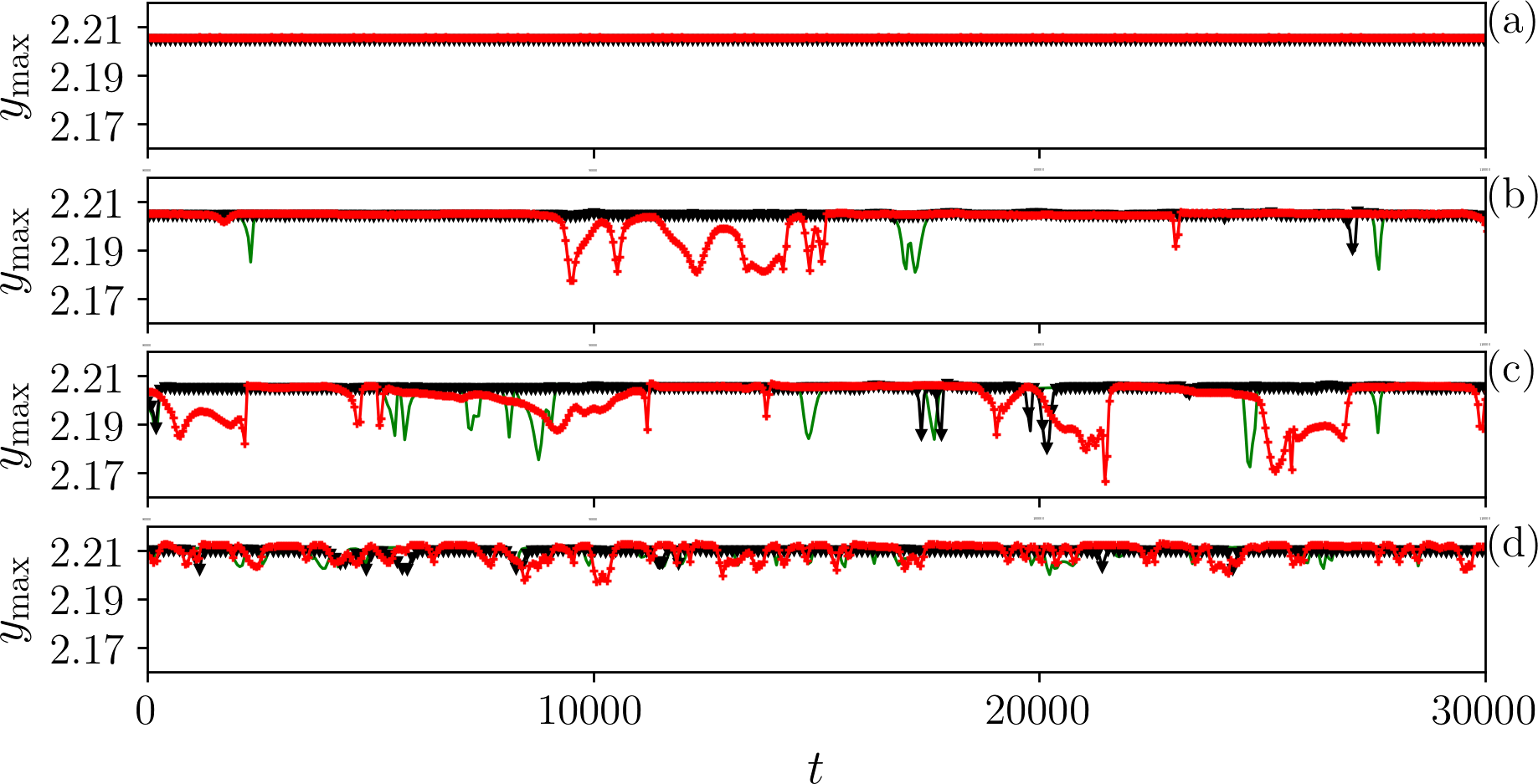}
\caption[Illustration of intermittent chaotic events.]{\textbf{Illustration of intermittent chaotic events.} Maxima values of the variable $y_{i,t}$ for three random neurons in the network. (a) $\varepsilon>\varepsilon^*$ where $y_\mathrm{max}$ is constant for the three neurons (periodic dynamics). (b), (c) and (d) $\varepsilon<\varepsilon^*$, (in descending order), the neurons present some escape events from the periodic orbit which occur intermittently.}
\label{fig_sela_no_1}
\end{center} 
\end{figure}

For the backward direction, the network transition from the synchronized periodic state to a non-synchronized chaotic one, that is, the network presents a periodic dynamics for $\varepsilon<\varepsilon^*(p_\mathrm{nw})$, the periodic orbit loses stability to a chaotic one due a saddle-node bifurcation \cite{ott2002chaos}. In such transition, for $\varepsilon\lesssim \varepsilon^*$ intermittency can be detected in the traces of periodic orbit due to the \textit{quasi}-stable character of the synchronized state before the saddle-node bifurcation of the periodic-synchronized state \cite{ott2002chaos}. Figure~\ref{fig_sela_no_1} presents an illustration of the intermittent chaotic interruptions for $\varepsilon<\varepsilon^*$. It is calculated the maxima values of the $y_{i,t} = y_\mathrm{max}$ variable for three random neurons in the network. At panel (a) $\varepsilon>\varepsilon^*$, as expected, the three neurons present periodic dynamics. In the other panels (b), (c), and (d), $\varepsilon<\varepsilon^*$ (in descending order), the neurons exhibit some escapes of the periodic orbits. The greater the distance between $\varepsilon$ and $\varepsilon^*$, the smaller is the time between the escapes. It is possible to define a characteristic time of intermittent events $\tau_\mathrm{int}$ where,
\begin{equation*}
    \lim_{\varepsilon\rightarrow\varepsilon^*}\tau_\mathrm{int} = +\infty.
\end{equation*}
As discussed in Chapter \ref{chap:sistemas}, if the transition is characterized by a saddle-node bifurcation, the occurrence of the escapes decays with a power-law exponent \cite{ott2002chaos} in which
\begin{equation}
    \langle \tau_\mathrm{int}\rangle \sim |\varepsilon-\varepsilon^*|^{-1/2}.
\end{equation}

\begin{figure}[t!]
\begin{center}
\includegraphics[width=0.95\columnwidth]{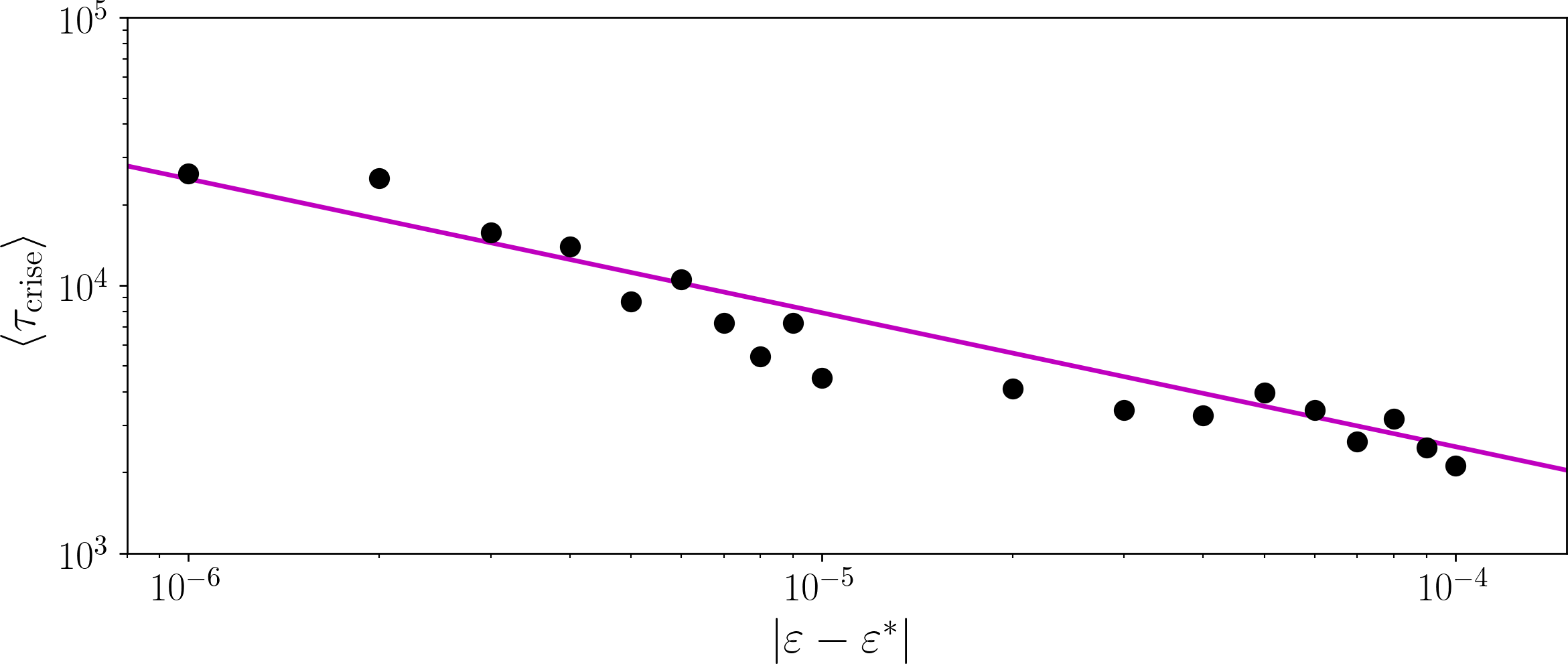}
\caption[Evidence of saddle node in the network.]{\textbf{Evidence of saddle node in the network.} Average intermittent time of the saddle-node bifurcation $\langle \tau_\mathrm{int}\rangle$ as a function of the distance from the critical coupling $|\varepsilon-\varepsilon^*|$ for different $50$ initial conditions. The magenta line represents the theoretic curve  $\propto |\varepsilon-\varepsilon^*|^{-1/2}$.
}
\label{fig_sela_no_2}
\end{center} 
\end{figure}

To study the mean time of the intermittent escapes $\langle \tau_\mathrm{int}\rangle$, it is computed the times in which $y_{i,\mathrm{max}}$ of the $i$-th neuron deviates from two standard deviations below the mean. In Fig.~\ref{fig_sela_no_2} it is presented the $\langle \tau_\mathrm{int}\rangle$ value  is an average for $50$ different initial conditions, as function of $|\varepsilon-\varepsilon^*|$. The magenta line represents the theoretic curve $\propto |\varepsilon-\varepsilon^*|^{-1/2}$, characterizing a saddle node bifurcation due to the intermittency transition of type I \cite{ott2002chaos}. 

\section{Discussions}

Throughout this chapter, it was explored different features of synchronization using a Newman-Watts complex network composed of non-identical spiking neurons. The dissimilitude is necessary to avoid the phase synchronization behavior detected in periodic neurons. While a network with lower connectivity no longer reaches a synchronized state $p_\mathrm{nw}\leq 0.1$, as the number of connections increases the synchronized state is achieved in different ways. For $0.1<p_\mathrm{nw}\lesssim0.4$ the transition is characterized by an explosive synchronization, and $p_\mathrm{nw}>0.4$ there is a continuous phase-transition scenario. In particular, the presence of the explosive synchronization is accompanied by a hysteresis loop on the network dynamics as the coupling parameter is adiabatically increased and decreased. It was shown that the abrupt synchronization transitions are associated with routes to chaos. The dynamical mechanisms for the bistability region, are given in terms of a saddle-node bifurcation and a boundary crisis \cite{ott2002chaos}. 

At last, different from the original findings of Gomez \textit{et al.} \cite{gomez2011explosive} where the explosive synchronization occurs due to the existence of hubs (sites with a high degree of connectivity), a characteristic of the topology of the scale-free network, the explosive synchronization can be achieved with a set of non-identical periodic neurons coupled in a complex network (in a small-world regime). It is concluded that this scenario is sufficient for the presence of explosive synchronization and opens a new approach with application in neural networks.

\chapter{Conclusions and future perspectives}\label{chap:cons}

\initial{T}{hroughout} this thesis, the cooperative behavior of dynamical systems which mimics the membrane potential of neuronal cells has been studied, in particular, it is explored how the individual dynamics of neurons affect the global behavior of phase synchronization of the network. In this sense, different approaches have been used with distinct neuronal models and coupling schemes. 
 %

The main results of this thesis have been presented in Chapter \ref{chap:role} where the emergence of phase synchronization in weakly-coupled neurons is explored. This phenomenon occurs when HH$\ell$ bursting neurons that individually exhibit periodic behavior are coupled. Despite the mechanisms of the non-monotonic transitions are still unknown, one of the possible explanations for this observation is due to the interplay of the neural activity patterns with the synaptic current because the collective dynamics induced by the synaptic currents are not strong enough to dismiss the periodicity of the neurons, but is sufficient to lead the network to a phase synchronized state (local maximum of phase synchronization). The phase synchronization is lost with the increment of the coupling parameter since the chaoticity induced by the synaptic current desynchronizes the network but is recovered for higher values of coupling. 

Particularly interesting, the results have shown a possibility to control the phase synchronization in weakly-coupled neurons which occurs by varying parameters of the model allowing the individual neuron dynamics to migrate from a regular to chaotic behavior and, consequently, changing the entire transition scenario of the phase synchronization even when just tiny changes in one ion-conductance is performed. In this sense, the ion conductance's variations can be understood as the blocking or activation of ion channels, and small changes in one specific ion conductance may act as a catalyst to the second conductance change. This is one of the working mechanisms of drugs used to normalize neuronal functioning to unhealthy neural behaviors. Such results may be of great interest in researches on drugs to control individual and collective behavior on the brain. This phenomenon has been explored with different approaches, by varying different parameters of the model and disturbing the neuron dynamics with the application of an external pulsed current. To corroborate these findings, the same non-monotonic transition has been found in another neuronal model under the same dynamical conditions, and in complex networks with different topologies. The hypothesis that such an effect can be found in networks of nonlinear oscillators models which not reproduces the neuronal dynamics is not discarded.
 

Moreover, the same HH$\ell$ model presents a range of parameters where a bistable state has been found. Hence, two identical neurons initialized with distinct initial conditions depict different dynamical behavior. The role of bistability in the phase synchronization of a network has been explored in Chapter \ref{chap:bistability}. The simplest coexistence of neurons in different states can lead the network to different states of phase synchronization, which depends only on the initialization of the system. In general, the existence of bistability at the network delays the occurrence of phase synchronization, since phase synchronization is only possible if the synaptic current is sufficiently strong to induce a transition from the less-stable state to a final and unique state. Otherwise, a local synchronization is achieved where neurons initialized in their respective states may synchronize with neurons in the same state preventing a global synchronized stable state. Furthermore, during the synaptic-induced transitions, interesting dynamical phenomena are observed such as chimera states and anti-phase synchronization.

Lastly, in Chapter \ref{chap:mechanism}, a discrete model of spiking neurons coupled in a small-world network has been studied. The dissimilitude among neurons allows a chaotic-non-synchronized regime which is kept with the adiabatic increment of the coupling until reaches a critical coupling and abruptly synchronizes. On the other hand, the adiabatic decrements of the coupling maintain the network in a periodic-synchronized state which is lost for a different critical coupling value, characterizing a hysteretic loop. This phenomenon, called explosive synchronization, has been explained with Kuramoto oscillators coupled in a scale-free topology of connections with the existence of hubs (one of the characteristics of this topology). However, the heterogeneity produced by the dissimilitude of the neurons is sufficient to achieve such a transition. In conclusion, the dynamical mechanisms for the bistability region are given in terms of a boundary crisis where the chaotic-non-synchronized loses stability when the chaotic attractor collides with a non-stable fixed point and the periodic-synchronized state loses stability to saddle-node bifurcation. 

The results of this thesis offer various directions for further research. It will be worthwhile to describe the mechanisms of the non-monotonic synchronization reported in neurons with individual non-chaotic behaviors but sensitive to transition to chaotic behaviors when coupled. In this sense, the use of the master stability functions \cite{pecora1998master} and other tools used in data analysis \cite{chollet2017deep,bandt2002permutation,marwan2007recurrence} allow the achievement of a general description to understand the dynamical properties of the synaptic current, which allows the non-monotonic synchronization. Also, this may offer a way of approaching similar observations in more general oscillators' systems. On the other hand, from the point of view of the bistability found in the neuron model, the analysis of the network with noise application can present an interesting result and will be explored in a future opportunity, since it is known that the noise may disrupt the synchronized state. However, the noise can induce transitions among neuronal states facilitating a possible state of phase synchronization. Lastly, the explosive synchronization found in the non-identical spiking neurons of Chialvo, future works will be devoted to the generalization of this behavior to other spiking neuronal models and oscillators to generalize this phenomenon.

\clearpage





\printbibliography[title=Bibliography,heading=bibintoc]


\appendix

\chapter{Kuramoto oscillators} \label{cap_a}

\initial{T}{he} Kuramoto model consist of the ensemble of $N$ oscillators with phase $\theta_i$, where the dynamics of each oscillator depends on a natural frequency and a nonlinear coupling \cite{kuramoto1984cooperative,acebron2005kuramoto}. The evolution of the phase of each oscillator is described by
\begin{equation} \label{Eq: kura_1}
    \frac{d\theta_i}{dt} = \omega_i + \frac{1}{N}\sum_{j=1}^N\varepsilon_{ij}\sin(\theta_j-\theta_i), \;\;\; i=1,2,\cdots,N,
\end{equation}
in which $\omega_i$ is the natural frequency of each oscillator, and $\varepsilon_{ij}$ is the parameter which controls the magnitude of the coupling between the $i$-th and the $j$-th oscillator. In the particular case of global coupling where $\varepsilon_{ij} = \varepsilon \; \forall \; i,j$,
\begin{equation} \label{eq_kura_2}
    \frac{d\theta_i}{dt} = \omega_i + \frac{\varepsilon}{N}\sum_{j=1}^N\sin(\theta_j-\theta_i), \;\;\; i=1,2,\cdots,N,
\end{equation}
where $\varepsilon$ is the intensity of the coupling. The natural frequencies \{$\omega_i$\} are distributed following a  probability density function $\varkappa(\omega)$, which is assumed to be symmetric for a given frequency $\Omega$, that is,
\begin{equation}
    \varkappa(\Omega-\omega) = \varkappa(\Omega+\omega).
\end{equation}
Kuramoto have defined an order parameter to quantify the phase synchronization of the oscillators
\begin{equation} \label{eq_parmed} 
    R(t)e^{i\psi(t)} = \frac{1}{N}\sum_{j=1}^N e^{i\theta_j(t)},
\end{equation} 
where $R(t)$ is the modulus of the order parameters and $\psi(t)$ is the average circular frequency of the oscillators. Multiplying both sides of Eq. (\ref{eq_parmed}) by $e^{-i\theta_i}$ 
\begin{equation*}
    Re^{i(\psi-\theta_i)} = \frac{1}{N}\sum_{j=1}^N e^{i(\theta_j -   \theta_i)},
\end{equation*}
and taking the imaginary part
\begin{equation} \label{eq_im}
    R\sin(\psi-\theta_i) = \frac{1}{N}\sum_{j=1}^N \sin(\theta_j-\theta_i).
\end{equation}
Replacing Eq.(\ref{eq_im}) at Eq. (\ref{eq_kura_2}), 
\begin{equation}  \label{eq_kura_3}
    \frac{d\theta_i}{dt} = \omega_i + \varepsilon R\sin(\psi-\theta_i), \;\;\; i=1,2,\cdots,N,    
\end{equation}
that is, the phase of each oscillator is attracted by the frequency $\psi$ and the intensity of the coupling is proportional to the order parameter $R$.

The Eq.  (\ref{eq_parmed}) can be rewritten as \cite{acebron2005kuramoto}
\begin{equation}
    R e^{i\psi} = \frac{1}{N}\sum_{j=1}^N e^{i\theta_j} = \int_{-\pi}^{\pi} e^{i\theta} \left ( \frac{1}{N} \sum_{j=1}^N \delta(\theta - \theta_j) \right)d\theta.
\end{equation}
Considering infinity oscillators $N\rightarrow \infty$,
\begin{equation} \label{eq_parmed_2}
     R e^{i\psi} = \int_{-\pi}^{\pi} \int_{-\infty}^{+\infty} e^{i\theta} \rho(\theta,\omega,t) \varkappa(\omega) d\omega d\theta.
\end{equation}
in which the system can be described in terms of the probability density $\rho(\theta,\omega,t)$. Where $\rho(\theta,\omega,t)d\theta$ represents the fraction of oscillators with frequency $\omega$ which are found between $\theta$ and $\theta+d\theta$ in a given time instant $t$.

Since the oscillators are moving in a unitary circle $[-\pi, \pi]$ with angular velocity $\dot{\theta_i}$ described by the Eq. (\ref{eq_kura_3}), the probability density $\rho$ obeys the continuity equation
\begin{equation} \label{densidade}
    \frac{\partial \rho}{\partial t} + \frac{\partial}{\partial \theta}(\upsilon \rho) = 0,
\end{equation}
where $\upsilon$ is the angular velocity  $\upsilon=\dot{\theta}=\omega+\varepsilon R\sin(\psi-\theta)$. The probability density presents the normalization condition described by 
\begin{equation} \label{densidade_2}
    \int_{-\pi}^{\pi} \rho(\theta,\omega,t)d\theta = 1.
\end{equation}

One of the possible solutions for this system is to consider $R=0$. This case corresponds to the incoherent solution where the infinite oscillators are uniformly distributed around the circle $ \rho = 1/(2 \pi) $. Thus, we say that for $ R = 0 $ we have complete desynchronization, in which for each $i$-th oscillator with a phase $\theta_i=\theta$ there will be a $j$-th oscillator in anti-phase $ \theta_j = \theta + \pi$. For the case of complete synchronization, where all oscillators have the same phase, that is, $ \theta_i = \psi $, and, consequently, the module $ R = 1 $ due
\begin{equation*}
 |R(t)|\cdot|e^{i\psi(t)}| = \left |\frac{1}{N}\sum_{j=1}^N e^{i\theta_j(t)} \right | = \left | \frac{1}{N}(e^{i\theta_1(t)}+ e^{i\theta_2(t)} + \cdots + e^{i\theta_N(t)}) \right | = \frac{1}{N}\cdot N|e^{i\psi(t)}| = 1.  
\end{equation*}

To study the cases where $0<R<1$, the stationary solution of the Eq. (\ref{eq_kura_3}),
\begin{equation*}
    \frac{d\theta_i}{dt} = \omega_i + \varepsilon R \sin(\psi - \theta_i) = 0, \Longrightarrow \varepsilon R\sin(\theta_i - \psi) = \omega_i.
\end{equation*}
For oscillators that have a natural frequency $ | \omega_i | <\varepsilon R$, these oscillators are locked at $ \theta_i-\psi = \arcsin({\omega_i / \varepsilon R})$. If $ | \omega_i |> \varepsilon R $ the oscillators rotate incoherently around the circle with a probability density $ \rho $ constant over time, that is $ \frac {\partial \rho} {\partial t} = 0 $. Therefore, from Eq. (\ref{densidade}) $ \upsilon \rho = C $ (constant), that is
\begin{equation*}
    \rho =\frac{C}{|\omega-\varepsilon R\sin(\theta_i - \psi)|}, 
\end{equation*}
where $C$ can be evaluated using the normalization condition of Eq. (\ref{densidade_2}),
\begin{equation*}
    \int_{-\pi}^{\pi} \rho(\theta,\omega,t)d\theta = 1, \Rightarrow \int_{-\pi}^{\pi} \frac{d\theta}{|\omega - \varepsilon R \sin(\theta-\psi)|} = \frac{1}{C},
\end{equation*}
being $\omega> \varepsilon R$, it is possible to remove the absolute value 
\begin{equation*}
    \int_{-\pi}^{\pi} \frac{d\theta}{\omega-\varepsilon R\sin (\theta-\psi)}, 
\end{equation*}
and is used the solution 
\begin{equation*}
    \int \frac{dx}{a-b\sin x} = \frac{-2\mathrm{arctg} \left(\frac{b-a\mathrm{tg}(\frac{x}{2})}{\sqrt{a^2-b^2}} \right)}{\sqrt{a^2-b^2}}+C_0,
\end{equation*}
where $a=\omega$, $b=\varepsilon R$, $x=\theta - \psi$ and $dx=d\theta$, so,
\begin{eqnarray*}
    \mathrm{if} \; x &=& +\pi \rightarrow \frac{-2\cdot \left ( \frac{-\pi}{2}\right)}  {\sqrt{\omega^2-
    (\varepsilon R)^2}} = \frac{\pi}{\sqrt{\omega^2-
    (\varepsilon R)^2}}, \\
    \mathrm{if} \; x &=& -\pi \rightarrow \frac{-2\cdot \left ( \frac{\pi}{2}\right)}  {\sqrt{\omega^2-
    (\varepsilon R)^2}} = \frac{-\pi}{\sqrt{\omega^2-
    (\varepsilon R)^2}},
\end{eqnarray*}
hence,
\begin{equation}
    C = \frac{\sqrt{\omega^2-
    (\varepsilon R)^2}}{2\pi},
\end{equation}
therefore, the probability density can be written as
\begin{eqnarray}
    \rho(\theta,\omega) = 
    \begin{dcases}\delta \left [ \theta - \psi - \arcsin\left( \frac{\omega}{\varepsilon R}\right) \right], \;\;\; |\omega|<\varepsilon R \\
    \frac{C}{|\omega - \varepsilon R\sin(\theta- \psi)|}, \;\;\;\;\;\;\;\;\;\;\;\;\mathrm{otherwise.}
    \end{dcases}
\end{eqnarray}
$\rho$ can be replaced using Eq. (\ref{eq_parmed_2}) for $R$,
\begin{equation} 
    \begin{split}
     R & = \int_{-\pi}^{\pi} \int_{|\omega|<\varepsilon R} e^{i\theta} \delta \left [ \theta - \psi - \arcsin\left( \frac{\omega}{\varepsilon R}\right) \right] \varkappa(\omega) d\omega d\theta \;+ \\
     & + \int_{-\pi}^{\pi} \int_{|\omega|>\varepsilon R} e^{i\theta} \frac{C}{|\omega - \varepsilon R\sin(\theta- \psi)|} \varkappa(\omega) d\omega d\theta.
     \end{split}
\end{equation}

And with the symmetry of the probability density of $\rho$
\begin{equation*}
    \begin{split}
    \rho(\theta+\pi,-\omega) & = \frac{C}{|-\omega - \varepsilon R\sin(\theta + \pi - \psi)|} = \frac{C}{|-\omega + \varepsilon R\sin(\theta - \psi)|} =\\
    & = \frac{C}{|-(\omega - \varepsilon R\sin(\theta - \psi))|} = \frac{C}{|\omega - \varepsilon R\sin(\theta - \psi)|} = \rho(\theta,\omega).
    \end{split}
\end{equation*}

Assuming that $\varkappa(\omega)=\varkappa(-\omega)$, and with $\rho(\theta + \pi, -\omega) = \rho(\theta,\omega)$ the second term of the integral (for $|\omega| > \varepsilon R$)
\begin{equation*} 
\begin{split}
      \int_{-\pi}^{\pi} \int_{|\omega|>\varepsilon R} e^{i\theta} \rho(\theta,\omega) \varkappa(\omega) d\omega d\theta & = \underbrace{\int_{-\pi}^{\pi} \int_{-\infty}^{-\varepsilon R} e^{i\theta} \rho(\theta,\omega) \varkappa(\omega) d\omega d\theta}_{\omega = -\omega, \; \theta = \theta + \pi} + \\ & + \int_{-\pi}^{\pi} \int_{\varepsilon R}^{+\infty} e^{i\theta} \rho(\theta,\omega) \varkappa(\omega) d\omega d\theta,
\end{split}
\end{equation*}
\begin{equation*}
\begin{split}
\int_{-\pi}^{\pi} \int_{|\omega|>\varepsilon R} e^{i\theta} \rho(\theta,\omega) \varkappa(\omega) d\omega d\theta & =
    \int_{-\pi}^{\pi} \int_{+\infty}^{\varepsilon R} e^{i\theta} e^{i\pi} \rho(\theta+\pi,-\omega) \varkappa(-\omega) (-d\omega) d\theta +
    \\ & + \int_{-\pi}^{\pi} \int_{\varepsilon R}^{+\infty} e^{i\theta} \rho(\theta,\omega) \varkappa(\omega) d\omega d\theta,
    \end{split}
\end{equation*}
after the substitutions, the first term of the integral gains three signals, one due the $e^{i\pi}=-1$, other from $-d\omega$, and the last from the inversion of the integral limits at $\omega$. Finally, %
\begin{equation*} 
    \begin{split}
      \int_{-\pi}^{\pi} \int_{|\omega|>\varepsilon R} e^{i\theta} \rho(\theta,\omega) \varkappa(\omega) d\omega d\theta  = & -\int_{-\pi}^{\pi} \int_{\varepsilon R}^{+\infty} e^{i\theta} \rho(\theta,\omega) \varkappa(\omega) d\omega d\theta \; + \\ & + \int_{-\pi}^{\pi} \int_{\varepsilon R}^{+\infty} e^{i\theta} \rho(\theta,\omega) \varkappa(\omega) d\omega d\theta = 0.
      \end{split}
\end{equation*}
Consequently, the oscillators with $|\omega|>\varepsilon R$ do not contribute to the synchronization of the system. Remaining
\begin{equation*} 
     R = \int_{-\pi}^{\pi} \int_{|\omega|<\varepsilon R} e^{i\theta} \delta \left [ \theta - \psi - \arcsin\left( \frac{\omega}{\varepsilon R}\right) \right] \varkappa(\omega) d\omega d\theta,
\end{equation*}
and rewriting,
\begin{equation*} 
     R  = \int_{-\varepsilon R}^{\varepsilon R} \varkappa(\omega) d\omega \int_{-\pi}^{\pi} (\cos\theta+i\sin\theta)\delta \left [ \theta - \psi - \arcsin\left( \frac{\omega}{\varepsilon R}\right) \right]  d\theta,
\end{equation*}
being $\delta(x)$ and $\cos(x)$ functions of even parity, and $\sin(x)$ with odd parity. By symmetry, the $\sin$ therm vanishes, hence,  
\begin{equation*} 
     R  = \int_{-\varepsilon R}^{\varepsilon R} \varkappa(\omega) d\omega \int_{-\pi}^{\pi} \cos\theta\delta \left [ \theta - \psi - \arcsin\left( \frac{\omega}{\varepsilon R}\right) \right]  d\theta
\end{equation*}
solving the integral in $\theta$, using the filtering property of the $\delta$, and translating the system to $\theta=\theta-\psi$
\begin{equation*} 
     R  = \int_{-\varepsilon R}^{\varepsilon R} \varkappa(\omega) d\omega \cos{ \left(\arcsin\left( \frac{\omega}{\varepsilon R}\right)  \right)},
\end{equation*}
and replacing
\begin{equation*}
    \begin{split}
    \omega & = \varepsilon R \sin \theta, \;\; \Rightarrow
    d\omega = \varepsilon R \cos \theta d\theta, \\
    \mathrm{If} & \;\; \omega = -\varepsilon R \rightarrow \theta = -\frac{\pi}{2}, \\
    \mathrm{If} & \;\; \omega = +\varepsilon R \rightarrow \theta = +\frac{\pi}{2}, \\
    \end{split}
\end{equation*}
the integral
\begin{equation} \label{rrr}
    R = \varepsilon R \int_{-\frac{\pi}{2}}^{\frac{\pi}{2}} \cos{^2\theta} \varkappa(\varepsilon R \sin \theta) d\theta, \;\;\;\; 0<R<1.
\end{equation}

Considering the critical coupling $\varepsilon^*$ in which
\begin{equation*}
    \lim_{R \rightarrow 0^+} \varkappa(\varepsilon R \sin \theta) = \varkappa(0),
\end{equation*}
hence, 
\begin{equation*}
    1 = \varepsilon^* \varkappa(0) \underbrace{\int_{-\frac{\pi}{2}}^{\frac{\pi}{2}} \cos{^2 \theta} d\theta}_{= \; \pi/2},
\end{equation*}
that is, there is an expression for the critical coupling that limits the region of desynchronization $ 0\leq \varepsilon \leq \varepsilon^*$. Region which depends only on the probability density of natural frequencies, which it is assumed to be unimodal and symmetrical
\begin{equation} \label{critico_1}
    \varepsilon^* = \frac{2}{\pi \varkappa(0)}.
\end{equation}

For $\varepsilon\gtrsim\varepsilon^*$, close to the synchronization transition point, using Eq. (\ref{rrr})
\begin{equation*}
    1 = \varepsilon \int_{-\frac{\pi}{2}}^{\frac{\pi}{2}}\cos{^2\theta} \varkappa(\varepsilon R \sin \theta) d\theta,
\end{equation*}
expanding the Taylor function $\varkappa(\omega)$ centered in $0$ 
\begin{equation*}
    \varkappa(\omega) = \varkappa(0) + \varkappa'(0)(\omega) + \frac{\varkappa''(0)}{2}(\omega^2) + \cdots,
\end{equation*}
it is know that $\varkappa(\omega)=\varkappa(-\omega)$, hence, the first derivative $\varkappa'(0)=0$, and disregarding the cubic therms, the integral
\begin{equation*}
 \begin{split}
    1 = \varepsilon \int_{-\frac{\pi}{2}}^{\frac{\pi}{2}} \cos{^2\theta} \left[\varkappa(0) + \frac{\varkappa''(0)}{2}(\varepsilon R \sin\theta)^2 \right] d\theta & = \varepsilon \varkappa(0)\overbrace{\int_{-\frac{\pi}{2}}^{\frac{\pi}{2}} \cos{^2\theta} d\theta}^{=\; \pi/2} \; + \\ & + \frac{\varepsilon^3R^2\varkappa''(0)}{2}\overbrace{\int_{-\frac{\pi}{2}}^{\frac{\pi}{2}} \cos{^2\theta}\sin{^2\theta} d\theta}^{=\;\pi/8},
    \end{split}
\end{equation*}
soon,
\begin{equation*}
    1 = \varepsilon \frac{\pi \varkappa(0)}{2} + \frac{\pi\varepsilon^3R^2\varkappa''(0)}{16}, 
\end{equation*}
and replacing Eq. (\ref{critico_1})
\begin{equation*}
    1 = \frac{\varepsilon}{\varepsilon^*} + \frac{\pi\varepsilon^3R^2\varkappa''(0)}{16} \Rightarrow \varepsilon^* = \varepsilon + \frac{\pi\varepsilon^3\varepsilon^* R^2\varkappa''(0)}{16} \Rightarrow \varepsilon-\varepsilon^* = -\frac{\pi\varepsilon^3\varepsilon^* R^2\varkappa''(0)}{16}.
\end{equation*}
Isolating $R$, finally the expression for the modulus of the order parameter
\begin{equation} \label{critico_2}
    R = \sqrt{\frac{-16}{\pi\varepsilon^2\varepsilon^*\varkappa''(0)}\left(1-\frac{\varepsilon^*}{\varepsilon}\right)},
\end{equation}
and since it is an approximation for $\varepsilon \gtrsim \varepsilon^*$, it is written the first therm as a $\varepsilon^*$ dependence only \cite{acebron2005kuramoto}
\begin{equation} \label{critico_3}
     R = \sqrt{\frac{-16}{\pi(\varepsilon^*)^{3}\varkappa''(0)}\left(1-\frac{\varepsilon^*}{\varepsilon}\right)}.
\end{equation}

\end{document}